\documentclass[aps,prx,reprint,superscriptaddress,longbibliography]{revtex4-2}

\usepackage{amssymb,amsfonts,mathtools}
\usepackage{bm}
\usepackage[normalem]{ulem}
\usepackage{graphicx}
\graphicspath{{figures/}{./}}
\usepackage{tikz}
\usepackage{braket}
\usepackage{xcolor}
\usepackage{microtype}
\usepackage{float}
\usepackage{hyperref}

\newcommand{\dd}{\mathrm d}

\newcommand{\parhead}[1]{\textbf{\textit{#1}}---}

\begin{document}

% Suppress main-paper entries from the Supplemental TOC
\let\savedaddcontentsline\addcontentsline
\renewcommand{\addcontentsline}[3]{}
	
	\title{Quantum-Geometric Length Scale for Long-distance Squeezing in Bosonic Bogoliubov Systems}
	
	\author{Sonu Verma}
	\email{sonu.vermaiitk@gmail.com}
	\affiliation{Institute for High Pressure, Department of Physics, Hanyang University, Seoul 04763, Republic of Korea }
	
	\author{Sangmo Cheon}
    \email{sangmocheon@hanyang.ac.kr}
    \affiliation{Department of Physics, Hanyang University, Seoul 04763, Korea}
    \affiliation{High Pressure Research Center, Hanyang University, Seoul 04763, Korea}
    \affiliation{Hanyang Institute for Quantum Science and Quantum Technology, Hanyang University, Seoul 04763, Korea}
	
	\author{Myung-Joong Hwang}
	\email{myungjoong.hwang@duke.edu}
	\affiliation{Division of Natural and Applied Sciences, Duke Kunshan University, Kunshan, Jiangsu 215300, China}

	\author{Moon Jip Park}
	\email{moonjippark@hanyang.ac.kr}
	\affiliation{Department of Physics, Hanyang University, Seoul 04763, Korea}

\begin{abstract}
Multimode squeezing is a key resource for continuous-variable quantum technologies, but its spatial range in bosonic lattices is usually tied to dispersive propagation. Here we show that parametric pairing can create an exactly flat Bogoliubov band spanned by compact Bogoliubov generators, while producing phase-sensitive anomalous correlations and sub-vacuum collective-mode squeezing that extend far beyond their finite support. Each compact generator mixes annihilation and creation operators, thereby
encoding the Bogoliubov squeezing structure, and neighboring translated
generators can have nonzero commutator overlap. Enforcing canonical
bosonic commutation relations therefore requires spatially extended
linear combinations of these translated compact generators, which define
the canonical Bogoliubov modes. The squeezing transformation of these
modes varies with momentum and is quantified by the squeezing-sector
symplectic quantum metric. A complex-momentum singularity of the analytically continued canonical
Bogoliubov modes sets both the anomalous-correlation decay length and the
momentum-space width of this metric, defining an intrinsic quantum-geometric length scale. This singularity can be tuned continuously while preserving exact flatness,
thereby controlling the spatial range of correlations and squeezing. Away from exact flatness, long-distance correlations persist through
multiple decay channels, while modes localized by defects and dimerized
boundaries provide complementary probes of the underlying quantum geometry.
In the weak-damping limit, the same quantum-geometric length scale can be
extracted from frequency-filtered two-port correlations. Our results establish the quantum geometry of canonical Bogoliubov modes as a mechanism for spatially extended quantum resources arising from compact Bogoliubov generators.
\end{abstract}

\maketitle

\begin{figure}[t]
	\centering
	\includegraphics[width=\columnwidth]{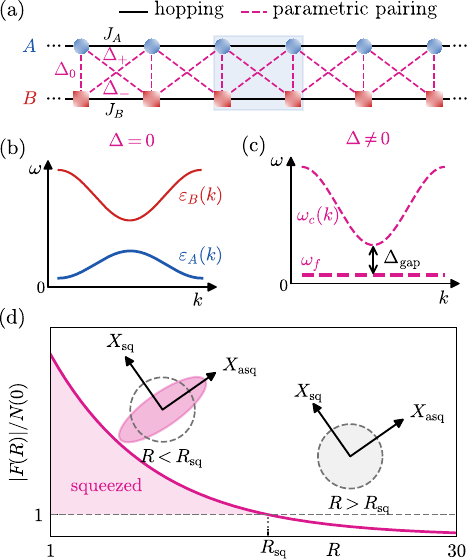}
\caption{\textbf{Pairing-induced Bogoliubov flat band and long-distance
squeezing.}
(a) Bosonic lattice with modes $A$ and $B$ in each unit cell.
Nearest-neighbor hoppings $J_A$ and $J_B$ act within the $A$ and $B$
sublattices, while onsite and intercell parametric pairings $\Delta_0$ and
$\Delta_{\pm}$ couple the two sublattices. The shaded region marks the finite
support of a compact Bogoliubov generator.
(b) Dispersive bands $\varepsilon_A(k)$ and
$\varepsilon_B(k)$ in the absence of pairing.
(c) For the pairing choice of Eq.~\eqref{eq:model}, an exactly flat
Bogoliubov band appears at $\omega_f$, separated by $\Delta_{\rm gap}$ from
the remaining positive-frequency dispersive band $\omega_c(k)$. The
corresponding negative-frequency Nambu partners are omitted.
(d) For a collective mode combining the $A$ mode in cell $0$
and the $B$ mode in cell $R$,
$\hat X_{\rm sq}$ and $\hat X_{\rm asq}$ denote the optimized squeezed
and antisqueezed quadratures. Here
$F(R)=\langle\delta\hat a_{A,0}\delta\hat a_{B,R}\rangle_G$ is the
anomalous correlation and $N(0)$ is the local occupation.
Sub-vacuum squeezing occurs when $|F(R)|>N(0)$.
The squeezing range $R_{\rm sq}$ is defined by
$|F(R_{\rm sq})|=N(0)$ and can extend far beyond the support of the
compact Bogoliubov generator.}
	\label{fig:concept}
\end{figure}

\parhead{Introduction.}
Distributing phase-sensitive correlations across bosonic lattices is central to continuous-variable quantum information~\cite{Weedbrook2012}. Multimode Gaussian correlations have recently been realized in acoustic, microwave, mechanical, and optical platforms~\cite{Andersson2022,Cai2017ReconfigurableGraph,Jia2025IntegratedMicrocomb,Jia2026MonolithicCluster,Eichler2011TwoModeMicrowave,Flurin2012EntangledMicrowave,Barzanjeh2019StationaryRadiation,Esposito2022TravelingWaveSqueezing,OckeloenKorppi2018MechanicalEntanglement,Kalash2026RealtimeMultimode}. Such correlations can be controlled through engineered dissipation~\cite{Kronwald2013DissipativeSqueezing,Pontula2024}, driven quadrature lattices~\cite{Lustig2025QuadratureMicrocombs}, and parametrically driven Bogoliubov dynamics~\cite{Peano2016,McDonald2018BosonicKitaev,Wanjura2023QuadratureNonreciprocity,Busnaina2024,Slim2024}. In many existing approaches, however, extending phase-sensitive correlations over large separations relies on dispersive propagation or parametric amplification. Whether such correlations can extend over large separations without dispersive propagation and thereby support sub-vacuum collective-mode squeezing remains an open question.

An exactly flat Bogoliubov band spanned by compact Bogoliubov generators
provides a natural setting to separate spatial correlation range from
dispersive propagation. Its vanishing group velocity rules out dispersive
transport, while the finite support of the generators provides no obvious
source of a long spatial scale. Parametric pairing makes each compact
generator a superposition of annihilation and creation operators. Because bosonic Bogoliubov modes must preserve the canonical commutation
relations through a symplectic normalization~\cite{Colpa1978,Lieu2018,
ChaudharyLevinClerk2021}, distinct translated generators with nonzero
commutator overlap do not themselves satisfy the canonical bosonic
commutation relations. Constructing canonical Bogoliubov modes that do
satisfy these relations therefore requires spatially extended
superpositions of these generators. Correspondingly, the canonical
Bogoliubov modes acquire a nontrivial momentum dependence characterized
by the symplectic quantum geometric tensor~\cite{Tesfaye2025}.

Flat-band systems without pairing can host compact localized states while
their Bloch eigenstates retain nontrivial momentum dependence
~\cite{Leykam2018,Maimaiti2017CLSGenerators,
Ramachandran2017ChiralFlatBands,Maimaiti2019UniversalGenerator,
Rhim2020QuantumDistance}. This momentum dependence gives rise to nontrivial
quantum geometry, which can govern superfluid weight and collective
response~\cite{Peotta2015,LiangEtAl2017,Julku2021,LiEtAl2025,
HuhtinenEtAl2026} as well as characteristic localization scales
~\cite{Oh2022BulkInterface,Kim2026,Lee2026EmbeddingIndependent,
Zhao2026,Ma2025}. In conventional paired systems, geometry inherited from the underlying normal-state bands can also set characteristic spatial scales, including Majorana localization and superconducting pair or coherence lengths~\cite{Guo2025,EldenIskin2026}. A bosonic Bogoliubov flat band is different because parametric pairing directly modifies the canonical Bogoliubov modes and thereby tunes their quantum geometry. The resulting pairing-induced quantum geometry has not previously been connected to an observable spatial decay length for anomalous correlations and sub-vacuum
collective-mode squeezing while preserving exact flatness and stability.

Here we establish this connection by showing that the quantum geometry of a pairing-induced bosonic Bogoliubov flat band defines an intrinsic decay length for anomalous correlations and determines the spatial range of sub-vacuum collective-mode squeezing. We demonstrate this mechanism in a two-sublattice bosonic lattice, where parametric pairing creates an isolated exactly flat Bogoliubov band from bands that are dispersive in the absence of pairing. We further show that the same length scale controls the momentum-space
concentration of the squeezing-sector symplectic quantum metric. We then examine the persistence of the long-distance correlations beyond exact flatness, where weak dispersion produces multiple decay channels, and
separately study how the quantum-geometric scale appears in modes localized by defects and dimerized boundaries. We finally show that, in the weak-damping limit, this quantum-geometric length scale can be extracted from frequency-filtered two-port correlations.

\parhead{Pairing-induced Bogoliubov flat band.}
We consider a one-dimensional bosonic lattice with two modes, $A$ and $B$,
per unit cell [Fig.~\ref{fig:concept}(a)]. The number-conserving sector $K$ contains onsite energies and hopping terms,  while the parametric-pairing terms
$\Delta$ create and annihilate pairs of excitations, thereby mixing annihilation and creation operators. After Fourier transformation, we use the Nambu basis
$\hat\Psi_k=(\hat a_{A,k},\hat a_{B,k},
\hat a_{A,-k}^{\dagger},\hat a_{B,-k}^{\dagger})^T$.
The Heisenberg equation is
$i\partial_t\hat\Psi_k=M(k)\hat\Psi_k$, with
$M(k)=\Sigma_zH_{\rm BdG}(k)$, where $H_{\rm BdG}(k)$ is the bosonic
Bogoliubov-de Gennes (BdG) Hamiltonian and
$\Sigma_z=\operatorname{diag}(I,-I)$, reflecting the opposite bosonic
commutator signs of the annihilation and creation sectors. Here $H_{\rm BdG}(k)$ contains the number-conserving block $K(k)$ and pairing block $\Delta(k)$. Bogoliubov modes are linear combinations of
annihilation and creation operators whose amplitudes $W_n(k)$ satisfy
$M(k)W_n(k)=\omega_n(k)W_n(k)$. Positive-frequency modes have
$W_n^\dagger(k)\Sigma_zW_n(k)>0$, with corresponding negative-frequency
Nambu partners~\cite{Colpa1978,Lieu2018}.

To construct an exactly flat Bogoliubov band, we choose the number-conserving block $K(k)$ and pairing block $\Delta(k)$ to be
\begin{equation}
	K(k)=
	\begin{pmatrix}
		\omega_f+|p_k|^2 & 0\\
		0 & |s_k|^2-\omega_f
	\end{pmatrix},
	~
	\Delta(k)=\eta
	\begin{pmatrix}
		0 & q_k\\
		q_k^* & 0
	\end{pmatrix},
	\label{eq:model}
\end{equation}
where $p_k=\sqrt{g}(\lambda_0+\lambda_1e^{ik})$,
$s_k=\sqrt{g}(c_0+c_1e^{-ik})$, and $q_k=p_ks_k$, with real parameters $\lambda_0,\lambda_1,c_0,c_1$. At $\eta=0$, the pairing block vanishes while $K(k)$ is unchanged,
giving the dispersive bands in Fig.~\ref{fig:concept}(b).
At $\eta=1$, Eq.~\eqref{eq:model} describes onsite energies and nearest-neighbor
hopping within each sublattice, together with onsite and nearest-neighbor
parametric pairings between the sublattices, as illustrated in
Fig.~\ref{fig:concept}(a). The explicit real-space Hamiltonian and couplings are given in Sec.~S.2.1 of the Supplemental
Material~\cite{SM}.
At $\eta=1$, the factorized pairing $q_k=p_ks_k$ yields the exact positive-frequency
eigenpairs
\begin{equation}
	\begin{aligned}
		\omega_f,\qquad
		&W_f(k)=(s_k,0,0,-p_k^*)^T,\\
		S(k)-\omega_f,\qquad
		&W_{\rm c}(k)=(0,s_k^*,-p_k,0)^T,
	\end{aligned}
	\label{eq:exact_eigenpairs}
\end{equation}
where $S(k)=|s_k|^2-|p_k|^2=S_0+2S_1\cos k$, with
$S_0=g(c_0^2+c_1^2-\lambda_0^2-\lambda_1^2)$ and
$S_1=g(c_0c_1-\lambda_0\lambda_1)$.
The first eigenpair forms an exactly flat Bogoliubov band at $\omega_f$,
while the other positive-frequency band remains dispersive, as shown in
Fig.~\ref{fig:concept}(c). Both eigenvectors have bosonic norm
$W_{f,c}^\dagger(k)\Sigma_zW_{f,c}(k)=S(k)$.
The conditions $\omega_f>0$ and $\min_kS(k)>2\omega_f$ ensure energetic
stability and isolate the flat band by the gap
$\Delta_{\rm gap}=\min_k[S(k)-2\omega_f]$.  A simpler choice $\lambda_0=\lambda_1=\lambda$, $c_0=c$, and $c_1=1$ already captures the essential mechanism.

\begin{figure*}[t]
	\centering
	\includegraphics[width=\textwidth]{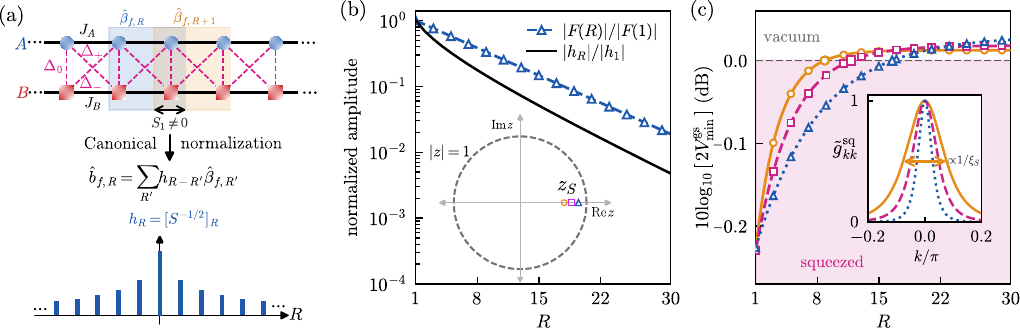}
\caption{\textbf{Extended canonical modes, anomalous correlations, and
ground-state squeezing.}
(a) Nearest-neighbor hoppings $J_A,J_B$ and onsite/intercell parametric
pairings $\Delta_0,\Delta_{\pm}$ generate compact Bogoliubov generators
$\hat\beta_{f,R}$. Neighboring generators have nonzero commutator overlap,
$[\hat\beta_{f,R},\hat\beta_{f,R+1}^{\dagger}]=S_1\neq0$.
Imposing canonical bosonic commutation relations gives
$\hat b_{f,R}=\sum_{R'}h_{R-R'}\hat\beta_{f,R'}$, where $h_R$ is the
Fourier coefficient of $S^{-1/2}(k)$, producing spatially extended
canonical Bogoliubov modes.
(b) For $c_1=1.77$, the normalized canonical-mode kernel
$|h_R|/|h_1|$ and anomalous correlation $|F(R)|/|F(1)|$ share the
exponential decay length $\xi_S=-a/\ln|z_S|$.
The inset shows the zeros $z_S$ of $S(z)$ inside the unit circle for
the three parameter sets used in (c).
(c) Optimized ground-state collective-quadrature noise
$10\log_{10}[2V_{\min}^{\rm gs}(R)]$. Negative values indicate
sub-vacuum squeezing. The inset shows the normalized squeezing-sector
symplectic quantum metric
$\tilde g_{kk}^{\rm sq}=g_{kk}^{\rm sq}/\max_k g_{kk}^{\rm sq}$,
whose momentum half-width
$\Delta k\simeq\sqrt{\sqrt{2}-1}\,a/\xi_S$ decreases as $\xi_S$
increases. The squeezing range $R_{\rm sq}$ is defined by
$|F(R_{\rm sq})|=N(0)$, where the optimized quadrature noise reaches
the vacuum level. Parameters are $\omega_f=1$, $g=32.5$,
$\lambda_0=0.017$, $\lambda_1=0.043$, $c_0=-2.037$, and
$c_1=1.40,1.60,1.77$ (orange, magenta, blue), giving
$\xi_S/a=2.68,4.18,7.30$, respectively.}
	\label{fig:pole}
\end{figure*}

\parhead{Compact generators and extended canonical Bogoliubov modes.}
For the isolated flat band, the Bloch eigenvector $W_f(k)$ spans the one-dimensional flat-band eigenspace at each momentum $k$. Since $W_f(k)=W_0+e^{-ik}W_1$ contains only two Fourier harmonics, its Fourier transform defines a compact Bogoliubov generator supported on two neighboring unit cells,
\begin{equation}
	\hat{\beta}_{f,R}
	=
	\sqrt g\left(
	c_0\hat a_{A,R}+c_1\hat a_{A,R+1}
	+\lambda_0\hat a_{B,R}^{\dagger}
	+\lambda_1\hat a_{B,R+1}^{\dagger}
	\right).
	\label{eq:compact_real_space_mode}
\end{equation}
The mixing of annihilation and creation operators in
$\hat{\beta}_{f,R}$ directly encodes the squeezing structure induced by
parametric pairing. Translations of this generator span the flat-band
subspace in real space. However, they are not themselves canonical
bosonic modes. Canonical Bogoliubov modes satisfy
$[\hat b_{f,R},\hat b_{f,R'}^\dagger]=\delta_{R,R'}$, whereas the compact
generators obey
$[\hat\beta_{f,R},\hat\beta_{f,R'}^\dagger]
=
S_0\delta_{R,R'}
+
S_1\!\left(\delta_{R,R'+1}+\delta_{R,R'-1}\right)$.
For $S_1\neq0$, neighboring compact generators therefore have a nonzero
commutator overlap. The corresponding canonical Bogoliubov modes have momentum-space profile $w_f(k)=W_f(k)/\sqrt{S(k)}$, or equivalently are represented in real space by
\begin{equation}
	\hat b_{f,R}
	=
	\sum_{R'}h_{R-R'}\hat\beta_{f,R'},
	\qquad
	[\hat b_{f,R},\hat b_{f,R''}^{\dagger}]
	=
	\delta_{R,R''},
	\label{eq:canonicalization}
\end{equation}
where $h_R$ is the Fourier coefficient of $S^{-1/2}(k)$. For $S_1\neq0$, these coefficients are nonzero at arbitrarily large separations, so the canonical Bogoliubov modes are spatially extended even though the compact generators have support on only two neighboring unit cells. The exact real-space form and asymptotic decay of the canonical modes are given in Sec.~S.2.2 of the Supplemental Material~\cite{SM}. As shown next, the same momentum-dependent commutator overlap sets the decay of anomalous correlations and thereby the spatial range of sub-vacuum collective-mode squeezing.

\parhead{Anomalous correlations and long-distance squeezing.}
The spatially extended canonical modes directly determine the
Bogoliubov ground-state covariance.
We denote the connected normal and anomalous correlations by
$N_{\alpha\beta}(k)=
\langle\delta\hat a^\dagger_{\alpha,k}\delta\hat a_{\beta,k}\rangle_G$
and
$F_{\alpha\beta}(k)=
\langle\delta\hat a_{\alpha,k}\delta\hat a_{\beta,-k}\rangle_G$,
respectively, where
$\delta\hat a=\hat a-\langle\hat a\rangle_G$ and
$\langle\cdots\rangle_G$ denotes the Bogoliubov ground-state expectation
value.
For the present model, the relevant components are
$N_{AA}(k)=N_{BB}(-k)=|p_k|^2/S(k)$ and
$F_{AB}(k)=-q_k/S(k)$, with the complementary positive-frequency band
making no contribution to the $AB$ anomalous channel, so $F_{AB}(k)$ is
exactly the flat-band contribution, as detailed in Secs.~S.2.3 and A.2
of the Supplemental Material~\cite{SM}.
Therefore, the anomalous correlation between mode $A$ in cell $0$
and mode $B$ in cell $R$ is
\begin{equation}
	F(R)\equiv
	\langle\delta\hat a_{A,0}\delta\hat a_{B,R}\rangle_G
	=
	-\int_{-\pi}^{\pi}\frac{dk}{2\pi}\,
	e^{-ikR}\frac{q_k}{S(k)} .
	\label{eq:anomalous_correlator}
\end{equation}
Although $q_k$ contains only finitely many Fourier harmonics,
the factor $S^{-1}(k)$ produces correlations that extend far beyond the
two-cell support of the compact generators
[Fig.~\ref{fig:concept}(d)].

Writing $z=e^{-ik}$, the zero $z_S$ of $S(z)$ inside the unit circle defines the intrinsic decay length
$\xi_S=-a/\ln|z_S|$ [Fig.~\ref{fig:pole}(b)].
For the parameter sets used in Fig.~\ref{fig:pole}, the residue of the
$(A,0)$--$(B,R)$ anomalous channel is nonzero, so
$|F(R)|\propto|z_S|^{R-1}$ for $R\geq1$ and hence
$\xi_F=\xi_S$, with details in Sec.~S.2.4 of the Supplemental
Material~\cite{SM}.
For $S(k)=S_0+2S_1\cos k$, this gives
$\xi_S/a=[\operatorname{arcosh}(S_0/2|S_1|)]^{-1}$.
The same complex-momentum point is a singularity of the flat-band
Bogoliubov projector, as shown in Sec.~A.1.1 of the Supplemental
Material~\cite{SM}, and controls the momentum-space concentration of the
squeezing-sector symplectic quantum metric discussed below. We therefore
identify $\xi_S$ as an intrinsic quantum-geometric length scale.
Varying the couplings within the exactly flat family shifts $z_S$ and
changes the correlation range while preserving exact flatness. The long-distance tail thus originates from the spatially extended canonical Bogoliubov modes
rather than from dispersive propagation within the flat band. Stability and spectral isolation bound how closely $z_S$ can approach the unit circle.

To establish the nonclassical consequence of these anomalous correlations, we consider the collective mode
$\delta\hat d_R(\phi)=
[\delta\hat a_{A,0}
+e^{-i\phi}\delta\hat a_{B,R}]/\sqrt2$
with an arbitrary relative phase $\phi$, and define its quadrature
$\hat X_R(\theta,\phi)=
[e^{-i\theta}\delta\hat d_R(\phi)
+e^{i\theta}\delta\hat d_R^\dagger(\phi)]/\sqrt2$.
For any $\phi$, minimizing its variance over the quadrature angle $\theta$ gives
\begin{equation}
	V_{\min}^{\rm gs}(R)
	\equiv
	\min_{\theta}
	\langle\hat X_R^2(\theta,\phi)\rangle_G
	=
	\frac{1}{2}+N(0)-|F(R)|,
	\label{eq:min_variance}
\end{equation}
where
$N(0)=
\langle\delta\hat a_{A,0}^{\dagger}\delta\hat a_{A,0}\rangle_G
=
\langle\delta\hat a_{B,0}^{\dagger}\delta\hat a_{B,0}\rangle_G$ is the local occupation. Thus $V_{\min}^{\rm gs}(R)<1/2$, equivalently $|F(R)|>N(0)$, certifies sub-vacuum squeezing of a collective mode formed from the spatially separated modes. In Fig.~\ref{fig:pole}(c), this condition persists far beyond the support of the compact Bogoliubov generators, demonstrating long-distance squeezing. Using the exact anomalous-correlation tail, the threshold distance for nonlocal sub-vacuum squeezing is
$R_{\rm sq}
=
1+\frac{\xi_S}{a}
\ln\!\left[\frac{|F(1)|}{N(0)}\right]$,
with squeezing occurring at lattice separations $R<R_{\rm sq}$.
For a fixed onset ratio $|F(1)|/N(0)$, the squeezing range grows linearly
with the intrinsic quantum-geometric length $\xi_S$, while its dependence
on the onset ratio is logarithmic. Exact real-space covariances, the long-distance squeezing criterion, and the general residue condition are given in Secs.~S.2.4, S.3, and A.2, respectively, of the Supplemental Material~\cite{SM}.

\parhead{Quantum geometry of the canonical Bogoliubov modes.}
We now connect the real-space decay length $\xi_S$ to the quantum geometry of the canonical Bogoliubov modes.
The symplectic quantum metric decomposes the momentum variation of a Bogoliubov mode into positive- and negative-frequency sectors~\cite{Tesfaye2025}. We focus on the positive magnitude of the negative-frequency contribution because it measures changes in the mixing between the annihilation and creation components of the Bogoliubov mode and therefore directly quantifies the momentum variation of its squeezing transformation. For the canonical flat-band mode $w_f(k)$, this squeezing-sector symplectic metric is
\begin{equation}
	g_{kk}^{\rm sq}(k)
	=
	\sum_{m<0}
	\left|
	w_m^\dagger(k)\Sigma_z\partial_k w_f(k)
	\right|^2
	=
	\frac{g^2(c_0\lambda_1-\lambda_0 c_1)^2}{S^2(k)},
	\label{eq:squeezing_metric}
\end{equation}
where the sum runs over the negative-frequency Bogoliubov modes. The relative weight and phase of the annihilation and creation components can equivalently be described by the complex two-mode squeezing parameter
$\zeta_k=-p_k/s_k^*=e^{i\phi_k}\tanh r_k$, where $r_k$ and $\phi_k$ specify the squeezing amplitude and phase, respectively. A momentum-dependent $\zeta_k$ therefore means that the squeezing transformation itself varies
across the Brillouin zone, and $g_{kk}^{\rm sq}(k)$ quantifies this variation. The squeezing-sector symplectic quantum metric and the vanishing transverse
positive-frequency contribution for the present two-band model are derived in Sec.~A.3 of the Supplemental Material~\cite{SM}. When $c_0\lambda_1=\lambda_0c_1$, $\zeta_k$ becomes momentum independent and $g_{kk}^{\rm sq}=0$, although the squeezing amplitude can remain finite. Thus it is the momentum-dependent squeezing geometry encoded in $g_{kk}^{\rm sq}(k)$, rather than the squeezing magnitude alone, that is linked to the long-distance anomalous correlations and collective-mode
squeezing.

This connection follows because the anomalous covariance and $g_{kk}^{\rm sq}(k)$ contain $S^{-1}(k)$ and $S^{-2}(k)$, respectively. The same zero $z_S$ of $S(z)$ that controls the anomalous-correlation tail therefore also controls the momentum-space concentration of the metric. As $|z_S|$ approaches the unit circle, the metric becomes more sharply concentrated in momentum space, with asymptotic half-width 
$\Delta k\simeq\sqrt{\sqrt{2}-1}\,a/\xi_S$
[Fig.~\ref{fig:pole}(c), inset].
Thus the real-space correlation decay and the momentum-space symplectic metric width are governed by the same quantum-geometric length scale $\xi_S$. The Brillouin-zone-integrated squeezing-sector metric, by contrast,
depends on the full momentum-space profile. Its relation to $\xi_S$,
together with the real-space decay of the squeezing-sector metric, is
given in Secs.~A.3.1 and A.3.2 of the Supplemental
Material~\cite{SM}.

\parhead{Persistence beyond exact flatness with defect and boundary probes.}
The exact flat-band construction relies on constrained couplings, so we next examine whether the long-distance correlations underlying the squeezing persist when exact flatness is weakly broken. A weak translationally invariant hopping $t_f$ gives the target band a
small dispersion without eliminating the long-distance anomalous correlations. The single flat-band decay structure associated with $z_S$ then splits into two nearby interior roots $z_{1,2}$, giving rise to two distinct spatial decay channels with characteristic lengths set by their moduli. For weak $t_f$, these roots remain close to $z_S$, and as $t_f\rightarrow0$ both continuously approach $z_S$, recovering the intrinsic length scale $\xi_S$ of the
exactly flat band. Thus weak dispersion preserves the long-distance correlations while resolving their decay into distinct channels, as detailed in Sec.~B.1 of the Supplemental Material~\cite{SM}.

Local defects and boundaries provide complementary probes of how the complex-momentum structure associated with $z_S$ and the intrinsic length scale $\xi_S$ appears in localized modes. A local defect, $\delta\hat H_d=v_d\hat a_{A,0}^{\dagger}\hat a_{A,0}$, can bind a mode at frequency $\omega_d$. At finite dispersion, when two interior roots $z_{d,\nu}$ of the
bulk resolvent at $\omega_d$ contribute, the long-distance defect
profile takes the form
\begin{equation}
\psi_{\alpha,R}
\simeq
\sum_{\nu=1,2}
\mathcal B_{\alpha,\nu}z_{d,\nu}^R,
\qquad
\xi_{d,\nu}=-\frac{a}{\ln|z_{d,\nu}|},
\label{eq:two_channels}
\end{equation}
where $\mathcal B_{\alpha,\nu}$ is the amplitude associated with each
root. Both decay channels contribute to a given component
$\psi_{\alpha,R}$ when the corresponding coefficients
$\mathcal B_{\alpha,1}$ and $\mathcal B_{\alpha,2}$ are nonzero, while
the root with the largest modulus among those with nonzero amplitude
controls the asymptotic decay. At finite dispersion and defect detuning,
these defect decay lengths evolve continuously from the flat-band
structure but need not equal $\xi_S$. At exact flatness, the defect root
approaches $z_S$ in the near-resonant limit, and correspondingly
$\xi_d\to\xi_S$. The role of compact-generator nonorthogonality in the localized profile
and the resulting defect decay channels are developed in Secs.~B.2 and
B.3 of the Supplemental Material~\cite{SM}.

The same geometric decay structure also appears in the separate
dimerized boundary construction. With an SSH-type coupling pattern between compact
generators~\cite{SSH1979}, the localized boundary mode contains
a dimerization-induced decay contribution together with a geometric
contribution inherited from the flat-band scale $\xi_S$. The former is set
by the boundary dimerization, whereas the latter carries the long-distance
quantum-geometric decay when its amplitude is nonzero. Thus defects and
dimerized boundaries select the spatial channels differently, but both can
retain the geometric decay structure that controls the bulk long-distance correlations. The geometric contribution to the boundary decay is derived in Sec.~B.4.3 of the Supplemental Material~\cite{SM}.

\begin{figure}[t]
	\centering
	\includegraphics[width=\columnwidth]{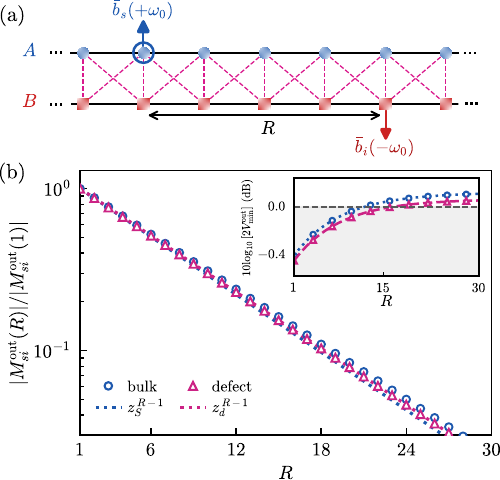}
\caption{\textbf{Frequency-filtered detection of long-distance squeezing.}
(a) The selected local output channels at $A,0$ and $B,R$ define the
conjugate filtered signal and idler modes $\bar b_s(+\omega_0)$ and $\bar b_i(-\omega_0)$,
where $s$ and $i$ denote signal and idler, respectively.
We take $\omega_0=\omega_f$ for the translation-invariant bulk response
and $\omega_0=\omega_d$ for the defect resonance.
(b) Normalized anomalous output correlation
$|M_{si}^{\rm out}(R)|/|M_{si}^{\rm out}(1)|$, where
$M_{si}^{\rm out}(R)=
\langle\delta\bar b_s\,\delta\bar b_i\rangle_{\rm ss}$.
Symbols show finite-bandwidth results, while dotted lines show the
closed-system envelopes $|z_S|^{R-1}$ for the bulk and $|z_d|^{R-1}$
for the defect, respectively.
At $t_f=0$, each response has a single exponential decay length in the
closed-system limit. The weak curvature of the bulk data arises from
the finite-linewidth output poles $S(z_\pm)=\pm i\kappa/2$, which
approach $z_S$ as $\kappa\to0$.
The inset shows the optimized output-quadrature noise
$10\log_{10}[2V_{\min}^{\rm out}(R)]$. Negative values indicate
sub-vacuum squeezing, which persists through $R=12$ in the bulk and
$R=16$ near the defect.
Parameters are as in Fig.~2 for $c_1=1.77$, with
$\kappa_{\rm ext}=0.02$, $\kappa_{\rm int}=0$,
$\sigma_f=0.05\kappa_{\rm ext}$, $v_d=-0.10$, $L=121$, and
$\omega_d\simeq0.900$.}
\label{fig:readout}
\end{figure}

\parhead{Frequency-filtered detection.}
To determine whether the intrinsic length scale $\xi_S$ can be accessed experimentally in an open device, we couple the lattice locally to Markovian output channels and select
the outputs at modes $A,0$ and $B,R$, filtered around the conjugate
sidebands $+\omega_0$ and $-\omega_0$ [Fig.~\ref{fig:readout}(a)]. We take $\omega_0=\omega_f$ for the
translation-invariant flat band and $\omega_0=\omega_d$ for a defect resonance. The resulting filtered modes, $\bar b_s$ and $\bar b_i$, with $s$ and $i$
denoting signal and idler, respectively, are characterized by the connected
moments
$M_{si}^{\rm out}(R)=
\langle\delta\bar b_s\,\delta\bar b_i\rangle_{\rm ss}$,
$N_{ss}=
\langle\delta\bar b_s^{\dagger}\delta\bar b_s\rangle_{\rm ss}$, and
$N_{ii}(R)=
\langle\delta\bar b_i^{\dagger}\delta\bar b_i\rangle_{\rm ss}$, where
$\delta\bar b_\mu=
\bar b_\mu-\langle\bar b_\mu\rangle_{\rm ss}$.
The input-output coupling and filtered-mode construction are given in
Secs.~C.1 and C.2.2 of the Supplemental Material~\cite{SM}. These moments probe a frequency-resolved steady-state covariance, rather than the equal-time ground-state covariance discussed above~\cite{Gardiner1985,Clerk2010}, and can be accessed through established two-mode microwave correlation
measurements~\cite{Eichler2011TwoModeMicrowave,Flurin2012EntangledMicrowave}.

At $t_f=0$, both the bulk and defect filtered responses reduce to a single exponential decay channel in the weak-damping limit [Fig.~\ref{fig:readout}(b)]. For the bulk response, finite linewidth shifts the output poles away from $z_S$, but as $\kappa\rightarrow0$ they continuously approach $z_S$ and the measured decay length approaches the intrinsic scale $\xi_S$. The frequency-filtered bulk correlation therefore provides a direct open-system probe of the same length scale that governs the ground-state anomalous correlations and squeezing. At the defect resonance $\omega_d$, the decay is instead governed by the defect root $z_d$, whose associated length remains continuously connected to the flat-band geometric scale discussed above. Weak dispersion can produce two decay contributions when both corresponding residues are nonzero. The exactly flat bulk response, its finite-linewidth pole structure, and the defect response are derived in Secs.~D.1, D.2, and D.4, respectively, of the Supplemental Material~\cite{SM}.

For the collective output mode
$\bar b_{\rm col}(\phi)=
[\bar b_s+e^{i\phi}\bar b_i]/\sqrt2$,
optimizing over the relative collection phase and the quadrature angle gives
\begin{equation}
	V_{\min}^{\rm out}(R)
	=
	\frac{1}{2}
	+\bar N(R)
	-\left|M_{si}^{\rm out}(R)\right|,
	\label{eq:Vminout}
\end{equation}
where $\bar N(R)=[N_{ss}+N_{ii}(R)]/2$ is the mean filtered occupation.
Thus $V_{\min}^{\rm out}(R)<1/2$ certifies sub-vacuum squeezing of a
collective mode formed from the two spatially separated ports.
For the translationally invariant bulk, where $N_{ss}=N_{ii}$, this
threshold also coincides with the positive-partial-transpose (PPT)
entanglement criterion for the selected filtered signal-idler pair.
The collective-mode construction, squeezing criterion, and general PPT
criterion are derived in Sec.~C.3 of the Supplemental Material~\cite{SM}.
In the weak-pairing regime, analytical expressions for the anomalous
output correlation, filtered occupation, and resulting squeezing are
given in Sec.~D.3 of the
Supplemental Material~\cite{SM}. A coherent probe can reveal the same spatial decay through its mean response, but does not by itself certify nonclassicality. The corresponding signal and phase-conjugate idler responses are given in Sec.~D.5 of the Supplemental Material~\cite{SM}.

\parhead{Discussion.}
The mechanism is not specific to the two-sublattice model considered here. More generally, a stable, isolated bosonic BdG flat band admitting finite-support compact Bogoliubov generators with momentum-dependent
commutator overlap can support spatially extended canonical Bogoliubov modes. The nearest nonremovable complex-momentum singularity of the flat-band Bogoliubov projector sets their asymptotic decay length, while
the anomalous covariance inherits this length only when the corresponding projected residue is nonzero. The same geometric scale can also appear in defect- and boundary-localized modes, while additional singularities with
nonzero channel weights generate additional decay lengths.

Compact localized states and projector localization are familiar from flat-band systems~\cite{Sathe2021CompactWannierProjectors,Kim2026}. The distinctive Bogoliubov feature is that parametric pairing can make the commutator overlap between translated compact generators momentum dependent.
Enforcing the canonical bosonic commutation relations then requires
spatially extended superpositions of the compact generators, so their finite
support does not determine the spatial range of the canonical Bogoliubov
modes and phase-sensitive correlations. The squeezing-sector symplectic quantum metric resolves the momentum
variation of the mixing between the annihilation and creation components
of the canonical Bogoliubov modes. This distinction is important because finite squeezing can persist when this mixing is
momentum independent, while the corresponding metric contribution vanishes and the anomalous covariance remains local or finite range. The intrinsic
quantum-geometric length scale therefore reflects the momentum-dependent Bogoliubov structure rather than the squeezing magnitude alone. The general construction relating compact generators, their commutator-overlap matrix, and canonical Bogoliubov modes is developed in Sec.~A.1 of the Supplemental Material~\cite{SM}. The complex-momentum projector singularity and residue
criterion, its inheritance by the anomalous covariance, the squeezing-sector symplectic quantum metric, and the common exponential-decay theorem are derived in Secs.~A.1.1, A.2, A.3, and A.4, respectively.

The closed- and open-system measurements access different manifestations of this same spatial structure. Equal-time covariance probes the anomalous correlations of the Bogoliubov ground state, while frequency-filtered
two-port measurements determine whether the corresponding length survives in a driven and damped device and whether the separated output modes exhibit
sub-vacuum squeezing. Their agreement in the weak-damping limit provides a direct way to
distinguish the intrinsic Bogoliubov length from a linewidth-dependent decay scale. The corresponding momentum-space geometry may be independently accessed
through periodic parameter modulation~\cite{Tesfaye2025}.

Programmable superconducting-resonator and optomechanical networks can implement the required hopping and parametric couplings
~\cite{Busnaina2024,Slim2024,Wanjura2023QuadratureNonreciprocity}, while nonlinear integrated photonic systems permit multimode covariance reconstruction~\cite{Jia2025IntegratedMicrocomb,Jia2026MonolithicCluster}. Magnonic systems provide another prospective setting, with
microwave-parametric single- and two-mode thermal magnon squeezing in YIG
films and quantum-level magnon squeezing in cavity-coupled YIG sphere already
demonstrated~\cite{Hioki2026MagnonSqueezing,Weng2026MagnonSqueezing}.
Realizing the present mechanism in such systems would additionally require
engineered hopping and parametric pair couplings. Extensions to higher-dimensional,
topological, and driven-dissipative systems will involve matrix-valued commutator-overlap kernels together with separate analyses of stability 
and channel weights. The momentum-dependent commutator overlap among compact Bogoliubov generators thus provides a mechanism for controlling the quantum-geometric decay length, phase-sensitive correlations, and long-distance collective-mode squeezing without dispersive propagation within the target band.

\begin{acknowledgments}
	This work was supported by the National Research Foundation of Korea
	(NRF) grant funded by the Korea government (MSIT) (Grants
	No.~RS-2025-25464760, RS-2026-25519864, RS-2025-25446099,
	RS-2023-NR119928, RS-2025-16070482, RS-2026-25607155, and RS-2025-03392969). This work was also supported
	by the BK21 FOUR (Fostering Outstanding Universities for Research)
	program through the National Research Foundation (NRF), funded by the
	Ministry of Education of Korea. S.V. acknowledges support from the Brain Pool Program funded by the Ministry of Science and ICT through the National Research Foundation of Korea (RS-2025-25446099).
	
\end{acknowledgments}

\section*{Data Availability}

The numerical data underlying the figures in this article are openly
available in Zenodo~\cite{VermaData2026}. The code used to generate
and analyze the data is available from the authors upon reasonable request.

\bibliography{references}
\clearpage
\clearpage
\onecolumngrid

\begin{center}
{\large\bfseries
Supplemental Material for
``Quantum-Geometric Length Scale for Long-distance Squeezing
in Bosonic Bogoliubov Systems''}

\vspace{0.6em}

Sonu Verma, Sangmo Cheon, Myung-Joong Hwang, and Moon Jip Park
\end{center}

\vspace{1em}

% ============================================================
% Supplemental numbering
% ============================================================

\setcounter{section}{0}
\setcounter{subsection}{0}
\setcounter{subsubsection}{0}
\setcounter{equation}{0}
\setcounter{figure}{0}
\setcounter{table}{0}

\setcounter{secnumdepth}{3}
\setcounter{tocdepth}{3}

\newcommand{\suppsectionlabel}{%
  \ifcase\value{section}%
  \or S%
  \or A%
  \or B%
  \or C%
  \or D%
  \else\arabic{section}%
  \fi%
}

\renewcommand{\thesection}{\suppsectionlabel}
\renewcommand{\thesubsection}{\thesection.\arabic{subsection}}
\renewcommand{\thesubsubsection}{%
  \thesubsection.\arabic{subsubsection}%
}

\numberwithin{equation}{section}
\renewcommand{\theequation}{\thesection\arabic{equation}}

\renewcommand{\thefigure}{S\arabic{figure}}
\renewcommand{\thetable}{S\arabic{table}}

% ============================================================
% Supplemental TOC formatting
% ============================================================

\makeatletter

\renewcommand{\p@subsection}{}
\renewcommand{\p@subsubsection}{}

\renewcommand*\l@section{%
  \@dottedtocline{1}{0em}{2.0em}%
}

\renewcommand*\l@subsection{%
  \@dottedtocline{2}{2.0em}{2.5em}%
}

\renewcommand*\l@subsubsection{%
  \@dottedtocline{3}{4.5em}{3.0em}%
}

\makeatother

% ============================================================
% Supplemental contents
% ============================================================

\begingroup
\tableofcontents
\endgroup

% Restore TOC writing only for Supplemental sections
\let\addcontentsline\savedaddcontentsline

\clearpage

\section{Exact flat-band model and ground-state squeezing}
\subsection{Bosonic BdG formalism and conventions}
\label{sec:bdg_formalism}

For bosonic modes $\hat a_i$, where $i=(R,\alpha)$ labels the unit
cell and the internal mode, a general finite-range quadratic
Hamiltonian is
\begin{equation}
\hat H
=
\sum_{i,j}
K_{ij}\hat a_i^\dagger\hat a_j
+
\frac{1}{2}
\sum_{i,j}
\left(
\Delta_{ij}\hat a_i^\dagger\hat a_j^\dagger
+
\Delta_{ij}^{*}\hat a_j\hat a_i
\right),
\label{eq:realSpaceHamiltonian}
\end{equation}
with $K=K^\dagger$ and $\Delta=\Delta^T$. The matrix $K$ contains
onsite energies and hopping amplitudes, while
$\Delta$ contains pair-creation and pair-annihilation processes. This real-space form will also serve as the device Hamiltonian for
the finite systems with defects and boundaries considered below,
before coupling them to external ports.

For a translationally invariant lattice with $N$ modes per unit cell, we
introduce
\[
\hat a_{\alpha,k}
=
L^{-1/2}\sum_R e^{-ikR}\hat a_{\alpha,R},
\qquad
\hat{\mathbf a}_k
=
(\hat a_{1,k},\ldots,\hat a_{N,k})^T,
\]
and the Nambu spinor
$\hat\Psi_k=(\hat{\mathbf a}_k,\hat{\mathbf a}_{-k}^{\dagger})^T$.
The Hamiltonian then takes the form
\begin{equation}
\hat H
=
\frac{1}{2}\sum_k
\hat\Psi_k^\dagger H_{\mathrm{BdG}}(k)\hat\Psi_k
+E_0,
\qquad
H_{\mathrm{BdG}}(k)
=
\begin{pmatrix}
K(k) & \Delta(k)\\
\Delta^\dagger(k) & K^T(-k)
\end{pmatrix},
\label{eq:BdGHamiltonian}
\end{equation}
where $H_{\mathrm{BdG}}(k)$ is the bosonic Bogoliubov-de Gennes
(BdG) Hamiltonian matrix, $K^\dagger(k)=K(k)$, and
$\Delta^T(-k)=\Delta(k)$. The $c$-number $E_0$ accounts for the
normal-ordering contribution introduced by the Nambu representation.

The Nambu commutator is
$[\hat\Psi_k,\hat\Psi_{k'}^\dagger]=\delta_{kk'}\Sigma_z$, with
$\Sigma_z=\operatorname{diag}(I_N,-I_N)$, where the opposite signs
reflect the bosonic commutators of the annihilation and creation sectors.
The linear Heisenberg dynamics of the Bogoliubov excitations are governed by
\begin{equation}
i\partial_t\hat\Psi_k=M(k)\hat\Psi_k,
\qquad
M(k)=\Sigma_zH_{\mathrm{BdG}}(k),
\label{eq:BdGDynamicalMatrix}
\end{equation}
where $M(k)$ is the bosonic BdG dynamical matrix.

Although $M(k)$ is generally non-Hermitian with respect to the ordinary
inner product, it obeys the bosonic pseudo-Hermiticity relation
$M^\dagger(k)\Sigma_z=\Sigma_zM(k)$. The doubled Nambu representation also
implies the redundancy
$H_{\mathrm{BdG}}(k)=\Sigma_xH_{\mathrm{BdG}}^*(-k)\Sigma_x$, or equivalently
$M(k)=-\Sigma_xM^*(-k)\Sigma_x$, where
$\Sigma_x=\begin{pmatrix}0&I_N\\ I_N&0\end{pmatrix}$ exchanges the
annihilation and creation sectors.

We impose energetic stability, meaning that the Hermitian BdG matrix
$H_{\mathrm{BdG}}(k)$ is positive definite for every real $k$,
$H_{\mathrm{BdG}}(k)>0$. Under this condition, the Bogoliubov excitation
frequencies are real and the eigenvectors form a complete basis that can be
normalized with respect to the bosonic metric $\Sigma_z$~\cite{Colpa1978}.
Collecting the normalized positive-frequency eigenvectors in $U_+(k)$ and
their negative-frequency creation partners
$U_-(k)=\Sigma_xU_+^*(-k)$, we define the Bogoliubov transformation
$\mathcal{T}(k)=(U_+(k),U_-(k))$. It satisfies
\begin{equation}
M(k)\mathcal{T}(k)
=
\mathcal{T}(k)
\begin{pmatrix}
\Omega(k)&0\\
0&-\Omega(-k)
\end{pmatrix},
\qquad
\mathcal{T}^\dagger(k)\Sigma_z\mathcal{T}(k)=\Sigma_z,
\label{eq:ParaunitaryTransformation}
\end{equation}
where $\Omega(k)=\operatorname{diag}[\omega_1(k),\ldots,\omega_N(k)]$
contains the positive excitation frequencies. The second relation makes
$\mathcal{T}(k)$ paraunitary and ensures that the transformed Bogoliubov
operators obey canonical bosonic commutation relations. The
negative-frequency modes are the creation partners required by the Nambu
representation rather than independent excitations.

For a normalized positive-frequency eigenvector
$w_n(k)=(u_n(k),v_n(k))^T$, with
$w_n^\dagger(k)\Sigma_z w_m(k)=\delta_{nm}$, the corresponding
canonical Bogoliubov annihilation operator is
$\hat b_{n,k}=w_n^\dagger(k)\Sigma_z\hat\Psi_k
=u_n^\dagger(k)\hat{\mathbf a}_k
-v_n^\dagger(k)\hat{\mathbf a}_{-k}^\dagger$.
The paraunitary normalization inherited from Eq.~\eqref{eq:ParaunitaryTransformation} ensures that the
$\hat b_{n,k}$ obey canonical bosonic commutation relations. The
Bogoliubov vacuum $|G\rangle$ is defined by
$\hat b_{n,k}|G\rangle=0$ for all positive-frequency modes. The
Hamiltonian then takes the diagonal form
$\hat H=E_{\rm vac}+\sum_{n,k}\omega_n(k)
\hat b_{n,k}^\dagger\hat b_{n,k}$, where
$E_{\rm vac}=E_0+\frac{1}{2}\sum_{n,k}\omega_n(k)$.
When $\Delta=0$, the positive-frequency modes contain only annihilation
operators. Finite parametric pairing mixes annihilation and creation
operators, so the Bogoliubov vacuum is a squeezed Gaussian state of the
physical bosonic modes.

For a pair of modes at opposite momenta, a canonical Bogoliubov mode can be written as
\begin{equation}
\hat b_k=\cosh r_k\,\hat a_{1,k}
-e^{i\phi_k}\sinh r_k\,\hat a^\dagger_{2,-k},
\qquad
\zeta_k=e^{i\phi_k}\tanh r_k .
\end{equation}
Here $r_k$ and $\phi_k$ determine the squeezing amplitude and phase. For the corresponding phase-matched collective mode
$\hat d_k=(\hat a_{1,k}+e^{-i\phi_k}\hat a_{2,-k})/\sqrt{2}$,
we define the quadrature
$\hat X_k(\theta)=[e^{-i\theta}\hat d_k+e^{i\theta}\hat d_k^\dagger]/\sqrt{2}$.
Minimizing and maximizing its ground-state variance
$\langle[\Delta\hat X_k(\theta)]^2\rangle_G$
over $\theta$ give the squeezed and antisqueezed variances
$e^{-2r_k}/2$ and $e^{2r_k}/2$, respectively, with the vacuum variance equal to $1/2$. Throughout the closed-system analysis, $\langle\cdots\rangle_G$ denotes expectation values in the Bogoliubov ground state $|G\rangle$, and $\delta\hat a=\hat a-\langle\hat a\rangle_G$. We denote the connected normal and anomalous correlations by $N_{\alpha\beta}(k)=\langle\delta\hat a^\dagger_{\alpha,k}\delta\hat a_{\beta,k}\rangle_G$ and $F_{\alpha\beta}(k)=\langle\delta\hat a_{\alpha,k}\delta\hat a_{\beta,-k}\rangle_G$.

Spatial decay is analyzed by analytically continuing the Bloch factor
$z=e^{-ik}$ away from the unit circle. For a finite Laurent matrix
$X(z)=\sum_n X_n z^n$, we define the reciprocal adjoint
$\widetilde X(z)=X^\dagger(1/z^*)=\sum_n X_n^\dagger z^{-n}$.
On the physical Brillouin zone, where $|z|=1$, this reduces to the
ordinary Hermitian adjoint of the Bloch matrix. Away from the unit
circle, the reciprocal adjoint preserves the relation between a
real-space coupling and its Hermitian-conjugate reverse process. The
bosonic metric $\Sigma_z$ is not included in this operation and is
inserted separately when forming bosonic inner products.

\subsection{Exact flat-band construction}
\label{sec:pairing_flat_band}
\subsubsection{Construction and exact spectrum}
\label{subsec:finite_range_construction}

The explicit model used in the main text is a one-dimensional lattice
with two bosonic modes, $A$ and $B$, per unit cell. In the Nambu basis
$(\hat a_{A,k},\hat a_{B,k},
\hat a_{A,-k}^{\dagger},\hat a_{B,-k}^{\dagger})^T$, we choose the
number-conserving block
$K(k)=\operatorname{diag}\!\bigl(\omega_f+|p_k|^2,\,
|s_k|^2-\omega_f\bigr)$
and the parametric-pairing block
$\Delta(k)=\eta\bigl(\begin{smallmatrix}0&q_k\\ q_k^*&0\end{smallmatrix}\bigr)$,
with $q_k=p_ks_k$,
$p_k=\sqrt g\,(\lambda_0+\lambda_1e^{ik})$, and
$s_k=\sqrt g\,(c_0+c_1e^{-ik})$.
Here $\omega_f>0$ is the target flat-band frequency, $g>0$ sets the coupling scale, and $\lambda_0,\lambda_1,c_0,c_1$ are real coefficients. For $\eta=0$, the pairing block vanishes while $K(k)$ is unchanged. At $\eta=1$, the factorization $q_k=p_ks_k$ yields the exactly flat Bogoliubov band derived below. We assume $q_k\not\equiv0$, so parametric pairing is present for $\eta\neq0$. For generic nonzero intercell coefficients, the two bands are dispersive before pairing is introduced. The simpler choice $\lambda_0=\lambda_1\equiv\lambda$, $c_0\equiv c$, and $c_1=1$ already captures the same flat-band mechanism and quantum-geometric decay scale.

The corresponding real-space Hamiltonian is
\begin{align}
\hat H_2={}&
\sum_R\left(
\epsilon_A\hat a_{A,R}^\dagger\hat a_{A,R}
+\epsilon_B\hat a_{B,R}^\dagger\hat a_{B,R}
\right)
\nonumber\\
&+\sum_R\left(
J_A\hat a_{A,R}^\dagger\hat a_{A,R+1}
+J_B\hat a_{B,R}^\dagger\hat a_{B,R+1}
+\mathrm{H.c.}
\right)
\nonumber\\
&+\eta\sum_R\left(
\Delta_0\hat a_{A,R}^\dagger\hat a_{B,R}^\dagger
+\Delta_+\hat a_{A,R}^\dagger\hat a_{B,R+1}^\dagger
+\Delta_-\hat a_{A,R}^\dagger\hat a_{B,R-1}^\dagger
+\mathrm{H.c.}
\right).
\label{eq:realSpaceModel}
\end{align}
The coefficients are
$\epsilon_A=\omega_f+g(\lambda_0^2+\lambda_1^2)$,
$\epsilon_B=g(c_0^2+c_1^2)-\omega_f$,
$J_A=g\lambda_0\lambda_1$,
$J_B=gc_0c_1$,
$\Delta_0=g(\lambda_0c_0+\lambda_1c_1)$,
$\Delta_+=g\lambda_1c_0$, and
$\Delta_-=g\lambda_0c_1$.
Thus the Hamiltonian contains only onsite and nearest-neighbor hopping and pairing terms.

At $\eta=1$, the factorization $q_k=p_ks_k$ makes the bosonic BdG dynamical matrix
exactly solvable. In the sector spanned by the $A$-annihilation and
$B$-creation components, acting with the dynamical matrix on
$(s_k,-p_k^*)^T$ cancels the
terms proportional to $|p_k|^2$ and $|s_k|^2$, leaving the
momentum-independent eigenvalue $\omega_f$. The complementary sector is
diagonalized analogously. The two positive-frequency eigenpairs are
\begin{align}
\omega_f,\qquad
W_f(k)&=(s_k,0,0,-p_k^*)^T,
\label{eq:compactFlatBandMode}\\
S(k)-\omega_f,\qquad
W_c(k)&=(0,s_k^*,-p_k,0)^T,
\label{eq:remainingPositiveMode}
\end{align}
where
$S(k)=|s_k|^2-|p_k|^2=S_0+2S_1\cos k$,
$S_0=g(c_0^2+c_1^2-\lambda_0^2-\lambda_1^2)$, and
$S_1=g(c_0c_1-\lambda_0\lambda_1)$.
Thus $W_f(k)$ forms an exactly flat positive-frequency Bogoliubov band,
whereas the other positive-frequency band is dispersive for $S_1\neq0$.
Both eigenvectors have bosonic norm
$W_{f,c}^\dagger(k)\Sigma_zW_{f,c}(k)=S(k)$, so positive norm requires
$S(k)>0$. Energetic stability further requires $\omega_f>0$ and
$\min_kS(k)>\omega_f$. In the regime used in the main text we impose the
stronger isolation condition
$\min_kS(k)=S_0-2|S_1|>2\omega_f$, which places the dispersive
positive-frequency band above the flat band and gives
$\Delta_{\rm gap}=S_0-2|S_1|-2\omega_f>0$.

Because $W_f(k)$ contains only the two Fourier harmonics $1$ and
$e^{-ik}$, its inverse Fourier transform is supported on two neighboring
unit cells. The corresponding compact Bogoliubov generator is
\begin{equation}
\hat\beta_{f,R}
=
\sqrt g\left(
c_0\hat a_{A,R}
+c_1\hat a_{A,R+1}
+\lambda_0\hat a^\dagger_{B,R}
+\lambda_1\hat a^\dagger_{B,R+1}
\right).
\label{eq:compactBogoliubovGenerator}
\end{equation}
The mixing of annihilation and creation components in
$\hat\beta_{f,R}$ directly encodes the squeezing structure induced by
parametric pairing. Translations of $\hat\beta_{f,R}$ span the exactly
flat-band subspace. Their finite spatial support, however, does not in
general determine the spatial extent of the corresponding Bogoliubov
modes. The origin of this distinction lies in the bosonic commutation
relations, as we show next.

\subsubsection{Canonical Bogoliubov modes}
\label{subsec:canonical_modes}

To obtain Bogoliubov modes that satisfy canonical bosonic commutation
relations, we first note that the translated compact generators obey
\[
[\hat\beta_{f,R},\hat\beta_{f,R'}^\dagger]
=
S_0\delta_{R,R'}
+
S_1\left(\delta_{R,R'+1}+\delta_{R,R'-1}\right),
\]
rather than $\delta_{R,R'}$. Equivalently, in momentum space,
$[\hat\beta_{f,k},\hat\beta_{f,k'}^\dagger]
=\delta_{kk'}S(k)$, where
$S(k)=S_0+2S_1\cos k$. For $S_1\neq0$, neighboring compact generators
therefore have nonzero commutator overlap, and the overlap function
$S(k)$ depends on momentum. Enforcing canonical bosonic commutation
relations then gives
$w_f(k)=W_f(k)/\sqrt{S(k)}$ and the corresponding Bogoliubov
annihilation operator
$\hat b_{f,k}
=[s_k^*\hat a_{A,k}+p_k\hat a_{B,-k}^\dagger]/\sqrt{S(k)}$.
The second positive-frequency band similarly gives
$\hat b_{c,k}
=[s_k\hat a_{B,k}+p_k^*\hat a_{A,-k}^\dagger]/\sqrt{S(k)}$.
Both sets satisfy canonical bosonic commutation relations.
The factor $S^{-1/2}(k)$ introduced by this construction is
nonpolynomial in momentum and gives the canonical flat-band modes their
extended real-space structure.

The real-space canonical modes follow from the same normalization,
$\hat b_{f,R_0}=\sum_{R'}h_{R_0-R'}\hat\beta_{f,R'}$, or equivalently
$w_{f,R_0}=\sum_{R'}h_{R_0-R'}W_{f,R'}$, where
$h_R=[S^{-1/2}]_R$ denotes the Fourier coefficient of $S^{-1/2}(k)$.
To determine its spatial decay, we use the analytic continuation
$z=e^{-ik}$ introduced in Sec.~S.1, for which
$S(z)=S_0+S_1(z+z^{-1})$. For $S_1\neq0$, define
$D_S=\sqrt{S_0^2-4S_1^2}$ and let
$z_S=(-S_0+D_S)/(2S_1)$ be the root of $S(z)$ inside the unit circle.
Stability, $S(k)>0$, implies $|z_S|<1$. With
$\Lambda_S=(S_0+D_S)/2$, the overlap factorizes as
$S(z)=\Lambda_S(1-z_Sz)(1-z_Sz^{-1})$. The exact Fourier coefficient
and its large-distance asymptotic form are
\begin{equation}
\begin{aligned}
h_R &=
\frac{z_S^{|R|}}{\sqrt{\Lambda_S}}
\frac{(1/2)_{|R|}}{|R|!}\,
{}_2F_1\!\left(
\frac12,|R|+\frac12;|R|+1;z_S^2
\right),\\
h_R &=
\frac{z_S^{|R|}}
{\sqrt{\pi |R|\Lambda_S(1-z_S^2)}}
\left[1+O(|R|^{-1})\right],
\qquad |R|\rightarrow\infty .
\end{aligned}
\label{eq:hR}
\end{equation}
Here $(x)_n=x(x+1)\cdots(x+n-1)$ denotes the Pochhammer symbol
(rising factorial), with $(x)_0=1$, and
${}_2F_1(a,b;c;x)$ denotes the Gauss hypergeometric function,
${}_2F_1(a,b;c;x)=\sum_{m=0}^{\infty}(a)_m(b)_m x^m/[(c)_m m!]$.

Since $S(k)$ is even, $h_R=h_{-R}$. For $S_1\neq0$, the Fourier
coefficient $h_R$ has infinite spatial support with an exponential
envelope controlled by $|z_S|$. For a canonical mode centered at the
origin, the annihilation component on sublattice $A$ in cell $R$ is
$\sqrt{g}\,[c_0h_R+c_1h_{R-1}]$, while the creation component on
sublattice $B$ is
$-\sqrt{g}\,[\lambda_0h_R+\lambda_1h_{R-1}]$.
When the corresponding coefficient at $z_S$ is nonzero, each component
has the asymptotic form $|R|^{-1/2}z_S^{|R|}$. The algebraic factor
arises from the square-root singularity of $S^{-1/2}(z)$, while the
exponential factor defines the decay length
$\xi_S=-a/\ln|z_S|$. If the coefficient of a particular component
vanishes at $z_S$, its leading contribution is suppressed, although
subleading terms associated with the same branch point can remain.

The resulting real-space modes satisfy
$w_{f,R}^\dagger\Sigma_z w_{f,R'}=\delta_{R,R'}$, or equivalently
$[\hat b_{f,R},\hat b_{f,R'}^\dagger]=\delta_{R,R'}$.
Thus an exactly flat band spanned by two-cell compact Bogoliubov
generators can nevertheless support spatially extended canonical
Bogoliubov modes when their commutator overlap depends on momentum.
Their exponential decay is set by $\xi_S$, the quantum-geometric
length introduced in the main text. Section~\ref{sec:intrinsic_bdg_length} develops the general formulation
in terms of compact generating frames and the flat-band projector, and
gives the conditions under which physical correlations inherit the same
decay length.

\subsubsection{Multimode squeezed ground state and momentum-space squeezing}
\label{subsec:ground_state}

The canonical Bogoliubov operators derived above determine the
Bogoliubov ground state. For each $k$, the modes
$\hat a_{A,k}$ and $\hat a_{B,-k}$ form a two-mode squeezed pair.
The common vacuum of all positive-frequency Bogoliubov operators can
therefore be written as
\begin{equation}
|G\rangle
=
\prod_{k\in\mathrm{BZ}}
\sqrt{1-|\zeta_k|^2}\,
\exp\!\left(
\zeta_k\hat a_{A,k}^\dagger\hat a_{B,-k}^\dagger
\right)|0\rangle,
\qquad
\zeta_k=-\frac{p_k}{s_k^*}
=e^{i\phi_k}\tanh r_k .
\label{eq:squeezedGroundState}
\end{equation}
Here $|0\rangle$ is the vacuum of the physical bosonic modes,
$\hat a_{\alpha,k}|0\rangle=0$,
$r_k=\operatorname{artanh}(|p_k|/|s_k|)$ is the squeezing amplitude,
and $\phi_k=\arg\zeta_k$ is the squeezing phase. Thus
$\zeta_k=e^{i\phi_k}\tanh r_k$ is the complex squeezing parameter.
The condition $S(k)=|s_k|^2-|p_k|^2>0$ implies
$|\zeta_k|<1$, ensuring that the squeezed state is normalizable.
For the present model,
$\zeta_k=-(\lambda_0+\lambda_1e^{ik})/(c_0+c_1e^{ik})$,
so its squeezing amplitude and phase are independent of the overall
coupling scale $g$.

For each momentum pair, Eq.~\eqref{eq:squeezedGroundState} has the number-state expansion
\[
\sqrt{1-|\zeta_k|^2}
\sum_{n=0}^{\infty}
\zeta_k^n
|n\rangle_{A,k}|n\rangle_{B,-k}.
\]
It follows directly that
$\hat a_{A,k}|G\rangle
=\zeta_k\hat a_{B,-k}^\dagger|G\rangle$ and
$\hat a_{B,k}|G\rangle
=\zeta_{-k}\hat a_{A,-k}^\dagger|G\rangle$.
Using
$s_k^*\zeta_k+p_k=0$ and
$s_k\zeta_{-k}+p_k^*=0$ in the canonical operators derived above gives
$\hat b_{f,k}|G\rangle=0$ and
$\hat b_{c,k}|G\rangle=0$.
Thus $|G\rangle$ is the common vacuum of both positive-frequency
Bogoliubov bands. In the energetically stable regime, where both
excitation branches have positive frequencies, this common vacuum is the
ground state of the quadratic Hamiltonian.

All first moments vanish in $|G\rangle$. From the number-state expansion,
the normal and anomalous momentum-space covariances are
$\langle\delta\hat a_{A,k}^\dagger\delta\hat a_{A,k}\rangle_G
=|\zeta_k|^2/(1-|\zeta_k|^2)$ and
$\langle\delta\hat a_{A,k}\delta\hat a_{B,-k}\rangle_G
=\zeta_k/(1-|\zeta_k|^2)$.
Using
$1-|\zeta_k|^2=S(k)/|s_k|^2$ and
$\zeta_k=-p_k/s_k^*$ gives
\[
N_{AA}(k)=N_{BB}(-k)=\frac{|p_k|^2}{S(k)},
\qquad
F_{AB}(k)=-\frac{q_k}{S(k)}.
\]
The cross-sublattice normal correlation and the same-sublattice
anomalous correlations vanish,
$N_{AB}(k)=F_{AA}(k)=F_{BB}(k)=0$.
The factor $S^{-1/2}(k)$ entering each canonical Bogoliubov mode
therefore appears as the common factor $S^{-1}(k)$ in the ground-state
correlations, which will determine their real-space decay.

To quantify the squeezing at fixed $k$, we define the collective mode
$\delta\hat d_k(\varphi)
=[\delta\hat a_{A,k}+e^{-i\varphi}\delta\hat a_{B,-k}]/\sqrt{2}$
and its quadrature
$\hat X_k(\theta,\varphi)
=[e^{-i\theta}\delta\hat d_k(\varphi)
+e^{i\theta}\delta\hat d_k^\dagger(\varphi)]/\sqrt{2}$.
Its normal and anomalous moments are
$\langle\delta\hat d_k^\dagger(\varphi)\delta\hat d_k(\varphi)\rangle_G
=N_{AA}(k)$ and
$\langle\delta\hat d_k(\varphi)\delta\hat d_k(\varphi)\rangle_G
=e^{-i\varphi}F_{AB}(k)$.
The minimum ground-state quadrature variance is therefore
\[
V_{\min}^{\rm gs}(k)
\equiv
\min_{\theta}
\left\langle
\hat X_k^2(\theta,\varphi)
\right\rangle_G
=
\frac{1}{2}+N_{AA}(k)-|F_{AB}(k)|
=
\frac{1}{2}e^{-2r_k}
=
\frac{1}{2}\frac{|s_k|-|p_k|}{|s_k|+|p_k|},
\]
which is independent of the relative phase $\varphi$.
The orthogonal quadrature has
$V_{\max}^{\rm gs}(k)=e^{2r_k}/2$.
Since $S(k)>0$ is equivalent to $|s_k|>|p_k|$,
$V_{\min}^{\rm gs}(k)$ is finite and positive.
For $p_k\neq0$, one has $r_k>0$ and hence
$V_{\min}^{\rm gs}(k)<1/2$, demonstrating sub-vacuum squeezing,
whereas $p_k=0$ gives $r_k=0$ and restores the vacuum value.
Thus the momentum-dependent Bogoliubov mixing encoded by $\zeta_k$
determines the squeezing amplitude and phase, while the overall
coupling scale $g$ drops out of $V_{\min}^{\rm gs}(k)$.

\subsubsection{Exact real-space covariances}
\label{subsec:exact_real_space_covariances}

Fourier transforming the momentum-space covariances gives their spatial
dependence. With
$\delta\hat a_{\alpha,R}
=\int_{-\pi}^{\pi}(dk/2\pi)\,
e^{ikR}\delta\hat a_{\alpha,k}$,
we define the normal covariance
$N(R)=
\langle\delta\hat a_{A,0}^{\dagger}
\delta\hat a_{A,R}\rangle_G$.
Since $q_k$ is not generally even in momentum, the anomalous correlation
depends on the ordering of the two sublattices. We therefore define
$F_{A0,BR}(R)=
\langle\delta\hat a_{A,0}\delta\hat a_{B,R}\rangle_G
=-\int_{-\pi}^{\pi}(dk/2\pi)\,
e^{-ikR}q_k/S(k)$
and
$F_{AR,B0}(R)=
\langle\delta\hat a_{A,R}\delta\hat a_{B,0}\rangle_G
=-\int_{-\pi}^{\pi}(dk/2\pi)\,
e^{ikR}q_k/S(k)$.
The notation used in the main text is
$F(R)\equiv F_{A0,BR}(R)$.

We use the analytic continuation $z=e^{-ik}$ introduced in
Sec.~\ref{subsec:canonical_modes}. For $S_1\neq0$, the two roots of
$S(z)=S_0+S_1(z+z^{-1})$ are reciprocal to one another. The root
$z_S$ lies inside the unit circle and $z_S^{-1}$ lies outside.
With $D_S=\sqrt{S_0^2-4S_1^2}$, the inverse overlap has the exact
Fourier expansion
$S^{-1}(k)=D_S^{-1}
[1+2\sum_{R=1}^{\infty}z_S^R\cos(kR)]$.
Equivalently, the same coefficients follow from the pole at $z_S$ in
the unit-circle contour integral. Combining this expansion with
$|p_k|^2$ and $q_k$ gives exact expression of real space covariances, for $R\geq0$,
\begin{equation}
\begin{aligned}
N(R)
&=
\frac{g}{D_S}
\begin{cases}
\lambda_0^2+\lambda_1^2
+2\lambda_0\lambda_1z_S,
& R=0,\\
(\lambda_1+\lambda_0z_S)
(\lambda_0+\lambda_1z_S)z_S^{R-1},
& R\geq1,
\end{cases}
\\[2pt]
F_{A0,BR}(R)
&=
-\frac{g}{D_S}
\begin{cases}
\lambda_0c_0+\lambda_1c_1
+z_S(\lambda_1c_0+\lambda_0c_1),
& R=0,\\
(\lambda_1+\lambda_0z_S)
(c_0+c_1z_S)z_S^{R-1},
& R\geq1,
\end{cases}
\\[2pt]
F_{AR,B0}(R)
&=
-\frac{g}{D_S}
\begin{cases}
\lambda_0c_0+\lambda_1c_1
+z_S(\lambda_1c_0+\lambda_0c_1),
& R=0,\\
(\lambda_0+\lambda_1z_S)
(c_1+c_0z_S)z_S^{R-1},
& R\geq1.
\end{cases}
\end{aligned}
\label{eq:exactRealSpaceCovariancesSM}
\end{equation}

Reversing the separation exchanges the two anomalous correlations,
$F_{A0,BR}(-R)=F_{AR,B0}(R)$, while
$N(-R)=N(R)^*$. For the real parameters considered here, $N(R)$ is
real and even.

For $R\geq1$,
every nonvanishing covariance contains the factor $z_S^{R-1}$ and
therefore has the exponential decay length
$\xi_S=-a/\ln|z_S|$.
The sign of $z_S$ determines the spatial phase pattern. A negative
$z_S$ produces an alternating sign between neighboring cells, while a
positive $z_S$ does not. For the parameter sets used in the main text,
$S_1<0$ and hence $z_S>0$.

The presence of the common denominator $S(k)$ does not guarantee that
every covariance contains this long-distance contribution. After
analytic continuation, the pole at $z_S$ contributes only when the
numerator of the selected covariance is nonzero at that root. For
$R\geq1$, the corresponding coefficients are
$(\lambda_1+\lambda_0z_S)(\lambda_0+\lambda_1z_S)$ for $N(R)$,
$(\lambda_1+\lambda_0z_S)(c_0+c_1z_S)$ for $F_{A0,BR}(R)$, and
$(\lambda_0+\lambda_1z_S)(c_1+c_0z_S)$ for $F_{AR,B0}(R)$.
If the corresponding coefficient vanishes, the residue of the pole at
$z_S$ vanishes and that contribution is absent from the covariance.
The observable decay length is therefore set by the nearest
complex-momentum singularity with nonzero residue in the correlation
being measured.

The value at $R=0$ contains additional local contributions and need not
follow the nonlocal exponential tail. The spatial decay length is
therefore determined from the $R\geq1$ behavior rather than from the
on-site value.

The role of the momentum-dependent commutator overlap is especially
clear when $S_1=0$, or equivalently
$c_0c_1=\lambda_0\lambda_1$. In this case $S(k)=S_0$ is momentum
independent and the canonical normalization introduces no
complex-momentum pole. The covariances then have finite spatial support.
For $R\geq0$, the nonzero terms are
$N(0)=g(\lambda_0^2+\lambda_1^2)/S_0$,
$N(1)=g\lambda_0\lambda_1/S_0$,
$F_{A0,BR}(0)=-g(\lambda_0c_0+\lambda_1c_1)/S_0$,
$F_{A0,BR}(1)=-g\lambda_1c_0/S_0$, and
$F_{AR,B0}(1)=-g\lambda_0c_1/S_0$.
Thus parametric pairing can produce anomalous correlations without an
extended spatial tail when the canonical normalization is momentum
independent.

The formal limit $|z_S|\rightarrow1$, for which $\xi_S$ diverges,
occurs when $S_0-2|S_1|\rightarrow0$. For $S_1<0$ this gives
$z_S\rightarrow1$, while for $S_1>0$ it gives
$z_S\rightarrow-1$. This limit lies outside the energetically stable
and spectrally isolated regime established in
Sec.~\ref{subsec:finite_range_construction}. The physically accessible
$\xi_S$ therefore remains finite. A long-distance covariance also
requires a nonzero residue at the corresponding root.

These exact covariances determine the spatial correlations used below
to establish the criterion for sub-vacuum collective-mode squeezing.

\subsection{Long-distance collective-mode squeezing}
\label{subsec:long_distance_collective_squeezing}

The anomalous correlation $F(R)\equiv F_{A0,BR}(R)$ derived above
produces phase-sensitive correlations between the spatially separated
modes $A,0$ and $B,R$. For $R\geq1$, we define the collective
mode
$\delta\hat d_R(\varphi)=
[\delta\hat a_{A,0}
+e^{-i\varphi}\delta\hat a_{B,R}]/\sqrt{2}$
and its quadrature
$\hat X_R(\theta,\varphi)=
[e^{-i\theta}\delta\hat d_R(\varphi)
+e^{i\theta}\delta\hat d_R^\dagger(\varphi)]/\sqrt{2}$.
The absence of cross-sublattice normal correlations and the equality of
the two local occupations give
$\langle\delta\hat d_R^\dagger(\varphi)
\delta\hat d_R(\varphi)\rangle_G=N(0)$ and
$\langle\delta\hat d_R(\varphi)
\delta\hat d_R(\varphi)\rangle_G=e^{-i\varphi}F(R)$.
Minimizing the ground-state quadrature variance over $\theta$ removes
the dependence on $\varphi$. For $S_1\neq0$, insertion of the exact
covariances from Sec.~\ref{subsec:exact_real_space_covariances} gives
\begin{equation}
\begin{aligned}
V_{\min}^{\rm gs}(R)
&\equiv
\min_\theta
\langle\hat X_R^2(\theta,\varphi)\rangle_G
=
\frac{1}{2}+N(0)-|F(R)|
\\
&=
\frac{1}{2}
+
\frac{g}{D_S}
\left[
\lambda_0^2+\lambda_1^2
+2\lambda_0\lambda_1z_S
-
\left|
(\lambda_1+\lambda_0z_S)
(c_0+c_1z_S)
\right|
|z_S|^{R-1}
\right],
\qquad R\geq1 .
\end{aligned}
\label{eq:groundStateSpatialSqueezing}
\end{equation}

The collective mode is squeezed below the vacuum level when
$V_{\min}^{\rm gs}(R)<1/2$, which is equivalent to
$|F(R)|>N(0)$. The exact condition is therefore
\[
\left|
(\lambda_1+\lambda_0z_S)
(c_0+c_1z_S)
\right|
|z_S|^{R-1}
>
\lambda_0^2+\lambda_1^2
+2\lambda_0\lambda_1z_S .
\]
Since $|F(R)|=|F(1)|\,|z_S|^{R-1}$ for $R\geq1$, squeezing at
$R=1$ implies a finite range of squeezed separations. We define
$R_{\rm sq}$ by $|F(R_{\rm sq})|=N(0)$, as in the main text. This gives
\[
R_{\rm sq}
=
1+\frac{\xi_S}{a}
\ln\!\left[
\frac{|F(1)|}{N(0)}
\right]
=
1+\frac{\xi_S}{a}
\ln\!\left[
\frac{
\left|
(\lambda_1+\lambda_0z_S)
(c_0+c_1z_S)
\right|
}{
\lambda_0^2+\lambda_1^2
+2\lambda_0\lambda_1z_S
}
\right].
\]
The squeezed integer separations satisfy $R<R_{\rm sq}$. If
$|F(1)|\leq N(0)$, no separation $R\geq1$ is squeezed. Thus $\xi_S$
sets the exponential spatial scale, while the squeezing range also
depends on the ratio of the first nonlocal anomalous correlation to the
local occupation.

For the simpler choice
$\lambda_0=\lambda_1\equiv\lambda$, $c_0\equiv c$, and $c_1=1$,
the squeezing condition reduces to
\[
|c+z_S|\,|z_S|^{R-1}>2|\lambda|,
\]
and the squeezing range becomes
\[
R_{\rm sq}
=
1+\frac{\xi_S}{a}
\ln\!\left[
\frac{|c+z_S|}{2|\lambda|}
\right].
\]
These expressions recover the closed-form result for the simpler
parametrization.

The squeezing depth at a prescribed separation is
$D_{\rm gs}(R)\equiv1/2-V_{\min}^{\rm gs}(R)=|F(R)|-N(0)$.
For the general four-coefficient model,
\[
D_{\rm gs}(R)
=
\frac{g}{D_S}
\left[
\left|
(\lambda_1+\lambda_0z_S)
(c_0+c_1z_S)
\right|
|z_S|^{R-1}
-
\lambda_0^2-\lambda_1^2
-2\lambda_0\lambda_1z_S
\right].
\]
The squeezing depth depends on both the decay length and the amplitudes
of the normal and anomalous correlations.

A simple analytical form follows from the same choice
$\lambda_0=\lambda_1\equiv\lambda$, $c_0\equiv c$, and $c_1=1$.
On the $c<-1$ branch and for $|\lambda|\ll1$,
$z_S=1/|c|+O(\lambda^2)$, which gives
\[
D_{\rm gs}(R)
=
|\lambda|(|c|+1)|c|^{-(R+1)}
-
\frac{2|\lambda|^2}{|c|(|c|-1)}
+O(|\lambda|^3).
\]
At fixed $c$ and $R$, the leading expression is maximized at
\[
|\lambda|_{\rm opt}
=
\frac{|c|^2-1}{4|c|^R},
\qquad
D_{\rm gs}^{\max}(R;c)
=
\frac{(|c|+1)^2(|c|-1)}
{8|c|^{2R+1}},
\]
provided that $|\lambda|_{\rm opt}\ll1$ and the resulting parameters
remain within the stable, spectrally isolated regime established in
Sec.~\ref{subsec:finite_range_construction}. These expressions show
explicitly that increasing the spatial range does not by itself increase
the squeezing depth. The depth also depends on the anomalous-correlation
amplitude and the local occupation. In the four-coefficient model used
for the main-text results, these quantities are not restricted to the
single tuning path of the simpler parametrization.

A long anomalous-correlation length therefore does not by itself
guarantee sub-vacuum squeezing. A long-distance contribution also
requires a nonzero residue in the selected anomalous correlation, as
discussed in Sec.~\ref{subsec:exact_real_space_covariances}. The
quadrature orthogonal to the minimizing one has variance
$V_{\max}^{\rm gs}(R)=1/2+N(0)+|F(R)|$.
The collective mode $\delta\hat d_R$ is generally not in a pure
single-mode Gaussian state, since it is embedded in the multimode
Bogoliubov ground state. Accordingly,
\[
V_{\min}^{\rm gs}(R)V_{\max}^{\rm gs}(R)
=
\left[\frac12+N(0)\right]^2-|F(R)|^2
\geq \frac14,
\]
with equality only when the reduced state of $\delta\hat d_R$ is pure.
Thus the two variances cannot in general be parametrized as
$e^{-2r}/2$ and $e^{2r}/2$ using a single squeezing parameter $r$.
%=====================================================================
\section{Quantum-geometric length, anomalous covariance, and squeezing-sector quantum metric}
\label{sec:intrinsic_bdg_length}
%=====================================================================

The explicit model of Sec.~\ref{sec:pairing_flat_band} shows that
momentum-dependent commutator overlap between compact Bogoliubov
generators requires spatially extended superpositions to construct
canonical Bogoliubov modes. The same structure can be formulated for
an arbitrary isolated positive-frequency flat subspace using a compact
Bogoliubov frame and its commutator-overlap matrix. The latter determines
the canonical normalization, while the associated flat-band projector
provides a frame-independent description of the flat subspace.
Complex-momentum singularities of this projector can appear in the
ground-state anomalous covariance and in the squeezing-sector quantum
metric when the corresponding singular contribution is not cancelled.

%---------------------------------------------------------------------
%---------------------------------------------------------------------
\subsection{Compact Bogoliubov frame and flat-band projector}
\label{subsec:compact_frame_projector}
%---------------------------------------------------------------------
Consider an energetically stable bosonic BdG lattice with an isolated
rank-$r$ positive-frequency flat subspace at frequency $\omega_f$.
Using the analytic continuation $z=e^{-ik}$ introduced above, let
$\mathcal W(z)$ be a $2N\times r$ matrix whose columns form a basis for
this flat subspace. We assume that these columns are finite Laurent
polynomials, so that $\mathcal W(z)$ defines a compact Bogoliubov
frame~\cite{Sathe2021CompactWannierProjectors}. For
$\mathcal W(z)=\sum_n\mathcal W_n z^n$, its reciprocal adjoint is
$\widetilde{\mathcal W}(z)
=\mathcal W^\dagger(1/z^*)
=\sum_n\mathcal W_n^\dagger z^{-n}$,
following the convention introduced above. We define the
commutator-overlap matrix of this frame as $\mathsf G(z)$ and write
\begin{equation}
\mathcal W(z)
=
\sum_{n=-n_-}^{n_+}\mathcal W_n z^n,
\qquad
M(z)\mathcal W(z)
=
\omega_f\mathcal W(z),
\qquad
\mathsf G(z)
=
\widetilde{\mathcal W}(z)
\Sigma_z
\mathcal W(z).
\label{eq:compactFrameGeneral}
\end{equation}
Because each column of $\mathcal W(z)$ is a finite Laurent polynomial,
its inverse Fourier transform has finite spatial support and defines a
compact Bogoliubov generator. Here $\mathsf G(z)$ is the
commutator-overlap matrix of the compact generating frame. We assume
that $\mathcal W(z)$ has rank $r$ everywhere on the physical
Brillouin-zone contour $z=e^{-ik}$.

On the physical Brillouin zone, the reciprocal adjoint reduces to the
ordinary Hermitian conjugate, giving
$\mathsf G(k)=\mathcal W^\dagger(k)\Sigma_z\mathcal W(k)$.
If $\hat\beta_{a,k}$ denotes the compact Bogoliubov generator associated
with column $a$ of $\mathcal W(k)$, then
$[\hat\beta_{a,k},\hat\beta_{b,k'}^\dagger]
=\delta_{kk'}\mathsf G_{ab}(k)$.
Thus $\mathsf G(k)$ is the Gram matrix of commutator overlaps between
the compact generators. For the compact rank-one frame introduced in
Sec.~\ref{subsec:canonical_modes}, this relation reduces to
$\mathsf G(k)=S(k)$ and, after analytic continuation,
$\mathsf G(z)=S(z)$. The matrix $\mathsf G$ is therefore the
multiband generalization of the overlap function $S$ appearing in the
explicit model. Because the selected positive-frequency subspace has
positive Krein signature, $\mathsf G(k)$ is positive definite for every
real $k$. Hence $\det\mathsf G(k)\neq0$ on the physical Brillouin zone,
although $\det\mathsf G(z)$ may vanish at complex momentum.

Because $\mathsf G(k)$ is positive definite, it has a unique positive
Hermitian square root on the physical Brillouin zone. We define the
canonical Bogoliubov frame $\mathcal U_f(k)$ and the associated
flat-band projector $P_f(z)$ by
\begin{equation}
\mathcal U_f(k)
=
\mathcal W(k)\mathsf G^{-1/2}(k),
\qquad
\mathcal U_f^\dagger(k)\Sigma_z\mathcal U_f(k)
=
\mathbb I_r,
\qquad
P_f(z)
=
\mathcal W(z)\mathsf G^{-1}(z)
\widetilde{\mathcal W}(z)\Sigma_z .
\label{eq:normalizedFrameProjectorGeneral}
\end{equation}
Indeed,
$\mathcal U_f^\dagger\Sigma_z\mathcal U_f
=
\mathsf G^{-1/2}
(\mathcal W^\dagger\Sigma_z\mathcal W)
\mathsf G^{-1/2}
=
\mathbb I_r$.
Thus $\mathsf G^{-1/2}(k)$ provides the normalization required for the
corresponding Bogoliubov operators to satisfy canonical bosonic
commutation relations. In the explicit rank-one model, this reduces to
the factor $S^{-1/2}(k)$.

For the complex-momentum analysis, $P_f(z)$ is more convenient than the
canonical frame $\mathcal U_f(k)$. Analytic continuation of
$\mathsf G^{-1/2}$ generally introduces square-root branch points,
whereas $\mathsf G^{-1}$ is meromorphic. Wherever $\mathsf G(z)$ is
invertible, Eq.~\eqref{eq:normalizedFrameProjectorGeneral} gives
\[
P_f^2
=
\mathcal W\mathsf G^{-1}
(\widetilde{\mathcal W}\Sigma_z\mathcal W)
\mathsf G^{-1}
\widetilde{\mathcal W}\Sigma_z
=
P_f,
\]
because
$\widetilde{\mathcal W}\Sigma_z\mathcal W=\mathsf G$.
It also gives
\[
P_f\mathcal W
=
\mathcal W\mathsf G^{-1}
(\widetilde{\mathcal W}\Sigma_z\mathcal W)
=
\mathcal W,
\]
so the image of $P_f$ is the flat subspace spanned by $\mathcal W$.
On the physical Brillouin zone,
$P_f(k)=\mathcal U_f(k)\mathcal U_f^\dagger(k)\Sigma_z$ is the
Krein-orthogonal projector onto the selected positive-frequency flat
subspace.

The compact generating frame is not unique. Consider another frame
$\mathcal W'(z)=\mathcal W(z)\mathcal A(z)$, where both
$\mathcal A(z)$ and its reciprocal adjoint
$\widetilde{\mathcal A}(z)$ are regular and invertible in the
complex-momentum region under consideration. The Gram matrix transforms
as
\[
\mathsf G'(z)
=
\widetilde{\mathcal A}(z)
\mathsf G(z)
\mathcal A(z).
\]
Using
$\widetilde{\mathcal W'}
=
\widetilde{\mathcal A}\widetilde{\mathcal W}$,
the corresponding projector becomes
\[
P_f'(z)
=
\mathcal W\mathcal A
(\widetilde{\mathcal A}\mathsf G\mathcal A)^{-1}
\widetilde{\mathcal A}
\widetilde{\mathcal W}\Sigma_z
=
P_f(z).
\]
The projector is therefore independent of the compact frame used to
span the flat subspace. A complex-momentum singularity of $P_f(z)$
cannot be introduced or removed by a regular and invertible change of
frame.

A zero of $\det\mathsf G(z)$ is only a candidate singularity of the
flat-band projector. Although $\mathsf G^{-1}(z)$ diverges at such a
zero, this divergence may be cancelled by the factors
$\mathcal W(z)$ and $\widetilde{\mathcal W}(z)$ in the complete
projector. A pole of $P_f(z)$ remains only when this cancellation does
not occur. The condition for such a pole to remain is derived in the
following subsubsection.

This distinction also determines whether the flat subspace admits
compact canonical Bogoliubov modes. If a Krein-orthonormal frame
$\mathcal V(z)$ with strictly finite real-space support existed, then
$\mathcal V(z)$ would be finite Laurent and its projector would have
the finite-Laurent form
$\mathcal V(z)\widetilde{\mathcal V}(z)\Sigma_z$. Such a projector
cannot have a pole at finite complex momentum. A finite
complex-momentum pole of $P_f(z)$ therefore rules out a strictly
compact canonical frame. Compact Bogoliubov generators can still span
the flat band, but imposing canonical bosonic commutation relations
then requires spatially extended linear combinations of those
generators.

%---------------------------------------------------------------------
%---------------------------------------------------------------------
\subsubsection{Complex-momentum zero and flat-band projector residue}
\label{subsec:projector_residue}
%---------------------------------------------------------------------

A complex-momentum zero of the commutator-overlap matrix can produce a
pole of the flat-band projector $P_f(z)$. Let $z_g$ be a simple zero
of $\det\mathsf G(z)$ inside the unit circle. At $z_g$,
$\mathsf G(z_g)$ is singular, so there exist right and left null
vectors $r_g$ and $\ell_g$ satisfying
$\mathsf G(z_g)r_g=0$ and
$\ell_g^\dagger\mathsf G(z_g)=0$.
The vector $r_g$ identifies the linear combination of compact
Bogoliubov generators whose analytically continued commutator overlap
vanishes at $z_g$. For the rank-one model of
Sec.~\ref{subsec:canonical_modes}, this condition reduces simply to
$\mathsf G(z_g)=S(z_g)=0$.

To characterize the behavior near $z_g$, define
\[
\alpha_g
=
\ell_g^\dagger\mathsf G'(z_g)r_g .
\]
For a simple zero, $\alpha_g\neq0$. With the normalization
$\ell_g^\dagger r_g=1$, let $\lambda_g(z)$ denote the eigenvalue of
$\mathsf G(z)$ that vanishes at $z_g$. Near the zero,
$\lambda_g(z)\simeq\alpha_g(z-z_g)$, so the corresponding eigenvalue of
$\mathsf G^{-1}(z)$ behaves as
$\lambda_g^{-1}(z)\simeq[\alpha_g(z-z_g)]^{-1}$. The position $z_g$ fixes the spatial decay
scale, while $\alpha_g$ determines the local strength with which the
inverse overlap diverges in this channel.

Expanding
$\mathsf G(z)=\mathsf G(z_g)+(z-z_g)\mathsf G'(z_g)
+O((z-z_g)^2)$ and using the one-dimensional left and right null
spaces gives
\begin{equation}
\mathsf G^{-1}(z)
=
\frac{
r_g\ell_g^\dagger
}{
\alpha_g(z-z_g)
}
+
\mathsf G_{\rm reg}^{-1}(z),
\label{eq:GramInversePole}
\end{equation}
where $\mathsf G_{\rm reg}^{-1}(z)$ is regular at $z_g$. The singular
part therefore acts only in the compact-generator channel selected by
$r_g$.

Substituting Eq.~\eqref{eq:GramInversePole} into the normalized
projector in Eq.~\eqref{eq:normalizedFrameProjectorGeneral} gives
\begin{equation}
P_f(z)
=
\frac{\mathcal R_g}{z-z_g}
+
P_{f,\rm reg}(z),
\label{eq:projectorPoleGeneral}
\end{equation}
with
\begin{equation}
\mathcal R_g
=
\frac{
\mathcal W(z_g)r_g
\ell_g^\dagger
\widetilde{\mathcal W}(z_g)\Sigma_z
}{
\alpha_g
}.
\label{eq:projectorResidueGeneral}
\end{equation}
The numerator lifts the singular compact-generator channel into the
full Nambu-orbital space, while the factor $1/\alpha_g$ sets the
strength of the inverse-overlap divergence. The residue
$\mathcal R_g$ therefore fixes both the weight and the internal
Nambu-orbital structure of the resulting projector pole. Consequently,
the zero at $z_g$ generates an actual, nonremovable singularity of
$P_f(z)$ only if
\begin{equation}
\mathcal R_g\neq0 .
\label{eq:activeProjectorPole}
\end{equation}
If $\mathcal R_g=0$, the numerator in
Eq.~\eqref{eq:projectorResidueGeneral} cancels the singularity of
$\mathsf G^{-1}(z)$

For positive separation and the convention $z=e^{-ik}$, the real-space
projector is
\begin{equation*}
P_f(R)
=
\oint_{|z|=1}\frac{dz}{2\pi i}\,
z^{R-1}P_f(z),
\qquad R>0.
\end{equation*}
Assume that $z_g$ is the interior pole of $P_f(z)$ with largest modulus
among those with nonzero residue.
Evaluating its residue gives
\begin{equation}
P_f(R)
=
z_g^{R-1}\mathcal R_g
+
O(\rho^R),
\qquad
\rho<|z_g| .
\label{eq:projectorAsymptoticGeneral}
\end{equation}
This expression separates the two physical roles of the pole data.
The modulus $|z_g|$ determines the exponential decay rate, whereas
$\mathcal R_g$ determines the amplitude and internal structure of the
long-distance projector tail. The corresponding projector decay length
is
\begin{equation}
\xi_P
=
-\frac{a}{\ln|z_g|}.
\label{eq:intrinsicProjectorLength}
\end{equation}
For the explicit rank-one model,
$\mathsf G(z)=S(z)$, so $z_g=z_S$ and
$\xi_P=\xi_S$. In that case
$\alpha_g=S'(z_S)$ and
\[
\mathcal R_g
=
\frac{
W_f(z_S)\widetilde W_f(z_S)\Sigma_z
}{
S'(z_S)
},
\]
so that $z_S$ determines the decay length while $\mathcal R_g$
determines the amplitude and Nambu structure of the projector tail.

When the pole structure is reciprocal, the reciprocal-conjugate
exterior pole controls the opposite spatial direction and gives the
same exponential decay length. More generally, a pole of order $m$ at $z_g$ contributes an
algebraic factor $R^{m-1}$ multiplying the same exponential factor
$z_g^R$. The pole location therefore fixes the exponential length,
while its residue and order determine the amplitude, internal
structure, and algebraic envelope.

A zero of $\det\mathsf G(z)$ alone is therefore insufficient to
determine the projector decay length. The corresponding singularity of
$\mathsf G^{-1}(z)$ must survive in the full projector $P_f(z)$.
The following subsection determines when such a projector pole is
inherited by the ground-state anomalous covariance.

%---------------------------------------------------------------------
%---------------------------------------------------------------------
\subsection{Flat-band subspace contribution to the ground-state anomalous covariance}
\label{subsec:ground_state_anomalous_covariance}
%---------------------------------------------------------------------

We now determine when a pole of the selected flat-band projector is
inherited by the ground-state anomalous covariance. Throughout this
subsection, $P_f$ denotes the Krein-orthogonal projector onto the
selected rank-$r$ positive-frequency flat subspace, constructed from
the canonically normalized frame introduced in Sec.~A.1. It is not the
projector onto the complete positive-frequency BdG sector.

Decompose the canonical flat-band frame into its annihilation and
creation components,
$\mathcal U_f(k)=(U_f(k),V_f(k))^T$, with
$\mathcal U_f^\dagger(k)\Sigma_z\mathcal U_f(k)=\mathbb I_r$.
The corresponding flat-band projector has the block form
\begin{equation}
P_f(k)
=
\mathcal U_f(k)\mathcal U_f^\dagger(k)\Sigma_z
=
\begin{pmatrix}
U_fU_f^\dagger & -U_fV_f^\dagger\\
V_fU_f^\dagger & -V_fV_f^\dagger
\end{pmatrix}.
\label{eq:projectorParticleHoleBlocks}
\end{equation}
The upper-right Nambu block is therefore $-U_fV_f^\dagger$.
As shown below, this block directly determines the contribution of the
selected flat subspace to the ground-state anomalous covariance.

Let $\mathcal U_+(k)$ contain a complete Krein-orthonormal set of
positive-frequency Bogoliubov modes, with annihilation and creation
components $u_n(k)$ and $v_n(k)$. The inverse Bogoliubov
transformation together with
$\hat b_{n,k}|G\rangle=0$ gives
$F^{(G)}(k)\equiv
\langle\delta\hat{\boldsymbol a}_{k}
\delta\hat{\boldsymbol a}_{-k}^{\mathsf T}\rangle_G
=\sum_{n>0}u_n(k)v_n^\dagger(k)$.
Restricting this sum to the selected flat subspace gives
\begin{equation}
F_f^{(G)}(k)
=
U_f(k)V_f^\dagger(k)
=
-\Pi_pP_f(k)\Pi_h^\dagger ,
\label{eq:flatSubspaceCovarianceProjector}
\end{equation}
where
$\Pi_p=(\mathbb I_N,0)$ and
$\Pi_h=(0,\mathbb I_N)$ select the annihilation and creation sectors,
respectively. Thus the anomalous covariance contributed by the flat
subspace is obtained directly from the off-diagonal Nambu block of its
projector.

Equation~\eqref{eq:flatSubspaceCovarianceProjector} also shows that
$F_f^{(G)}$ depends only on the selected flat subspace and not on the
choice of canonical basis within it. Under
$\mathcal U_f\rightarrow\mathcal U_fQ_f$, with
$Q_f^\dagger Q_f=\mathbb I_r$, the two blocks transform as
$U_f\rightarrow U_fQ_f$ and $V_f\rightarrow V_fQ_f$, leaving
$U_fV_f^\dagger$ unchanged. Individual mode contributions can
depend on the chosen canonical basis, whereas their sum over the
selected flat subspace does not.

Using the projector pole obtained in
Eq.~\eqref{eq:projectorPoleGeneral}, the meromorphic
continuation of the flat-subspace contribution is
\begin{equation}
F_f^{(G)}(z)
=
-\frac{
\Pi_p\mathcal R_g\Pi_h^\dagger
}{
z-z_g
}
+
F_{f,\mathrm{reg}}^{(G)}(z).
\label{eq:flatSubspaceAnomalousPole}
\end{equation}
The matrix $\Pi_p\mathcal R_g\Pi_h^\dagger$ is the 
residue obtained by projecting the full projector residue onto the
annihilation-creation Nambu block. A pole of $P_f(z)$ therefore appears in a
particular anomalous-covariance component $(\alpha,\beta)$ only when
\begin{equation}
\left[
\Pi_p\mathcal R_g\Pi_h^\dagger
\right]_{\alpha\beta}
\neq0 .
\label{eq:selectedAnomalousResidue}
\end{equation}
This condition is stronger than $\mathcal R_g\neq0$. The flat-band
projector can contain the pole while a particular anomalous channel is
insensitive to it because the corresponding projected residue
vanishes.

If $z_g$ is the interior pole of largest modulus with a nonzero residue
in the selected anomalous channel, its real-space contribution is
\begin{equation}
\left[
F_f^{(G)}(R)
\right]_{\alpha\beta}
=
-
\left[
\Pi_p\mathcal R_g\Pi_h^\dagger
\right]_{\alpha\beta}
z_g^{R-1}
+
O(\rho^R),
\qquad
\rho<|z_g|.
\label{eq:flatSubspaceAnomalousAsymptotic}
\end{equation}
The pole position therefore fixes the exponential decay length of
this flat-subspace covariance contribution,
$\xi_F=\xi_P=-a/\ln|z_g|$, while the projected residue fixes its amplitude and the physical
orbital channel in which the tail appears. Whether this pole also
controls the full physical anomalous covariance requires accounting
for the remaining positive-frequency bands.

The full physical ground-state covariance also contains the remaining
positive-frequency bands,
$F^{(G)}(z)=F_f^{(G)}(z)+F_{\bar f}^{(G)}(z)$, where $\bar f$ denotes the positive-frequency complement of the
selected flat subspace.
If $F_{\bar f}^{(G)}(z)$ is regular at $z_g$, the full covariance
inherits the pole in Eq.~\eqref{eq:flatSubspaceAnomalousPole}. If
another positive-frequency band is singular at the same point, its
residue must be included and the pole survives only when the total
residue is nonzero. The asymptotic decay is controlled by $z_g$ only
when no other covariance pole with nonzero residue lies closer to the
unit circle.

The channel selection is particularly transparent in the explicit
two-band model of Sec.~\ref{sec:pairing_flat_band}. The canonical flat-band mode has an $A$ annihilation component and a $B$ creation component, but
no $A$ creation component. Its same-sublattice anomalous $AA$ residue
therefore vanishes, even though the flat-band projector itself has a
nonzero pole. By contrast, the $AB$ anomalous residue is nonzero for
the parameter regime considered in the main text. The second
positive-frequency mode has no $A$ annihilation and $B$ creation
components and hence does not contribute to this $AB$ channel.
Consequently, the physical $AB$ ground-state covariance is exactly the
flat-subspace contribution,
\begin{equation}
F_{AB}^{(G)}(k)
=
\left[
F_f^{(G)}(k)
\right]_{AB}
=
-\frac{q_k}{S(k)} .
\label{eq:explicitCovarianceFromFlatProjector}
\end{equation}
Since $\mathsf G(z)=S(z)$ for the compact rank-one frame,
the relevant projector pole occurs at $z_g=z_S$. Equation~
\eqref{eq:flatSubspaceAnomalousAsymptotic} then gives
$\xi_F=\xi_P=\xi_S$ for the $AB$ anomalous correlation whenever its
residue is nonzero. The next
subsection establishes when the squeezing-sector quantum metric is
controlled by the same complex-momentum zero.

%---------------------------------------------------------------------
\subsection{Squeezing-sector quantum metric}
\label{subsec:squeezing_sector_metric}
%---------------------------------------------------------------------

We next isolate the part of the Bogoliubov geometry associated with a
change of the squeezing transformation. For a normalized
positive-frequency BdG mode
$w_m=(u_m,v_m)^{\mathsf T}$ with
$w_m^\dagger\Sigma_z w_n=\delta_{mn}$, the corresponding canonical
quasiparticle operator is
$\hat b_m=w_m^\dagger\Sigma_z\hat\Psi
=u_m^\dagger\hat{\boldsymbol a}
-v_m^\dagger\hat{\boldsymbol a}^{\dagger}$.
The Gaussian ground state $|G\rangle$ is the common quasiparticle
vacuum, $\hat b_m|G\rangle=0$. The Nambu-conjugate
negative-frequency vector
$\bar w_m=(v_m^*,u_m^*)^{\mathsf T}$ has Krein norm $-1$ and satisfies
$\bar w_m^\dagger\Sigma_z\hat\Psi=-\hat b_m^\dagger$.
Thus positive- and negative-frequency directions correspond,
respectively, to quasiparticle annihilation and creation operators.
For Bloch modes the Nambu conjugation also reverses momentum, which is
left implicit in the following intermediate expressions.

Let $w_f$ denote a normalized mode in the selected positive-frequency
flat band, and let $\vartheta^\mu$ denote a parameter of the BdG
Hamiltonian. It may be crystal momentum or an external control
parameter. Completeness of the bosonic BdG basis gives
\[
\partial_\mu w_f
=
\sum_{m>0}w_m A_{\mu,mf}
-
\sum_{m>0}\bar w_m B_{\mu,mf},
\]
where
$A_{\mu,mf}=w_m^\dagger\Sigma_z\partial_\mu w_f$ and
$B_{\mu,mf}=\bar w_m^\dagger\Sigma_z\partial_\mu w_f$.
The same decomposition follows directly for the quasiparticle
operator. Under an infinitesimal displacement $d\vartheta^\mu$,
\[
\delta\hat b_f
=
d\vartheta^\mu
\sum_{m>0}
\left(
A_{\mu,mf}^*\hat b_m
+
B_{\mu,mf}^*\hat b_m^\dagger
\right).
\]
The terms proportional to $A_{\mu,mf}$ remain within the
positive-frequency annihilation sector. They change the quasiparticle
basis but continue to annihilate the same vacuum. The terms
proportional to $B_{\mu,mf}$ instead introduce creation operators,
\[
\delta_{\rm sq}\hat b_f
=
d\vartheta^\mu
\sum_{m>0}B_{\mu,mf}^*\hat b_m^\dagger .
\]
They measure the part of the infinitesimal deformation that changes
the Bogoliubov transformation defining the vacuum.

Acting on the original quasiparticle vacuum gives
$\delta_{\rm sq}\hat b_f|G\rangle
=d\vartheta^\mu\sum_{m>0}
B_{\mu,mf}^*\hat b_m^\dagger|G\rangle$.
The one-quasiparticle states
$\hat b_m^\dagger|G\rangle$ are orthonormal, and hence
\[
\left\|
\delta_{\rm sq}\hat b_f|G\rangle
\right\|^2
=
d\vartheta^\mu d\vartheta^\nu
\operatorname{Re}
\sum_{m>0}
B_{\mu,mf}^*B_{\nu,mf}.
\]
The positive quadratic form on the right measures how strongly the
displaced annihilation operator fails to annihilate the original
Gaussian vacuum through the appearance of creation-operator
components. This is the local squeezing deformation associated with
the selected Bogoliubov mode.

For a rank-$r$ flat-band subspace, let $\mathcal U_f$ collect its
normalized positive-frequency modes and let $\mathcal U_-$ collect a
complete negative-frequency frame. We define the squeezing-sector quantum metric as the positive magnitude of the
negative-frequency contribution to the symplectic quantum metric,
\begin{equation}
\mathcal B_\mu(k)
=
\mathcal U_-^\dagger(k)\Sigma_z
\partial_\mu\mathcal U_f(k),
\qquad
g_{\mu\nu}^{\rm sq}(k)
=
\operatorname{Re}\operatorname{Tr}
\left[
\mathcal B_\mu^\dagger(k)
\mathcal B_\nu(k)
\right].
\label{eq:squeezingMetricGeneral}
\end{equation}
For one selected mode,
$g_{\mu\nu}^{\rm sq}
=\operatorname{Re}\sum_mB_{\mu,mf}^*B_{\nu,mf}$, reproducing the
vacuum deformation above. The tensor is invariant under unitary changes of the canonical frames.
Under
$\mathcal U_f\to\mathcal U_fQ_f$ and
$\mathcal U_-\to\mathcal U_-Q_-$,
with $Q_f$ and $Q_-$ unitary, the derivative of $Q_f$ drops out by
$\mathcal U_-^\dagger\Sigma_z\mathcal U_f=0$, leaving
$\mathcal B_\mu\to Q_-^\dagger\mathcal B_\mu Q_f$.
The trace in Eq.~\eqref{eq:squeezingMetricGeneral} is therefore
unchanged. Its construction also makes
$g_{\mu\nu}^{\rm sq}$ real, symmetric, and positive semidefinite.

The distinction from the conventional symplectic quantum metric is
important. Because negative-frequency vectors carry negative Krein
signature, they enter the standard symplectic quantum metric with the
opposite sign~\cite{Tesfaye2025}.If
$\mathcal U_{\bar f}$ denotes the positive-frequency complement of the
selected flat-band subspace and
$\mathcal A_\mu
=\mathcal U_{\bar f}^\dagger\Sigma_z\partial_\mu\mathcal U_f$, then
\[
g_{\mu\nu}^{\rm symp}
=
\operatorname{Re}\operatorname{Tr}
\left[
\mathcal A_\mu^\dagger\mathcal A_\nu
\right]
-
g_{\mu\nu}^{\rm sq}.
\]
Equation~\eqref{eq:squeezingMetricGeneral} is therefore the positive magnitude of the
negative-frequency contribution to the symplectic quantum metric.
We isolate this sector because it is precisely the part that changes
the annihilation and creation composition of the quasiparticle modes.

Its meaning becomes transparent for a single
two-mode squeezing channel. Consider
$\hat b_1
=\cosh r\,\hat a_1
-e^{i\phi}\sinh r\,\hat a_2^\dagger$ and
$\hat b_2
=\cosh r\,\hat a_2
-e^{i\phi}\sinh r\,\hat a_1^\dagger$.
Their common vacuum is
$|G_{12}\rangle
=\sqrt{1-|\zeta|^2}
\exp(\zeta\hat a_1^\dagger\hat a_2^\dagger)|0\rangle$,
with
$\zeta=e^{i\phi}\tanh r$.
The parameter $r$ fixes the squeezing amplitude, while $\phi$ fixes
the phase of the squeezed quadrature. The principal collective-quadrature variances are
$e^{-2r}/2$ and $e^{2r}/2$.

An infinitesimal change of $r$ and $\phi$, expressed again in the
quasiparticle basis, gives
\[
\delta\hat b_1
=
-i\sinh^2r\,d\phi\,\hat b_1
-
e^{i\phi}
\left(
dr+i\sinh r\cosh r\,d\phi
\right)
\hat b_2^\dagger .
\]
The first term is confined to the annihilation sector. The second is
the squeezing deformation identified in the general construction.
Its positive coefficient norm is
$dr^2+\sinh^2r\cosh^2r\,d\phi^2$.
Writing
$dr=(\partial_\mu r)d\vartheta^\mu$ and
$d\phi=(\partial_\mu\phi)d\vartheta^\mu$ therefore gives
\begin{equation}
g_{\mu\nu}^{\rm sq}
=
(\partial_\mu r)(\partial_\nu r)
+
\frac14\sinh^2(2r)
(\partial_\mu\phi)(\partial_\nu\phi)
=
\operatorname{Re}
\frac{
(\partial_\mu\zeta)^*
(\partial_\nu\zeta)
}{
(1-|\zeta|^2)^2
}.
\label{eq:rankOneSqueezingMetric}
\end{equation}
The first term measures the deformation of the squeezing amplitude.
The second measures the deformation of the squeezing phase and
vanishes at $r=0$, where the vacuum has no preferred squeezing
quadrature. Equation~\eqref{eq:rankOneSqueezingMetric} therefore
quantifies the change of the squeezing transformation rather than its
magnitude at a fixed parameter value. A squeezed state can have
$g_{\mu\nu}^{\rm sq}=0$ when $r$ and $\phi$ are independent of the
parameter being varied.

For the flat-band model of
Sec.~\ref{sec:pairing_flat_band}, the same squeezing coordinate follows
directly from the canonical flat-band operator,
$\hat b_{f,k}
=[s_k^*\hat a_{A,k}
+p_k\hat a_{B,-k}^\dagger]/\sqrt{S(k)}$.
Writing
$e^{i\chi_k}=s_k^*/|s_k|$ gives
\[
\hat b_{f,k}
=
e^{i\chi_k}
\left[
\cosh r_k\,\hat a_{A,k}
-
e^{i\phi_k}\sinh r_k\,
\hat a_{B,-k}^\dagger
\right],
\]
where
$\zeta_k=e^{i\phi_k}\tanh r_k=-p_k/s_k^*$.
The irrelevant phase $e^{i\chi_k}$ does not affect the vacuum.
Accordingly, the momentum pair $(A,k)$ and $(B,-k)$ is in the
two-mode squeezed factor
$\sqrt{1-|\zeta_k|^2}
\exp(\zeta_k
\hat a_{A,k}^\dagger\hat a_{B,-k}^\dagger)|0\rangle$,
consistent with Eq.~\eqref{eq:squeezedGroundState}.
The quantity $g_{kk}^{\rm sq}(k)$ therefore measures how rapidly this
physical squeezing transformation changes as the momentum is varied.

For the reduced parametrization
$\lambda_0=\lambda_1=\lambda$, $c_0=c$, and $c_1=1$, one has
$\zeta_k=-\lambda(1+e^{ik})/(c+e^{ik})$.
Its derivative is
$\partial_k\zeta_k
=-i\lambda(c-1)e^{ik}/(c+e^{ik})^2$, while the canonical
normalization gives
$1-|\zeta_k|^2
=S(k)/(g|c+e^{ik}|^2)$.
Substitution into
Eq.~\eqref{eq:rankOneSqueezingMetric} yields
\begin{equation}
g_{kk}^{\rm sq}(k)
=
\frac{
|\partial_k\zeta_k|^2
}{
\left(1-|\zeta_k|^2\right)^2
}
=
\frac{
g^2\lambda^2(c-1)^2
}{
S^2(k)
}
\equiv
\frac{C}{S^2(k)},
\qquad
C=g^2\lambda^2(c-1)^2 .
\label{eq:explicitSqueezingMetric}
\end{equation}
For the general coefficients of the main text, the numerator is
$g^2(c_0\lambda_1-\lambda_0c_1)^2$.
When $c_0\lambda_1=\lambda_0c_1$, the squeezing coordinate
$\zeta_k=-p_k/s_k^*$ is momentum independent and
$g_{kk}^{\rm sq}(k)=0$, even though the squeezing amplitude can remain
finite. Thus the metric measures the momentum variation of the squeezing
transformation rather than the squeezing magnitude alone.
The enhancement of the metric
when $S(k)$ becomes small therefore has a direct physical origin.
The canonical flat-band annihilation operator becomes increasingly
sensitive to momentum through its creation-operator component, or
equivalently through the momentum dependence of the ground-state
squeezing amplitude and squeezing phase.

Finally, the positive-frequency transverse contribution vanishes for
this particular two-band model. The two normalized
positive-frequency modes are
$w_f=S^{-1/2}(s_k,0,0,-p_k^*)^{\mathsf T}$ and
$w_c=S^{-1/2}(0,s_k^*,-p_k,0)^{\mathsf T}$.
Their nonzero Nambu components are complementary, so
$w_c^\dagger\Sigma_z\partial_kw_f=0$ and hence
$g_{kk}^{++}=0$. The nontrivial transverse variation of the flat-band
mode is therefore entirely in the negative-frequency sector. Its
positive magnitude is $g_{kk}^{\rm sq}$ in
Eq.~\eqref{eq:explicitSqueezingMetric}, while the conventional
Krein-signed symplectic quantum metric obeys
$g_{kk}^{\rm symp}=-g_{kk}^{\rm sq}$ for this model.
In a general multiband bosonic BdG system,
positive-frequency transverse components need not vanish, and the
full symplectic quantum metric contains both the positive-frequency
and negative-frequency contributions with their respective Krein
signs.
%---------------------------------------------------------------------
%---------------------------------------------------------------------
%---------------------------------------------------------------------
%---------------------------------------------------------------------
\subsubsection{Negative-frequency channel and double-pole corollary}
\label{subsec:double_pole_corollary}
%---------------------------------------------------------------------

We now determine how the squeezing-sector metric inherits the
singularities associated with canonical normalization of the compact
flat-band frame. On the physical Brillouin zone,
$\mathcal U_f=\mathcal W\mathsf G^{-1/2}$, where
$\mathsf G=\mathcal W^\dagger\Sigma_z\mathcal W$ is positive definite.
Differentiation gives
$\partial_\mu\mathcal U_f
=(\partial_\mu\mathcal W)\mathsf G^{-1/2}
+\mathcal W\partial_\mu\mathsf G^{-1/2}$.
The second term has no projection onto the negative-frequency sector
because
$\mathcal U_-^\dagger\Sigma_z\mathcal W
=\mathcal U_-^\dagger\Sigma_z
\mathcal U_f\mathsf G^{1/2}=0$.
Hence
$\mathcal B_\mu=\mathcal C_\mu\mathsf G^{-1/2}$, where
$\mathcal C_\mu
=\mathcal U_-^\dagger\Sigma_z\partial_\mu\mathcal W$.
Equation~\eqref{eq:squeezingMetricGeneral} then becomes
\begin{equation}
g_{\mu\nu}^{\rm sq}(k)
=
\operatorname{Re}\operatorname{Tr}
\left[
\mathsf G^{-1}(k)\mathcal K_{\mu\nu}(k)
\right],
\qquad
\mathcal K_{\mu\nu}
=
\mathcal C_\mu^\dagger\mathcal C_\nu .
\label{eq:metricGramInverse}
\end{equation}
The inverse Gram matrix supplies the normalization factors responsible
for the singular behavior, while $\mathcal K_{\mu\nu}$ determines
whether the corresponding variation of the compact frame has a
nonzero projection onto negative-frequency creation directions.

Equation~\eqref{eq:metricGramInverse} is invariant under regular invertible changes of the
compact generating frame. Under a regular invertible transformation
$\mathcal W\rightarrow\mathcal W\mathcal A$, Krein orthogonality
removes the term proportional to $\partial_\mu\mathcal A$, giving
$\mathcal C_\mu\rightarrow\mathcal C_\mu\mathcal A$.
On the physical Brillouin zone,
$\mathsf G\rightarrow\mathcal A^\dagger\mathsf G\mathcal A$ and
$\mathcal K_{\mu\nu}\rightarrow
\mathcal A^\dagger\mathcal K_{\mu\nu}\mathcal A$.
Cyclicity of the trace then leaves
Eq.~\eqref{eq:metricGramInverse} unchanged.

For the complex-momentum analysis we use $z=e^{-ik}$ and the
reciprocal adjoint introduced above. We first assume that the
negative-frequency frame can be chosen locally analytic and
nonsingular near the Gram-matrix zero under consideration. In this neighborhood, the meromorphic continuation of the
squeezing-sector quantum metric kernel is
\begin{equation}
\mathfrak g_{\mu\nu}^{\rm sq}(z)
=
\operatorname{Tr}
\left[
\mathsf G^{-1}(z)\mathcal K_{\mu\nu}(z)
\right],
\qquad
\mathcal K_{\mu\nu}(z)
=
\widetilde{\mathcal C}_\mu(z)
\mathcal C_\nu(z).
\label{eq:metricMeromorphicContinuation}
\end{equation}
Its restriction to the unit circle gives the corresponding physical
metric after taking the real part.

Let $z_g$ be the simple Gram-matrix zero introduced in
Sec.~\ref{subsec:projector_residue}, with
$\mathsf G(z_g)r_g=0$,
$\ell_g^\dagger\mathsf G(z_g)=0$, and
$\alpha_g=\ell_g^\dagger\mathsf G'(z_g)r_g\neq0$.
If $\mathcal K_{\mu\nu}(z)$ is regular at $z_g$,
Eq.~\eqref{eq:GramInversePole} gives
\begin{equation}
\mathfrak g_{\mu\nu}^{\rm sq}(z)
=
\frac{
\ell_g^\dagger
\mathcal K_{\mu\nu}(z_g)
r_g
}{
\alpha_g(z-z_g)
}
+
\mathfrak g_{\mu\nu}^{\rm sq,reg}(z).
\label{eq:metricSimplePoleGeneral}
\end{equation}
The analytically continued Gram matrix therefore develops a null
direction at $z_g$, but the selected squeezing-sector component
inherits the corresponding pole only when
\begin{equation}
\ell_g^\dagger
\mathcal K_{\mu\nu}(z_g)
r_g
\neq0 .
\label{eq:metricResidueCondition}
\end{equation}
The singular Gram-matrix direction must therefore acquire a nonzero
negative-frequency component under the variation
$\partial_\mu\mathcal W$. If this projection vanishes, the leading
normalization singularity is absent from that metric component.

A higher pole order can occur when the negative-frequency mode
entering the projection has a normalization factor that
vanishes at the same complex momentum. Consider a rank-one channel with compact
positive-frequency and negative-frequency representatives $W_+(z)$
and $W_-(z)$. Define
$S_+(z)=\widetilde W_+(z)\Sigma_zW_+(z)$ and
$S_-(z)=-\widetilde W_-(z)\Sigma_zW_-(z)$.
On the physical Brillouin zone these quantities are positive and the
normalized vectors
$w_+=W_+/\sqrt{S_+}$ and
$w_-=W_-/\sqrt{S_-}$ have Krein norms $+1$ and $-1$,
respectively. Since
$\widetilde W_-\Sigma_zW_+=0$, differentiation of the normalization
does not contribute to the intersector projection. Defining
$\Gamma_\mu(z)
=\widetilde W_-(z)\Sigma_z\partial_\mu W_+(z)$ gives
\begin{equation}
\mathfrak g_{\mu\nu}^{\rm sq,(-)}(z)
=
\frac{
\widetilde\Gamma_\mu(z)\Gamma_\nu(z)
}{
S_-(z)S_+(z)
}.
\label{eq:partnerMetricGeneral}
\end{equation}
If $S_+(z)$ has a simple zero at $z_g$, while
$S_-(z_g)\neq0$ and the projection numerator is nonzero, this channel
has a simple pole.

For the case relevant to the explicit model, the positive-frequency
vector and the active negative-frequency vector carry the same
normalization,
$S_-(z)=S_+(z)\equiv S(z)$. Equation~\eqref{eq:partnerMetricGeneral}
then reduces to
\begin{equation}
\mathfrak g_{\mu\nu}^{\rm sq,(-)}(z)
=
\frac{
\widetilde\Gamma_\mu(z)\Gamma_\nu(z)
}{
S^2(z)
}.
\label{eq:sharedNormalizationMetric}
\end{equation}
Let $z_g$ be a simple zero of $S(z)$ and assume that the projection
numerator is nonzero there. Expansion about $z_g$ gives
\begin{equation}
\mathfrak g_{\mu\nu}^{\rm sq,(-)}(z)
=
\frac{\mathcal Q_{\mu\nu,g}}{(z-z_g)^2}
+
\frac{\mathcal L_{\mu\nu,g}}{z-z_g}
+
O(1),
\label{eq:metricDoublePoleExpansion}
\end{equation}
where
\begin{equation}
\mathcal Q_{\mu\nu,g}
=
\frac{
\widetilde\Gamma_\mu(z_g)\Gamma_\nu(z_g)
}{
[S'(z_g)]^2
}.
\label{eq:metricDoublePoleCoefficient}
\end{equation}
Writing
$N_{\mu\nu}(z)
=\widetilde\Gamma_\mu(z)\Gamma_\nu(z)$,
the coefficient of the simple-pole term is
$\mathcal L_{\mu\nu,g}
=N_{\mu\nu}'(z_g)/[S'(z_g)]^2
-N_{\mu\nu}(z_g)S''(z_g)/[S'(z_g)]^3$.

We use the same real-space Fourier convention as for the anomalous
covariance,
$\mathfrak g(R)
=\int_{-\pi}^{\pi}\dd k\,
e^{-ikR}\mathfrak g(k)/(2\pi)$.
With $z=e^{-ik}$ this becomes
\[
\mathfrak g_{\mu\nu}^{\rm sq,(-)}(R)
=
\oint_{|z|=1}
\frac{\dd z}{2\pi i}\,
z^{R-1}
\mathfrak g_{\mu\nu}^{\rm sq,(-)}(z),
\]
where the contour is counterclockwise. The factor $z^{R-1}$ is the
product of the Fourier factor $z^R$ and the contour measure
$\dd z/z$. If $z_g$ is the unique contributing interior singularity
of largest modulus, residue evaluation gives
\begin{equation}
\mathfrak g_{\mu\nu}^{\rm sq,(-)}(R)
=
(R-1)\mathcal Q_{\mu\nu,g}z_g^{R-2}
+
\mathcal L_{\mu\nu,g}z_g^{R-1}
+
O(\rho^R),
\qquad
\rho<|z_g|.
\label{eq:metricDoublePoleFourier}
\end{equation}
If several active singularities have the same largest modulus, their
contributions must be summed. The double pole changes the algebraic
prefactor but not the exponential scale. For
$\mathcal Q_{\mu\nu,g}\neq0$, the contribution associated with
$z_g$ decays with
$\xi_{\rm sq}=-a/\ln|z_g|$.

For the explicit model of Sec.~\ref{sec:pairing_flat_band}, the
negative-frequency projection can be evaluated directly before making
the reduced parametrization. The selected compact
positive-frequency vector is
$W_+=W_f=(s_k,0,0,-p_k^*)^{\mathsf T}$.
Its own negative-frequency partner
$W_-^{(f)}=(0,-p_k^*,s_k,0)^{\mathsf T}$
has complementary Nambu support and satisfies
$[W_-^{(f)}]^\dagger\Sigma_z\partial_kW_f=0$.
Momentum variation of $W_f$ therefore has no projection onto this
creation direction.

The nonzero negative-frequency direction is the Nambu-conjugate
partner of the second positive-frequency mode,
$W_-=\Sigma_xW_c^*(-k)
=(-p_k,0,0,s_k^*)^{\mathsf T}$.
Both $W_f$ and $W_-$ have normalization $S(k)$. Their unnormalized
projection amplitude is
$\Gamma_k
=W_-^\dagger\Sigma_z\partial_kW_f
=s_k\partial_kp_k^*
-p_k^*\partial_ks_k$.
For the general coefficients,
\[
p_k^*
=
\sqrt g\left(\lambda_0+\lambda_1e^{-ik}\right),
\qquad
s_k
=
\sqrt g\left(c_0+c_1e^{-ik}\right),
\]
and hence
$\partial_kp_k^*
=-i\sqrt g\,\lambda_1e^{-ik}$ and
$\partial_ks_k
=-i\sqrt g\,c_1e^{-ik}$.
It follows that
\[
\Gamma_k
=
-i g
\left(c_0\lambda_1-\lambda_0c_1\right)e^{-ik}.
\]
With $z=e^{-ik}$,
$\Gamma_k(z)
=-ig(c_0\lambda_1-\lambda_0c_1)z$ and
$\widetilde\Gamma_k(z)
=ig(c_0\lambda_1-\lambda_0c_1)z^{-1}$, so
\begin{equation}
\widetilde\Gamma_k(z)\Gamma_k(z)
=
g^2
\left(c_0\lambda_1-\lambda_0c_1\right)^2 .
\label{eq:explicitConversionProduct}
\end{equation}
Since the two participating vectors have the same normalization,
Eq.~\eqref{eq:partnerMetricGeneral} gives
\begin{equation}
\mathfrak g_{kk}^{\rm sq,(-)}(z)
=
\frac{
g^2(c_0\lambda_1-\lambda_0c_1)^2
}{
S^2(z)
},
\qquad
g_{kk}^{\rm sq}(k)
=
\frac{
g^2(c_0\lambda_1-\lambda_0c_1)^2
}{
S^2(k)
}.
\label{eq:explicitMetricFromConversionChannel}
\end{equation}
This reproduces Eq.~\eqref{eq:explicitSqueezingMetric} from the
negative-frequency projection itself. The normalized flat-band
projector and the selected anomalous covariance contain one inverse
power of $S(z)$, whereas the squared negative-frequency projection
contains two. A simple active zero of $S(z)$ therefore produces
simple poles in the former quantities and a double pole in the
squeezing-sector metric when the projection numerator is nonzero.
Their algebraic real-space prefactors differ, while the exponential
length is fixed by the same root.

For the reduced parametrization
$\lambda_0=\lambda_1=\lambda$, $c_0=c$, and $c_1=1$,
Eq.~\eqref{eq:explicitConversionProduct} becomes
\[
\widetilde\Gamma_k(z)\Gamma_k(z)
=
g^2\lambda^2(c-1)^2
\equiv C,
\]
and
$g_{kk}^{\rm sq}(k)=C/S^2(k)$.

The simple-zero analysis does not apply at $c=1$. At this
point $C=0$, while
\[
S(z)
=
g(1-\lambda^2)\frac{(z+1)^2}{z},
\]
so the reciprocal zeros coalesce into a double zero at $z=-1$ on the
physical contour. In particular, $S'(z_S)$ vanishes together with the
numerator entering
Eq.~\eqref{eq:metricDoublePoleCoefficient}, and the simple-zero
Laurent expansion is no longer applicable. Moreover, $S(\pi)=0$, so the
positive-definite Gram condition required for the canonical
flat-band construction fails at $c=1$. The limit
$c\to1$ must therefore be taken from within the positive-norm regime
and analyzed separately and the point $c=1$ cannot be treated as an
ordinary numerator cancellation within the simple-zero result.

%---------------------------------------------------------------------
\subsubsection{Integrated squeezing-sector quantum metric}
%---------------------------------------------------------------------

The momentum-resolved squeezing-sector quantum metric also admits a real-space
interpretation. Let $\ket{w_{f,a;R}}$ and $\ket{w_{-,m;R}}$ denote
canonical Wannier vectors obtained by Fourier transforming the columns
of $\mathcal U_f(k)$ and $\mathcal U_-(k)$, respectively. The ket
notation here refers to single-mode Nambu vectors rather than
many-body states. For smooth periodic frames,
$\ket{w_{f,a;R}}
=L^{-1/2}\sum_k e^{-ikR}
\ket{k}\otimes\ket{\mathcal U_{f,a}(k)}$, with the analogous
definition for the negative-frequency sector.

Let $\hat N$ be the dimensionless cell-position operator,
$\hat N\ket{R,\alpha;\sigma}=R\ket{R,\alpha;\sigma}$.
Its matrix element between the positive- and negative-frequency
Wannier sectors is
\begin{equation}
X_{ma}(R)
\equiv
\bra{w_{-,m;R}}
\Sigma_z\hat N
\ket{w_{f,a;0}}
=
\frac{i}{L}
\sum_k e^{ikR}
\left[
\mathcal B_k(k)
\right]_{ma}.
\label{eq:WannierConversionMatrixElement}
\end{equation}
Terms proportional to the overlap of the two frequency sectors vanish
by Krein orthogonality. The same matrix
$\mathcal B_k=\mathcal U_-^\dagger\Sigma_z\partial_k\mathcal U_f$
that measures the local creation-sector deformation in momentum space
therefore determines the corresponding position matrix elements
between canonical Wannier vectors.

Parseval's identity gives
\begin{equation}
\mathcal G
\equiv
\frac{1}{2\pi}
\int_{-\pi}^{\pi}\dd k\,
g_{kk}^{\rm sq}(k)
=
\sum_{R,a,m}
\left|
\bra{w_{-,m;R}}
\Sigma_z\hat N
\ket{w_{f,a;0}}
\right|^2,
\qquad
\Omega_{\rm sq}\equiv a^2\mathcal G .
\label{eq:squeezingWannierSpread}
\end{equation}
Thus $\mathcal G$ measures the total squared position-matrix-element weight
connecting the positive-frequency flat-band Wannier sector to the
negative-frequency sector. Equivalently, it is a dimensionless
geometric second moment, while
$\Omega_{\rm sq}=a^2\mathcal G$ is the corresponding quantity with dimensions
of length squared. This geometric spread is not a
probability variance and is distinct from the ground-state anomalous
covariance. A shift
$\hat N\rightarrow\hat N-N_0$ leaves
Eq.~\eqref{eq:squeezingWannierSpread} unchanged because the added term
is proportional to the vanishing Krein overlap between the two
frequency sectors.

For the general four-coefficient model,
\[
S(k)=S_0+2S_1\cos k,
\qquad
S_0
=
g(c_0^2+c_1^2-\lambda_0^2-\lambda_1^2),
\qquad
S_1
=
g(c_0c_1-\lambda_0\lambda_1),
\]
and Sec.~\ref{subsec:double_pole_corollary} gives
$g_{kk}^{\rm sq}(k)=C_{\rm gen}/S^2(k)$, where
$C_{\rm gen}=g^2(c_0\lambda_1-\lambda_0c_1)^2$.
Introducing
$D_S=\sqrt{S_0^2-4S_1^2}$, the standard identity
$(2\pi)^{-1}\int_{-\pi}^{\pi}\dd k/S(k)=D_S^{-1}$
implies, after differentiation with respect to $S_0$,
$(2\pi)^{-1}\int_{-\pi}^{\pi}\dd k/S^2(k)=S_0/D_S^3$.
Hence
\begin{equation}
\mathcal G
=
\frac{
g^2(c_0\lambda_1-\lambda_0c_1)^2S_0
}{
\left(S_0^2-4S_1^2\right)^{3/2}
}.
\label{eq:integratedSqueezingMetric}
\end{equation}
Equation~\eqref{eq:integratedSqueezingMetric} is valid throughout the
positive-norm regime $S_0>2|S_1|$.

The relation to the exponential length follows from the interior zero
$z_S$ of $S(z)$. Writing
$|z_S|=e^{-\kappa_S}$ with $\kappa_S=a/\xi_S$ gives
$S_0=2|S_1|\cosh\kappa_S$ and
$D_S=2|S_1|\sinh\kappa_S$. Equation~\eqref{eq:integratedSqueezingMetric}
can therefore be written as
\[
\mathcal G
=
\frac{
g^2(c_0\lambda_1-\lambda_0c_1)^2
}{
4S_1^2
}
\frac{\cosh\kappa_S}{\sinh^3\kappa_S}.
\]
For $\kappa_S\ll1$,
\begin{equation}
\mathcal G
\simeq
\frac{
g^2(c_0\lambda_1-\lambda_0c_1)^2
}{
4S_1^2
}
\left(
\frac{\xi_S}{a}
\right)^3 .
\label{eq:metricGenericAsymptotic}
\end{equation}
This is the generic cubic growth of the integrated squeezing-sector
metric when a simple active zero approaches the unit circle while the
dimensionless prefactor
$g^2(c_0\lambda_1-\lambda_0c_1)^2/S_1^2$ remains finite and nonzero.
The power law can change when the negative-frequency projection
vanishes simultaneously with the approach of the root to the unit
circle.

We now specialize to the reduced parametrization
$\lambda_0=\lambda_1=\lambda$, $c_0=c$, and $c_1=1$, for which
\[
S_0=g(c^2+1-2\lambda^2),
\qquad
S_1=g(c-\lambda^2),
\qquad
C=g^2\lambda^2(c-1)^2 .
\]
Two different approaches to the positive-norm boundary are then
possible.

For $S_1<0$, the minimum of $S(k)$ occurs at $k=0$, and the interior
zero approaches $z_S=+1$ as
$S_0+2S_1\rightarrow0$. The exact identity
$S_0-2S_1=g(c-1)^2$ gives, on approaching this boundary,
$g(c-1)^2\rightarrow4|S_1|$. Consequently
$C/(4S_1^2)\rightarrow g\lambda^2/|S_1|$, so the
negative-frequency projection retains finite weight and
\begin{equation}
\mathcal G
\simeq
\frac{g\lambda^2}{|S_1|}
\left(
\frac{\xi_S}{a}
\right)^3,
\qquad
S_1<0 .
\label{eq:metricCubicBranch}
\end{equation}

The approach to $c=1$ is qualitatively different. In this regime
$S_1>0$, the minimum of $S(k)$ occurs at $k=\pi$, and the interior
zero approaches $z_S=-1$. The exact relation
$S_0-2S_1=g(c-1)^2$, together with
$S_0=2S_1\cosh\kappa_S$, gives
\[
g(c-1)^2
=
2S_1(\cosh\kappa_S-1)
=
4S_1\sinh^2\!\left(\frac{\kappa_S}{2}\right).
\]
Hence, for $\kappa_S\ll1$,
$\kappa_S\simeq|c-1|/\sqrt{1-\lambda^2}$ and
$S_1\rightarrow g(1-\lambda^2)$. The negative-frequency projection
weight simultaneously vanishes as
$C=g\lambda^2S_1\kappa_S^2+O(\kappa_S^4)$.
Substitution into the general expression for $\mathcal G$ removes two
powers of the generic $\kappa_S^{-3}$ divergence and gives
\begin{equation}
\mathcal G
\simeq
\frac{g\lambda^2}{4S_1}
\left(
\frac{\xi_S}{a}
\right),
\qquad
S_1>0 .
\label{eq:metricLinearBranch}
\end{equation}
Equation~\eqref{eq:metricLinearBranch} describes the approach to
$c=1$ from within the positive-norm regime. It does not apply at the
this point itself. At $c=1$, the two reciprocal zeros coalesce at
$z=-1$, $S(\pi)=0$, and the canonical isolated-band construction
ceases to satisfy the positive-definite Gram condition, as discussed
in Sec.~\ref{subsec:double_pole_corollary}.

The length associated with the integrated second moment is
\begin{equation}
\ell_{\rm sq}
\equiv
\sqrt{\Omega_{\rm sq}}
=
a\sqrt{\mathcal G}.
\label{eq:squeezingSpreadLength}
\end{equation}
The quantities $\ell_{\rm sq}$ and $\xi_S$ therefore probe different
aspects of the same Bogoliubov geometry. The exponential length
$\xi_S$ is fixed solely by the position of the complex zero of
$S(z)$ and controls the asymptotic spatial envelope. In contrast,
$\ell_{\rm sq}=a\sqrt{\mathcal G}$ is an integrated second-moment
length and depends both on this exponential scale and on the weight of
the negative-frequency projection. Consequently, the two lengths need
not scale linearly. For a finite projection weight,
$\ell_{\rm sq}/a\propto(\xi_S/a)^{3/2}$ in the long-length limit,
whereas on the $c\rightarrow1$ branch
$\ell_{\rm sq}/a\propto(\xi_S/a)^{1/2}$. These different powers
reflect changes in the geometric weight carried by the same long-distance geometric structure rather than different exponential decay lengths.
%---------------------------------------------------------------------
%---------------------------------------------------------------------
%---------------------------------------------------------------------
\subsection{Common exponential decay length: theorem and proof}
\label{subsec:common_pole_theorem}
%---------------------------------------------------------------------

We now establish the model-independent relation between the
localization of the canonical flat-band projector, the decay of the
ground-state anomalous covariance, and the real-space kernel of the
squeezing-sector quantum metric.

\paragraph{\emph{Theorem.}}
Consider a stable bosonic BdG lattice with an isolated
positive-frequency flat-band subspace generated by a finite-Laurent
frame $\mathcal W(z)$. Let $z_g$, with $0<|z_g|<1$, be a simple zero of
$\det\mathsf G(z)$ for which $\mathsf G(z_g)$ has one-dimensional
right and left null spaces,
$\mathsf G(z_g)r_g=0$ and
$\ell_g^\dagger\mathsf G(z_g)=0$, with
$\alpha_g=\ell_g^\dagger\mathsf G'(z_g)r_g\neq0$.
Suppose that the zero is nonremovable from the canonical flat-band
projector, so that $\mathcal R_g\neq0$, and that the selected anomalous
covariance has nonzero projected residue,
$[\Pi_p\mathcal R_g\Pi_h^\dagger]_{\alpha\beta}\neq0$.
For the full physical anomalous covariance, assume additionally that the
remaining positive-frequency bands do not cancel this residue at $z_g$.
For the squeezing-sector quantum metric, assume that the negative-frequency
projection associated with the same Gram-matrix direction is nonzero.
When $\mathcal K_{\mu\nu}(z)$ is regular at $z_g$, this requires
$\ell_g^\dagger\mathcal K_{\mu\nu}(z_g)r_g\neq0$. For the rank-one
same-normalization channel of
Eq.~\eqref{eq:sharedNormalizationMetric}, the corresponding condition
is
$\widetilde\Gamma_\mu(z_g)\Gamma_\nu(z_g)\neq0$.
Finally, assume that no singularity with nonzero weight and modulus
larger than $|z_g|$ contributes to any of the three selected kernels.
Then their asymptotic exponential decay lengths are identical,
\begin{equation}
\xi_P
=
\xi_F
=
\xi_{\rm sq}
=
-\frac{a}{\ln|z_g|}.
\label{eq:commonPoleTheorem}
\end{equation}
Here $\xi_P$ is the localization length of the canonical flat-band
projector, $\xi_F$ is the decay length of the selected component of the full
physical ground-state anomalous covariance, and $\xi_{\rm sq}$ is the
decay length of the real-space squeezing-sector quantum metric kernel.

\paragraph{\emph{Proof.}}
Near a simple nonremovable Gram-matrix zero,
Eq.~\eqref{eq:projectorPoleGeneral} gives a simple pole of the
normalized flat-band projector with residue $\mathcal R_g$.
Its Fourier transform therefore contains a nonzero term proportional
to $z_g^{R-1}$, as shown in
Eq.~\eqref{eq:projectorAsymptoticGeneral}, and hence
$\xi_P=-a/\ln|z_g|$.

The flat-band contribution to the anomalous covariance is obtained by
projecting the same normalized projector between the annihilation and
creation Nambu sectors,
$F_f^{(G)}=-\Pi_pP_f\Pi_h^\dagger$.
Equation~\eqref{eq:flatSubspaceAnomalousPole} therefore carries the
same pole at $z_g$. When
$[\Pi_p\mathcal R_g\Pi_h^\dagger]_{\alpha\beta}\neq0$, the selected
covariance component contains a nonzero term proportional to
$z_g^{R-1}$ and consequently
$\xi_F=-a/\ln|z_g|$.

For the squeezing-sector quantum metric, the pole order depends on the
negative-frequency channel. If
$\mathcal K_{\mu\nu}(z)$ is regular at $z_g$,
Eq.~\eqref{eq:metricSimplePoleGeneral} gives a simple pole and the
corresponding Fourier coefficient again scales as $z_g^{R-1}$.
If the positive- and negative-frequency vectors entering the
projection carry the same normalization, the kernel instead contains
the double pole of
Eq.~\eqref{eq:metricDoublePoleExpansion}. Its Fourier coefficient is
\[
(R-1)\mathcal Q_{\mu\nu,g}z_g^{R-2}
+
\mathcal L_{\mu\nu,g}z_g^{R-1}.
\]
The additional power of $R$ is algebraic and does not alter the
exponential factor. A nonzero negative-frequency projection therefore
gives
$\xi_{\rm sq}=-a/\ln|z_g|$ in either case. Since no contributing
singularity lies closer to the unit circle, $z_g$ controls the
long-distance behavior of all three kernels, proving
Eq.~\eqref{eq:commonPoleTheorem}.

For the full ground-state anomalous covariance,
$F^{(G)}=F_f^{(G)}+F_{\bar f}^{(G)}$, the equality
$\xi_F=-a/\ln|z_g|$ additionally requires that the remaining
positive-frequency bands do not cancel the residue at $z_g$ and do
not introduce a contributing singularity of larger modulus. In the
explicit two-band model of Sec.~\ref{sec:pairing_flat_band}, the
second positive-frequency mode does not contribute to the selected
$AB$ anomalous channel, so the full $AB$ covariance inherits the
flat-band decay length directly.

The theorem identifies a common exponential scale, not a common pole
order or a common residue. The canonical flat-band projector and the selected
anomalous covariance generally contain
one inverse power of the Gram matrix. In the explicit rank-one model
this gives the factor $S^{-1}(z)$. The active squeezing-sector channel
instead contains $S^{-2}(z)$. Their real-space kernels can
therefore differ by algebraic factors while retaining the same
exponential decay length.

The result also does not identify the exponential length with the
integrated squeezing-sector length
$\ell_{\rm sq}=a\sqrt{\mathcal G}$. The former is determined by the
position of the dominant complex-momentum singularity. The latter is
a second-moment quantity and also depends on the weight carried by the
negative-frequency projection. As shown above, this distinction leads
to different scaling relations between $\ell_{\rm sq}$ and $\xi_S$
along different approaches to the positive-norm boundary.

A zero of $\det\mathsf G(z)$ is therefore a candidate source of
long-distance structure rather than, by itself, a statement about a
particular observable. It controls a selected covariance or
squeezing-sector channel only when the corresponding projected
residue is nonzero. If several active singularities have the same
largest modulus, their contributions must be summed and may produce
oscillatory or additional algebraic structure without changing the
common exponential scale. A singularity with larger modulus instead
sets the asymptotic decay.

For the pairing-generated model,
$\mathsf G(z)=S(z)$, the selected $AB$ anomalous covariance is
proportional to $S^{-1}(z)$, and the active squeezing-sector kernel is
proportional to $S^{-2}(z)$. Away from zeros of their respective
numerators, the same interior root $z_S$ therefore gives
$\xi_P=\xi_F=\xi_{\rm sq}=\xi_S$. The point $c=1$ is excluded from
this simple-zero theorem. There the reciprocal roots coalesce at
$z=-1$, $S(\pi)=0$, and the positive-definite Gram condition required for the canonical
flat-band construction fails at $c=1$.
%---------------------------------------------------------------------
\section{Weak dispersion, defects, and boundary localization}
%---------------------------------------------------------------------

%---------------------------------------------------------------------
\subsection{Persistence beyond exact flatness under weak dispersion}
\label{sec:SM_two_bulk_lengths}
%---------------------------------------------------------------------

We first determine how the decay scale of the exactly flat model
evolves when the target band acquires a weak dispersion. We add a
translationally invariant nearest-neighbor hopping
$t_f$ on the $A$ sublattice, without introducing a defect or
boundary. Writing $x_k=2\cos k$, $P_k=|p_k|^2$, and
$T_k=|s_k|^2$, the number-conserving block becomes
$K_{t_f}(k)=
\operatorname{diag}(\omega_f+P_k+t_fx_k,T_k-\omega_f)$,
while the parametric-pairing block is unchanged. In the
$A$-annihilation, $B$-creation sector, the dynamical matrix is
\begin{equation}
M_{t_f}^{(AB)}(k)
=
\begin{pmatrix}
\omega_f+P_k+t_fx_k & q_k\\
-q_k^* & \omega_f-T_k
\end{pmatrix}.
\label{eq:SM_Mactive}
\end{equation}

Define $S(k)=T_k-P_k$ and $H(k)=T_k+P_k$. The eigenfrequency splitting
of Eq.~\eqref{eq:SM_Mactive} is
\begin{align}
\mathcal D_{t_f}^2(k)
&=
\left[H(k)+t_fx_k\right]^2-4|q_k|^2
\nonumber\\
&=
S^2(k)+2t_fx_kH(k)+t_f^2x_k^2 ,
\label{eq:SM_Dt}
\end{align}
where the second equality follows from
$|q_k|^2=P_kT_k$. The two eigenfrequencies in this Nambu sector are
$\omega_\pm(k)
=\omega_f+[-S(k)+t_fx_k\pm\mathcal D_{t_f}(k)]/2$.
The branch continuously connected to the flat positive-frequency band
is therefore
$\omega_{f,t_f}(k)
=\omega_f+
[\mathcal D_{t_f}(k)-S(k)+t_fx_k]/2$.
For weak $t_f$,
$\mathcal D_{t_f}
=S+t_fx_kH/S+O(t_f^2)$, giving
$\omega_{f,t_f}(k)
=\omega_f+t_fx_kT_k/S(k)+O(t_f^2)$.
Thus the perturbation generates a finite bandwidth while continuously
deforming the flat-band eigenvectors. Since the unperturbed BdG
Hamiltonian is strictly positive and the flat band is spectrally
isolated, energetic stability and band isolation persist for
sufficiently small $|t_f|$.

The ground-state anomalous covariance follows from the corresponding
two-mode Bogoliubov transformation. Writing
$q_k=\rho_ke^{i\phi_k}$, the squeezing parameter satisfies
$\tanh(2r_k)=2\rho_k/[H(k)+t_fx_k]$ and hence
$\sinh(2r_k)=2\rho_k/\mathcal D_{t_f}(k)$. The physical $AB$
anomalous covariance is therefore
\begin{equation}
F_{AB}^{(t_f)}(k)
=
-\frac{1}{2}e^{i\phi_k}\sinh(2r_k)
=
-\frac{q_k}{\mathcal D_{t_f}(k)} .
\label{eq:SM_Fexact}
\end{equation}
At $t_f=0$, positivity of $S(k)$ fixes the physical branch
$\mathcal D_0(k)=S(k)$, and
Eq.~\eqref{eq:SM_Fexact} reduces to the flat-band result
$F_{AB}^{(G)}(k)=-q_k/S(k)$.

For the general four-coefficient model,
$S(k)=S_0+S_1x_k$ and $H(k)=H_0+H_1x_k$, with
\[
S_0
=
g(c_0^2+c_1^2-\lambda_0^2-\lambda_1^2),
\qquad
S_1
=
g(c_0c_1-\lambda_0\lambda_1),
\]
and
\[
H_0
=
g(c_0^2+c_1^2+\lambda_0^2+\lambda_1^2),
\qquad
H_1
=
g(c_0c_1+\lambda_0\lambda_1).
\]
Equation~\eqref{eq:SM_Dt} is quadratic in
$x=z+z^{-1}$,
\[
\mathcal D_{t_f}^2(x)
=
d_0+d_1x+d_2x^2,
\]
where
$d_0=S_0^2$,
$d_1=2(S_0S_1+t_fH_0)$, and
$d_2=S_1^2+2t_fH_1+t_f^2$.
Using the same complex-momentum convention as in the flat-band
analysis, $z=e^{-ik}$, multiplication by $z^2$ gives the reciprocal
quartic
\begin{equation}
z^2\mathcal D_{t_f}^2(z)
=
d_2z^4+d_1z^3+(d_0+2d_2)z^2+d_1z+d_2 .
\label{eq:SM_quartic}
\end{equation}
The palindromic coefficients imply reciprocal roots. Equivalently, the
two solutions in the $x$ plane are
\begin{equation}
x_{\pm}
=
\frac{
-d_1\pm\sqrt{d_1^2-4d_0d_2}
}{
2d_2
},
\label{eq:SM_xroots}
\end{equation}
and each $x_\nu$ generates the reciprocal pair
\begin{equation}
z_{\nu,\pm}
=
\frac{x_\nu\pm\sqrt{x_\nu^2-4}}{2},
\qquad
z_{\nu,+}z_{\nu,-}=1 .
\label{eq:SM_zroots}
\end{equation}
We denote by $z_\nu$ the member of each pair lying inside the unit
circle.

For the general model, the quadratic discriminant is
\begin{equation}
d_1^2-4d_0d_2
=
8t_fS_0\left(S_1H_0-S_0H_1\right)
+
4t_f^2\left(H_0^2-S_0^2\right).
\label{eq:SM_discriminant}
\end{equation}
When the quantity in Eq.~\eqref{eq:SM_discriminant} is positive and
both $x_\pm$ lie outside $[-2,2]$, the two solutions generate real
interior roots with generically different moduli. Their associated
bulk decay lengths are
\begin{equation}
\xi_1
=
-\frac{a}{\ln|z_1|},
\qquad
\xi_2
=
-\frac{a}{\ln|z_2|}.
\label{eq:SM_twoxi}
\end{equation}
When the discriminant is negative, $x_+$ and $x_-$ form a
complex-conjugate pair. The corresponding interior roots have equal
modulus and give oscillatory contributions with a common exponential
envelope. Two split complex roots therefore need not imply two
different decay lengths.

For the reduced parametrization
$\lambda_0=\lambda_1=\lambda$, $c_0=c$, and $c_1=1$,
Eq.~\eqref{eq:SM_discriminant} becomes
\[
d_1^2-4d_0d_2
=
16g^2\lambda^2t_f
\left[
2t_f(c^2+1)-S_0(c-1)^2
\right].
\]
For $S_0>0$, $\lambda\neq0$, and $c\neq1$, the term linear in $t_f$
controls the sign sufficiently close to the flat-band limit.
Consequently, sufficiently small negative $t_f$ produces two real
split $x$ roots, whereas sufficiently small positive $t_f$ produces a
complex-conjugate pair. This sign relation is specific to the reduced
parametrization.

The origin of the two roots is particularly transparent at
$t_f=0$. Equation~\eqref{eq:SM_quartic} then factorizes as
\[
z^2\mathcal D_0^2(z)
=
\left(
S_1z^2+S_0z+S_1
\right)^2 .
\]
The two interior zeros coincide at the parent overlap zero $z_S$, and
the two exterior zeros coincide at $z_S^{-1}$. Weak dispersion
resolves these doubled zeros into the two reciprocal pairs of
Eq.~\eqref{eq:SM_zroots}. The resulting decay channels therefore
evolve continuously from the single exponential scale $\xi_S$ of the
exactly flat model.

The analytic structure of the anomalous covariance changes when
$t_f\neq0$. A simple zero $z_j$ of
$\mathcal D_{t_f}^2(z)$ is a square-root branch point of
$\mathcal D_{t_f}^{-1}(z)$. Provided $q(z_j)\neq0$,
\[
F_{AB}^{(t_f)}(z)
\simeq
-q(z_j)
\left[
\left.
\partial_z\mathcal D_{t_f}^2(z)
\right|_{z_j}
\right]^{-1/2}
(z-z_j)^{-1/2}.
\]
Using the same Fourier convention as for the flat-band anomalous
covariance,
$F(R)=\int_{-\pi}^{\pi}\dd k\,
e^{-ikR}F(k)/(2\pi)$, the branch-cut contribution has the
large-distance form
\begin{equation}
F_{A0,BR}^{(t_f)}(R)
\sim
\frac{\mathcal A_1}{\sqrt R}\,z_1^R
+
\frac{\mathcal A_2}{\sqrt R}\,z_2^R,
\qquad
R\rightarrow+\infty .
\label{eq:SM_Fasymptotic}
\end{equation}
The $R$-independent powers of $z_j$ associated with the precise
contour convention are absorbed into $\mathcal A_j$. A decay channel
is present only when its branch-point coefficient is nonzero. If
$|z_1|\neq|z_2|$, the root of larger modulus controls the ultimate
long-distance tail, while the second root produces a shorter-range
contribution. The opposite spatial direction contains the same
exponential scales, although its amplitudes need not be identical.

The two finite-dispersion decay channels should not be confused with the
separate $R=0$ contribution of the exactly flat covariance. This local
contribution does not define an exponential decay length. At
finite $t_f$, $z_1$ and $z_2$ are distinct nonanalytic bulk
singularities and therefore define genuine decay channels whenever
their coefficients are nonzero.

The same roots also govern the squeezing-sector metric. Define
$A(z)=H(z)+t_fx(z)$,
$Q(z)=q(z)\widetilde q(z)$, and
$\mathcal D_{t_f}^2(z)=A^2(z)-4Q(z)$. On the physical Brillouin zone,
$Q(k)=|q_k|^2=\rho_k^2$. From
$\tanh(2r)=2\rho/A$ one obtains
\[
(\partial_kr)^2
=
\frac{
\left(A\partial_kQ-2Q\partial_kA\right)^2
}{
4Q\left(\mathcal D_{t_f}^2\right)^2
}.
\]
Moreover,
$2i\,\partial_k\phi
=(\partial_kq)/q
-(\partial_k\widetilde q)/\widetilde q$ and
$\sinh(2r)=2\sqrt Q/\mathcal D_{t_f}$.
Substitution into
Eq.~\eqref{eq:rankOneSqueezingMetric} gives the analytically continued
positive squeezing-sector kernel
\begin{equation}
\mathfrak g_{kk}^{\rm sq}(z)
=
\frac{
\left[
A\,\partial_kQ-2Q\,\partial_kA
\right]^2
}{
4Q\left[\mathcal D_{t_f}^2\right]^2
}
-
\frac{Q}{4\mathcal D_{t_f}^2}
\left(
\frac{\partial_kq}{q}
-
\frac{\partial_k\widetilde q}{\widetilde q}
\right)^2 ,
\label{eq:SM_metricdisp}
\end{equation}
where
$\partial_k=-iz\partial_z$ for $z=e^{-ik}$. Its restriction to the
unit circle reproduces the physical squeezing-sector metric after
taking the real part.

At a simple zero $z_j$ of $\mathcal D_{t_f}^2(z)$ with
$Q(z_j)\neq0$, the first term in
Eq.~\eqref{eq:SM_metricdisp} contains a double pole, while
the second contains a simple pole. Zeros of their respective
numerators can remove either contribution. When the corresponding
coefficients are nonzero, Fourier transformation gives
\begin{equation}
\mathfrak g_R^{\rm sq}
\sim
\sum_{j=1,2}
\left(
\alpha_jR+\beta_j
\right)z_j^R,
\qquad
R\rightarrow+\infty .
\label{eq:SM_metricasymptotic}
\end{equation}
The anomalous covariance and the squeezing-sector metric therefore
have different algebraic envelopes, respectively
$R^{-1/2}$ and generically $R$, while inheriting the same exponential
scales $\xi_1$ and $\xi_2$ whenever the relevant coefficients are
nonzero.

In the limit $t_f\rightarrow0$, the two branch points coalesce at
$z_S$ and
$\mathcal D_{t_f}^2(z)\rightarrow S^2(z)$. Selecting the physical
branch $\mathcal D_0(z)=S(z)$ recovers the flat-band covariance
$F_{AB}^{(G)}=-q/S$ and its simple pole at $z_S$. The two dispersive
decay channels thereby collapse continuously onto the single
flat-band exponential length $\xi_S$.

\begin{figure}[H]
  \centering
  \includegraphics[width=0.98\linewidth]{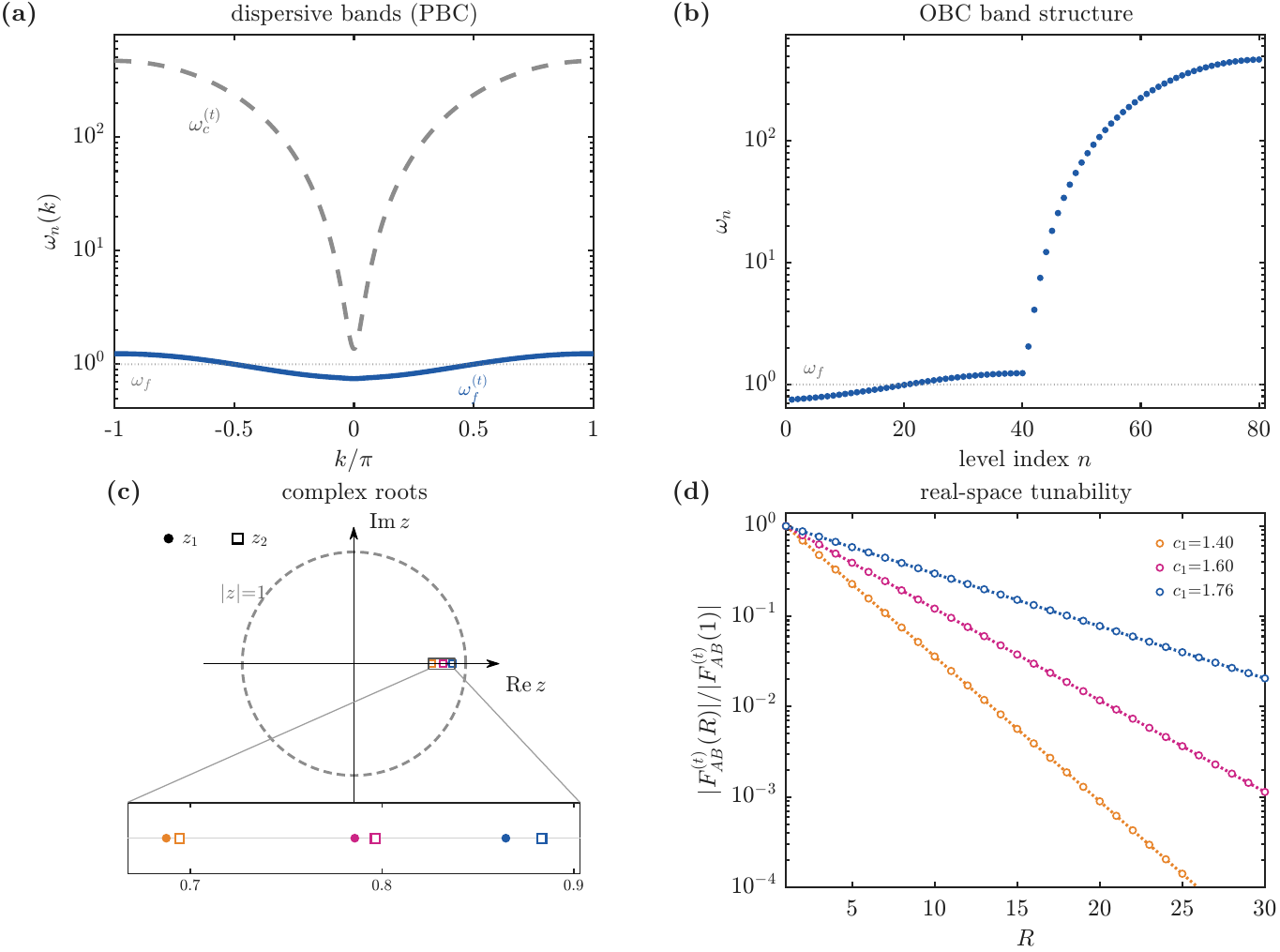}
  \caption{\textbf{Weak-dispersion splitting of the flat-band
singularity.} The general four-coefficient model of the main text
($c_0=-2.037$, $\lambda_0=0.017$, $\lambda_1=0.043$, $g=32.5$)
is perturbed by a weak translationally invariant $A$-sublattice
hopping $t_f$, with no defect.
(a) Positive-frequency PBC spectrum for $c_1=1.76$ and
$t_f=-0.12$. The target band is weakly dispersive, energetically
stable, and isolated from the remaining positive-frequency band.
(b) OBC spectrum of the same translation-invariant chain, showing the
bulk gap and the absence of an in-gap edge state.
(c) Interior roots for $c_1=1.40,1.60,1.76$. The parent flat-band
root $z_S$ splits into $z_1$ (filled) and $z_2$ (open), while the
exterior reciprocal roots are not shown.
(d) Normalized anomalous covariance
$|F_{AB}^{(t_f)}(R)|/|F_{AB}^{(t_f)}(1)|$ (markers, $c_1=1.40,1.60,1.76$)
with two-root fits $A_1 z_1^{R}+A_2 z_2^{R}$ (dotted) built from the split
roots $z_1,z_2$ of panel (c). The larger-modulus root sets the asymptotic
tail. At $c_1=1.76$ the split roots give $\xi_1=6.87a$ and $\xi_2=8.06a$. The
strict asymptotic form carries an additional $R^{-1/2}$ factor
[Eq.~\eqref{eq:SM_Fasymptotic}].}
  \label{fig:Sdisp}
\end{figure}
Figure~\ref{fig:Sdisp} shows the weakly dispersive spectrum, the split
complex roots, and the resulting anomalous-correlation decay.
%---------------------------------------------------------------------
%---------------------------------------------------------------------
%---------------------------------------------------------------------
\subsection{Compact-generator projection and localized decay channels}
\label{sec:localized_decay_channels}
%---------------------------------------------------------------------

The defect and boundary problems considered below require one further
step beyond the bulk singularity analysis of
Sec.~\ref{sec:SM_two_bulk_lengths}. A localized perturbation has direct overlaps with the compact flat-band
generators, while the corresponding state in their span is obtained
after accounting for their nonorthogonality. Let $\mathbb W=(\ldots,W_{-1},W_0,W_1,\ldots)$ be
the real-space compact frame. Any vector in the parent flat-band
subspace can be written as
$\ket{\psi_f}=\mathbb Wc=\sum_Rc_R\ket{W_R}$, and the overlaps of the
compact vectors form the Gram matrix
$G=\mathbb W^\dagger\Sigma_z\mathbb W$.

Consider a Nambu vector $\ket{\chi}$ localized on finitely many cells.
Its Krein-orthogonal projection onto the flat-band subspace is the
vector $\ket{\psi_f}=\mathbb Wc$ for which
\[
\mathbb W^\dagger\Sigma_z
\left(\ket{\chi}-\mathbb Wc\right)=0.
\]
Defining
$u=\mathbb W^\dagger\Sigma_z\ket{\chi}$ gives
$Gc=u$. The component
$u_R=\bra{W_R}\Sigma_z\ket{\chi}$ is the direct overlap of the
localized vector with the compact generator centered at $R$.
Consequently,
\[
c=G^{-1}u,
\qquad
\ket{\psi_f}
=
\mathbb W G^{-1}\mathbb W^\dagger\Sigma_z\ket{\chi}
=
P_f\ket{\chi}.
\]
Thus $G^{-1}$ is not an additional assumption. It converts the direct
overlaps with a nonorthogonal compact frame into the coefficients of
the physical projected state.

For the rank-one band of Sec.~\ref{sec:pairing_flat_band},
$\mathsf G(k)=S(k)$ and, after
analytic continuation, $\mathsf G(z)=S(z)$. A localized vector has a
finite-Laurent overlap $u(z)$, while its projected coefficients are
$c(z)=u(z)/S(z)$. When the projector pole at $z_S$ is nonremovable and
$u(z_S)\neq0$, the interior zero $S(z_S)=0$ produces a simple-pole
contribution to the projected response proportional to $z_S^{R-1}$.
Its exponential length is
$\xi_P=\xi_S=-a/\ln|z_S|$. A source with $u(z_S)=0$ does not couple to this leading projector-pole
contribution. 

This projected response should be distinguished from an individual
canonical Bogoliubov mode. Canonical normalization contains
$G^{-1/2}=S^{-1/2}$ and gives the
$R^{-1/2}z_S^R$ envelope derived in
Eq.~\eqref{eq:hR}. Projection of a localized
vector contains $G^{-1}=S^{-1}$ and therefore a simple-pole
contribution. The square-root and pole singularities arise from the
same compact-generator overlap zero but describe different physical
objects.

When the parent band is dispersive, spatial response is governed by
the full bulk resolvent rather than by the isolated flat-band
projector. For a finite-range source $b_R$ at frequency $\omega$,
let $b(z)$ be its finite-Laurent transform. The response to the right
of the source is
\[
\psi_R
=
\oint_{|z|=1}\frac{\dd z}{2\pi i}\,
z^{R-1}
[\omega\mathbb I-M(z)]^{-1}b(z).
\]
At exact flatness, the selected-band contribution to the resolvent is
$P_f(z)/(\omega-\omega_f)$, so the inverse-Gram pole appears directly
in the frequency-domain response. At finite dispersion, the relevant
spatial roots are instead the singularities of the full resolvent at
the frequency of the physical response.

For an isolated simple interior root $z_j$, define
$A(\omega,z)=\omega\mathbb I-M(z)$ and choose right and left null
vectors satisfying
$A(\omega,z_j)r_j=0$ and
$\ell_j^\dagger A(\omega,z_j)=0$. If
$\ell_j^\dagger\left.\partial_zA(\omega,z)\right|_{z=z_j}r_j\neq0$,
the pole of the resolvent gives
\begin{equation}
\psi_R
=
\sum_{|z_j|<1}
z_j^{R-1}
\frac{
r_j\ell_j^\dagger b(z_j)
}{
\ell_j^\dagger
\left.\partial_z[\omega\mathbb I-M(z)]\right|_{z=z_j}
r_j
}.
\label{eq:rootExpansion}
\end{equation}
Only roots with nonzero source overlap
$\ell_j^\dagger b(z_j)$ and nonzero weight in the measured component
contribute to that response. Their exponential lengths are
$\xi_j=-a/\ln|z_j|$. If the physical resolvent-source product has a
higher-order pole, the Fourier coefficient acquires the corresponding
polynomial factor in $R$. Such factors are fixed by the actual Laurent
pole order of the response and not by the multiplicity of the
characteristic determinant alone.

Equation~\eqref{eq:rootExpansion} is the bulk rule used below. A local
defect fixes the response frequency self-consistently and then selects
the residues of the allowed bulk roots. An open boundary instead
imposes a homogeneous matching condition. In the dimerized boundary
construction, that matching condition is most transparent in the
nonorthogonal compact-generator basis.
%---------------------------------------------------------------------
%---------------------------------------------------------------------
%---------------------------------------------------------------------
\subsection{Local defect bound state and spatial decay}
\label{sec:defect_state}
%---------------------------------------------------------------------

We now apply the bulk-resolvent construction to a local defect. The
translation-invariant system retains the weak $A$-sublattice hopping
$t_f$ introduced in Sec.~\ref{sec:SM_two_bulk_lengths}, and we add the
local onsite potential
\begin{equation}
\delta\hat H_d
=
v_d\hat a_{A,0}^\dagger\hat a_{A,0}.
\label{eq:defect2}
\end{equation}
Let $M_0$ denote the defect-free dynamical matrix including $t_f$ and
let $V_d$ be the corresponding defect contribution. A bound-state
Nambu eigenvector $\Phi_d$ at frequency $\omega_d$ obeys
\[
(M_0+V_d)\Phi_d=\omega_d\Phi_d,
\]
or
\[
(\omega_d\mathbb I-M_0)\Phi_d=V_d\Phi_d.
\]
For a bound-state frequency in a bulk spectral gap, the defect-free
resolvent
$\mathcal R_0(\omega)=(\omega\mathbb I-M_0)^{-1}$ is nonsingular on
the physical Brillouin zone. The eigenvalue equation is therefore
exactly equivalent to
\[
\Phi_d=\mathcal R_0(\omega_d)V_d\Phi_d.
\]

The model separates into the $A$-annihilation, $B$-creation sector and
its Nambu-conjugate partner. In the former,
$V_d=v_de_Ae_A^\dagger$ with $e_A=(1,0)^{\mathsf T}$. Hence
$V_d\Phi_d=v_d\Phi_{d,A}(0)e_A$, and
\[
\Phi_d
=
v_d\Phi_{d,A}(0)\mathcal R_0(\omega_d)e_A.
\]
Taking the $A$ component at the defect cell gives
\[
\Phi_{d,A}(0)
=
v_d\mathcal R_{AA}^{(0)}(\omega_d;0)\Phi_{d,A}(0),
\]
where
\[
\mathcal R_{AA}^{(0)}(\omega;0)
\equiv
\bra{A,0}\mathcal R_0(\omega)\ket{A,0}
=
\int_{-\pi}^{\pi}\frac{\dd k}{2\pi}
\left[
\omega\mathbb I-M_{t_f}^{(AB)}(k)
\right]^{-1}_{AA}.
\]
A nonzero bound-state amplitude therefore requires
\begin{equation}
1-v_d\mathcal R_{AA}^{(0)}(\omega_d;0)=0.
\label{eq:Tmatrix}
\end{equation}
Equation~\eqref{eq:Tmatrix} determines the allowed defect frequencies. A physical
bound-state solution must be normalizable, have positive frequency, and
carry positive Krein norm.

The spatial decay is determined by the defect-free bulk equation at
this same frequency. With $z=e^{-ik}$, $x=z+z^{-1}$, and
$\delta=\omega-\omega_f$, the analytically continued active-sector
matrix is
\[
\omega\mathbb I-M_{t_f}^{(AB)}(z)
=
\begin{pmatrix}
\delta-P(z)-t_fx & -q(z)\\
\widetilde q(z) & \delta+T(z)
\end{pmatrix}.
\]
Using $q(z)\widetilde q(z)=P(z)T(z)$ and
$S(z)=T(z)-P(z)$ gives
\begin{equation}
D_2(\omega,z)
\equiv
\det[\omega\mathbb I-M_{t_f}^{(AB)}(z)]
=
\delta[\delta+S(z)]
-
t_fx[\delta+T(z)].
\label{eq:D2}
\end{equation}
The possible defect decay factors are therefore the interior roots of
$D_2(\omega_d,z)=0$. The frequency must first be fixed by
Eq.~\eqref{eq:Tmatrix}. Evaluating the characteristic equation at the
parent frequency $\omega_f$ would in general give different spatial
roots.

Let $\phi_d(R)$ denote the two-component active-sector amplitude of
$\Phi_d$ in cell $R$. Away from the defect,
\[
\phi_d(R)
=
v_d\Phi_{d,A}(0)
\oint_{|z|=1}\frac{\dd z}{2\pi i}\,
z^{R-1}
[\omega_d\mathbb I-M_{t_f}^{(AB)}(z)]^{-1}e_A.
\]
Writing $\delta_d=\omega_d-\omega_f$, direct inversion gives
\[
[\omega_d\mathbb I-M_{t_f}^{(AB)}(z)]^{-1}e_A
=
\frac{1}{D_2(\omega_d,z)}
\begin{pmatrix}
\delta_d+T(z)\\
-\widetilde q(z)
\end{pmatrix}.
\]
Thus the zeros of $D_2(\omega_d,z)$ determine the available
exponential factors, while the numerator determines their weights in
the two Nambu components. For simple contributing roots,
Eq.~\eqref{eq:rootExpansion} becomes
\[
\phi_d(R)
=
v_d\Phi_{d,A}(0)
\sum_{|z_j|<1}
z_j^{R-1}
\frac{
\begin{pmatrix}
\delta_d+T(z_j)\\
-\widetilde q(z_j)
\end{pmatrix}
}{
\partial_zD_2(\omega_d,z_j)
}.
\]
A root is absent from the $A$-annihilation component when
$\delta_d+T(z_j)=0$ and from the $B$-creation component when
$\widetilde q(z_j)=0$. The asymptotic decay of a selected component is
set by the largest-modulus interior root with nonzero residue in that
component.

The connection to the parent Gram singularity is explicit at exact
flatness. Setting $t_f=0$ gives
$D_2(\omega,z)=\delta[\delta+S(z)]$ and
\[
[\omega\mathbb I-M_0^{(AB)}(z)]^{-1}e_A
=
\frac{1}{\delta}
\frac{1}{S(z)}
\begin{pmatrix}
T(z)\\
-\widetilde q(z)
\end{pmatrix}
+
\frac{1}{\delta+S(z)}
\frac{1}{S(z)}
\begin{pmatrix}
-P(z)\\
\widetilde q(z)
\end{pmatrix}.
\]
The first vector is the flat-band projection of a local $A$
excitation,
\[
P_f^{(AB)}(z)e_A
=
\frac{1}{S(z)}
\begin{pmatrix}
T(z)\\
-\widetilde q(z)
\end{pmatrix}.
\]
Its factor $S^{-1}(z)$ is the rank-one inverse Gram matrix of
Sec.~\ref{subsec:projector_residue}. The second term is the
contribution of the complementary active-sector mode.

At finite detuning, the apparent $S^{-1}(z)$ singularities of the two
spectral terms cancel in the complete resolvent. The physical spatial
pole of the exactly flat defect satisfies
\[
S(z_d)=-\delta_d.
\]
For $S_1\neq0$, this equation fixes one value of
$x=z+z^{-1}$ and hence one reciprocal pair of roots. The exactly flat
defect therefore has one interior decay factor. In the resonant limit
$\delta_d\rightarrow0$,
\[
z_d\rightarrow z_S,
\qquad
\xi_d=-\frac{a}{\ln|z_d|}\rightarrow\xi_S.
\]
The intrinsic Gram zero is recovered as the resonant limit of the
physical defect root. It is not an additional pole of the full
resolvent at finite detuning.

Finite dispersion can produce a second defect channel. For the
general four-coefficient model,
$S(z)=S_0+S_1x$ and $T(z)=T_0+T_1x$, with
$T_0=g(c_0^2+c_1^2)$ and $T_1=gc_0c_1$. At the self-consistent defect
frequency, Eq.~\eqref{eq:D2} becomes
\[
-t_fT_1x^2
+
[\delta_dS_1-t_f(\delta_d+T_0)]x
+
\delta_d(\delta_d+S_0)
=
0.
\]
For $t_fT_1\neq0$, the equation is quadratic in $x$. Each solution
generates a reciprocal pair through $z^2-xz+1=0$, so two interior
evanescent roots can contribute to the defect profile. These roots are
poles of the frequency-resolved bulk resolvent at $\omega_d$. They
need not coincide with the branch points controlling the equal-time
bulk anomalous covariance in Sec.~\ref{sec:SM_two_bulk_lengths}, which
are instead determined by zeros of $\mathcal D_{t_f}^2(z)$.

The parent geometric root is recovered in the joint near-flat and
near-resonant limit. For
$\delta_d\neq0$, Eq.~\eqref{eq:D2} can be written as
\[
\delta_d+S(z)
-
\frac{t_f}{\delta_d}
x[\delta_d+T(z)]
=
0.
\]
When $\delta_d\rightarrow0$ and
$t_f/\delta_d\rightarrow0$, one solution approaches $S(z)=0$ and
therefore $z_S$. If $S_1T_1\neq0$, the second solution is driven to
large $|x|$, so its interior reciprocal root approaches the origin
and becomes short ranged. 
%---------------------------------------------------------------------
%---------------------------------------------------------------------
%--------------------------------------------------------------------

%---------------------------------------------------------------------
\subsection{Dimerized boundary mode and geometric decay}
\label{sec:dimerized_boundary}
%---------------------------------------------------------------------

A localized boundary mode can be generated without a local defect by
alternating the couplings between neighboring compact flat-band
Bogoliubov generators. The alternating coupling has the SSH form in the
basis of these generators~\cite{SSH1979}. Because the compact
generators are nonorthogonal, the boundary eigenvector in this basis is
not itself the physical BdG wave function. The bulk dimerization opens
a gap around the parent flat-band frequency, while the physical
open-boundary profile is obtained after accounting for the Gram
matrix. This produces a geometric contribution to the boundary decay.

%---------------------------------------------------------------------
\subsubsection{Dimerized bulk continuation and PBC spectrum}
\label{sec:dimerized_bulk}
%---------------------------------------------------------------------

We use the reduced parametrization
$\lambda_0=\lambda_1=\lambda$, $c_0=c$, and $c_1=1$. The compact
flat-band Bogoliubov generators are
\begin{equation}
\hat\beta_R
=
\sqrt g\left[
c\hat a_{A,R}
+\hat a_{A,R+1}
+\lambda\hat a_{B,R}^\dagger
+\lambda\hat a_{B,R+1}^\dagger
\right].
\label{eq:betaDimerR}
\end{equation}
Let $W_R$ be the corresponding compact Nambu vector, defined by
$\hat\beta_R=W_R^\dagger\Sigma_z\hat\Psi$. It satisfies
$M_0W_R=\omega_fW_R$, and its translations span the parent flat-band
subspace. Their noncanonical commutators are
\begin{equation}
G_{RR'}
\equiv
[\hat\beta_R,\hat\beta_{R'}^\dagger]
=
S_0\delta_{RR'}
+
S_1
\left(
\delta_{R,R'+1}
+
\delta_{R,R'-1}
\right),
\label{eq:GramRealTop}
\end{equation}
where
$S_0=g(c^2+1-2\lambda^2)$ and
$S_1=g(c-\lambda^2)$.

We add the alternating coupling
\begin{equation}
\hat H_{\rm dim}
=
\epsilon\sum_m
\left[
t_1\hat\beta_{2m}^\dagger\hat\beta_{2m+1}
+
t_2\hat\beta_{2m+1}^\dagger\hat\beta_{2m+2}
+
\mathrm{H.c.}
\right].
\label{eq:HdimGamma}
\end{equation}
With the normalization of Eq.~\eqref{eq:betaDimerR}, the factor
$\epsilon$ sets the overall strength of the perturbation, while
$t_1$ and $t_2$ specify its alternating pattern. The ratio $t_1/t_2$ determines the decay factor of the boundary
solution derived below. Because each
$\hat\beta_R$ has support on two neighboring primitive cells,
Eq.~\eqref{eq:HdimGamma} remains a finite-range quadratic
perturbation.

Group two neighboring compact generators into a doubled cell,
$\hat\beta_{x,m}=\hat\beta_{2m}$ and
$\hat\beta_{y,m}=\hat\beta_{2m+1}$, and denote the doubled-cell
quasimomentum by $K$. Then
\begin{equation}
\hat H_{\rm dim}
=
\epsilon\sum_K
\hat{\bm\beta}_K^\dagger
\mathcal T_{\rm d}(K)
\hat{\bm\beta}_K,
\qquad
\mathcal T_{\rm d}(K)
=
\begin{pmatrix}
0&h(K)\\
h^*(K)&0
\end{pmatrix},
\label{eq:HdimK}
\end{equation}
with $h(K)=t_1+t_2e^{-iK}$. Let
$W_d(K)=(W_x(K),W_y(K))$ be the compact frame in the doubled cell, so
$\hat{\bm\beta}_K=W_d^\dagger(K)\Sigma_z\hat\Psi_{d,K}$. The
positive-frequency contribution to the dynamical matrix is
\[
\delta M_+(K)
=
\epsilon
W_d(K)\mathcal T_{\rm d}(K)
W_d^\dagger(K)\Sigma_z.
\]
Its Nambu-conjugate completion is
$\delta M_-(K)=-\Sigma_x\delta M_+^*(-K)\Sigma_x$, giving
\begin{equation}
M_{\rm bdry}(K)
=
M_{0,d}(K)+\delta M_+(K)+\delta M_-(K),
\qquad
H_{\rm BdG,bdry}(K)=\Sigma_zM_{\rm bdry}(K).
\label{eq:MbdryKfull}
\end{equation}

The doubled-cell Gram matrix is
\begin{equation}
G_d(K)
=
W_d^\dagger(K)\Sigma_zW_d(K)
=
\begin{pmatrix}
S_0&\eta_G(K)\\
\eta_G^*(K)&S_0
\end{pmatrix},
\qquad
\eta_G(K)=S_1(1+e^{-iK}).
\label{eq:GdoubleK}
\end{equation}
The positive-norm condition of the parent flat band implies
$G_d(K)>0$ for real $K$ and therefore
$\det G_d(K)=S_0^2-|\eta_G(K)|^2>0$.

For a state $\ket{\Psi_K}=W_d(K)c_K$ in the doubled flat-band
subspace, the parent contribution is
$M_{0,d}W_dc_K=\omega_fW_dc_K$. The dimerized contribution acts as
\[
\delta M_+W_dc_K
=
\epsilon W_d\mathcal T_{\rm d}G_dc_K,
\]
while the Nambu-conjugate contribution annihilates this
positive-frequency subspace by Krein orthogonality. The projected
bulk eigenproblem is therefore
\begin{equation}
M_{\rm bdry}(K)W_d(K)c_K
=
W_d(K)
\left[
\omega_f\mathbb I_2
+
\epsilon\mathcal T_{\rm d}(K)G_d(K)
\right]c_K.
\label{eq:projectedDispersion}
\end{equation}
Although $\mathcal T_{\rm d}G_d$ is generally non-Hermitian in this
nonorthogonal coefficient basis, it is similar to the Hermitian matrix
$G_d^{1/2}\mathcal T_{\rm d}G_d^{1/2}$. Its eigenvalues are therefore
real.

Defining
$r(K)=\operatorname{Re}[h(K)\eta_G^*(K)]$, the two shifts of the
parent flat-band frequency are
\begin{equation}
\mu_\pm(K)
=
r(K)
\pm
\sqrt{
r^2(K)
+
|h(K)|^2
\left[S_0^2-|\eta_G(K)|^2\right]
}.
\label{eq:muPMexact}
\end{equation}
The two frequencies are
$\omega_\pm(K)=\omega_f+\epsilon\mu_\pm(K)$. Since
$\det[\mathcal T_{\rm d}(K)G_d(K)]
=-|h(K)|^2\det G_d(K)<0$ whenever $h(K)\neq0$, the two shifts have
opposite signs,
\begin{equation}
\omega_-(K)<\omega_f<\omega_+(K).
\label{eq:gapAroundWf}
\end{equation}
For real positive $t_1$ and $t_2$, the gap closes when
$t_1=t_2$ at $K=\pi$. Since $\eta_G(\pi)=0$, the two frequencies at
$K=\pi$ are
\begin{equation}
\omega_\pm(\pi)
=
\omega_f
\pm
\epsilon S_0|t_1-t_2|.
\label{eq:piGapTop}
\end{equation}
The loop $h(K)=t_1+t_2e^{-iK}$ winds once around the origin for
$|t_1|<|t_2|$ and does not enclose it for $|t_1|>|t_2|$, as in the
SSH chain~\cite{SSH1979}. This winding characterizes the projected
two-component dimerized problem. It is not by itself a topological
invariant of the complete microscopic bosonic BdG Hamiltonian.

The other positive-frequency bands are the folded copies of the parent
dispersive band,
\begin{equation}
\omega_{{\rm c},j}(K)
=
S(k_j)-\omega_f,
\qquad
k_0=\frac{K}{2},
\qquad
k_1=\frac{K}{2}+\pi.
\label{eq:foldedDispersiveBands}
\end{equation}
They are unaffected because the perturbation acts entirely within the
parent flat-band quasiparticle subspace. For sufficiently small
$\epsilon$, the two bands derived from the parent flat band remain separated from
these folded bands and the full bosonic BdG Hamiltonian remains energetically
stable.

%---------------------------------------------------------------------
\subsubsection{Exact OBC boundary state at the parent frequency}
%---------------------------------------------------------------------

For an open chain, let $\mathbb W$ collect the compact flat-band
vectors that fit inside the sample and define
$G_{\rm OBC}=\mathbb W^\dagger\Sigma_z\mathbb W$. For the independent
compact vectors retained here, the parent positive-norm condition
makes this finite Gram matrix positive definite, so
$G_{\rm OBC}^{-1}$ exists. A state in the parent flat-band subspace
has the form $\ket{\Psi}=\mathbb Wc$. Its direct overlaps with the
compact vectors are
\[
d
\equiv
\mathbb W^\dagger\Sigma_z\ket{\Psi}
=
G_{\rm OBC}c.
\]
Thus $c$ gives the coefficients of the physical state in the compact
frame, whereas $d$ gives its overlaps with that nonorthogonal frame.

At the parent frequency $\omega_f$, the projected open-chain
eigenvalue equation reduces to
\[
\mathcal T_{\rm OBC}G_{\rm OBC}c=0,
\]
and hence
\[
\mathcal T_{\rm OBC}d=0.
\]
For an open chain with an odd number of dimerization sites and
termination on the $x$ sublattice, $\mathcal T_{\rm OBC}$ has a zero vector
with $d_{y,m}=0$. The remaining equation
on each $y$ site is
$t_1d_{x,m}+t_2d_{x,m+1}=0$, so
\begin{equation}
d_{x,m}=\zeta_t^m,
\qquad
d_{y,m}=0,
\qquad
\zeta_t=-\frac{t_1}{t_2}.
\label{eq:dSSHedge}
\end{equation}
For $|t_1|<|t_2|$, this overlap vector is localized at the exposed
boundary.

The physical flat-band coefficients follow from
$c=G_{\rm OBC}^{-1}d$, giving
\begin{equation}
\ket{\Psi_{\rm edge}}
=
\mathbb Wc
=
\mathbb W G_{\rm OBC}^{-1}d.
\label{eq:PsiEdgeExact}
\end{equation}
The overlap vector $d$ and the physical BdG boundary state are
therefore different objects. The physical coefficients are obtained
from $c=G_{\rm OBC}^{-1}d$, which accounts for the nonorthogonality of
the compact frame.

For this termination, the boundary state remains exactly at
$\omega_f$. The parent gives
$M_{0,\rm OBC}\ket{\Psi_{\rm edge}}
=\omega_f\ket{\Psi_{\rm edge}}$, while
\[
\delta M_+\ket{\Psi_{\rm edge}}
=
\epsilon
\mathbb W\mathcal T_{\rm OBC}G_{\rm OBC}c
=
\epsilon
\mathbb W\mathcal T_{\rm OBC}d
=
0.
\]
The Nambu-conjugate contribution also vanishes by Krein orthogonality.
Therefore
\begin{equation}
M_{\rm bdry}\ket{\Psi_{\rm edge}}
=
\omega_f\ket{\Psi_{\rm edge}}.
\label{eq:edgeEnergyExact}
\end{equation}
The zero vector of $\mathcal T_{\rm OBC}$ therefore gives a positive-frequency
BdG boundary mode at $\omega_f$. By Eq.~(B24), this mode lies between
the two bulk bands derived from the parent flat band.

%---------------------------------------------------------------------
\subsubsection{Geometric contribution to the boundary decay}
%---------------------------------------------------------------------

The two possible boundary decay factors are already visible in
Eq.~\eqref{eq:PsiEdgeExact}. The overlap vector $d$ contains the
dimerization factor $\zeta_t^m$, while $G_{\rm OBC}^{-1}$ can introduce a second
singularity. To analyze it, define the doubled-cell continuation
$\zeta=e^{-iK}$. Since the doubled-cell lattice constant is $2a$, a separation of $m$
doubled cells corresponds to the physical distance $2ma$. Under
unfolding, $K=2k$ modulo $2\pi$, and therefore
\[
\zeta=e^{-iK}=e^{-2ik}=z^2,
\]
where $z=e^{-ik}$ is the primitive-cell Bloch factor.

With this convention, the analytically continued doubled-cell Gram
matrix is
\begin{equation}
G_d(\zeta)
=
\begin{pmatrix}
S_0&S_1(1+\zeta)\\
S_1(1+\zeta^{-1})&S_0
\end{pmatrix},
\label{eq:GzetaTop}
\end{equation}
with determinant
\begin{equation}
D_G(\zeta)
=
S_0^2
-
S_1^2\left(2+\zeta+\zeta^{-1}\right).
\label{eq:detGzeta}
\end{equation}
The two zeros are reciprocal. In the positive-norm regime
$S_0>2|S_1|$, the interior root is
\begin{equation}
\zeta_g
=
\frac{\mu_G-\sqrt{\mu_G^2-4}}{2},
\qquad
\mu_G
=
\frac{S_0^2}{S_1^2}-2.
\label{eq:zetaGexact}
\end{equation}
The primitive projector root satisfies
$S(z_g)=S_0+S_1(z_g+z_g^{-1})=0$. Squaring this relation gives
$z_g^2+z_g^{-2}=\mu_G$, and hence the interior doubled-cell root is
\[
\zeta_g=z_g^2.
\]
Its physical decay length is therefore
\[
-\frac{2a}{\ln|\zeta_g|}
=
-\frac{a}{\ln|z_g|}
=
\xi_P.
\]

At the boundary-state frequency $\omega_f$, the projected
characteristic determinant factorizes as
\begin{equation}
\begin{aligned}
D_{\rm edge}(\zeta)
&=
\det[-\epsilon\mathcal T_{\rm d}(\zeta)G_d(\zeta)]
\\
&=
-\epsilon^2
(t_1+t_2\zeta)
(t_1+t_2\zeta^{-1})
D_G(\zeta).
\end{aligned}
\label{eq:edgeFactorization}
\end{equation}
The interior roots are the dimerization root
$\zeta_t=-t_1/t_2$ and the geometric root
$\zeta_g=z_g^2$.
Their contributions to a physical boundary-state component require
nonzero amplitudes.
\begin{equation}
\xi_t
=
-\frac{2a}{\ln|\zeta_t|}
=
-\frac{2a}{\ln|t_1/t_2|},
\qquad
\xi_P
=
-\frac{2a}{\ln|\zeta_g|}.
\label{eq:twoEdgeLengths}
\end{equation}
The overall strength $\epsilon$ controls the spectral splitting but
drops out of these root positions.

The relative weights follow directly from
$c=G_{\rm OBC}^{-1}d$. For a semi-infinite chain,
Eq.~\eqref{eq:dSSHedge} has the generating function
\[
d(\zeta)
=
\sum_{m=0}^{\infty}d_m\zeta^{-m}
=
\frac{\zeta}{\zeta-\zeta_t}e_x,
\qquad
e_x=(1,0)^{\mathsf T},
\]
so the physical coefficient function is
\begin{equation}
c(\zeta)=G_d^{-1}(\zeta)d(\zeta).
\label{eq:cGenerating}
\end{equation}
The pole at $\zeta_t$ is inherited from $d(\zeta)$, while the pole at
$\zeta_g$ arises from $G_d^{-1}(\zeta)$.

Let $r_g$ and $\ell_g$ be right and left null vectors of
$G_d(\zeta_g)$ and define
$\alpha_g=\ell_g^\dagger[\partial_\zeta G_d(\zeta_g)]r_g$. For
distinct simple roots, contour evaluation gives
\begin{equation}
c_m
\simeq
\zeta_t^mG_d^{-1}(\zeta_t)e_x
+
\zeta_g^m
\frac{
r_g\ell_g^\dagger e_x
}{
\alpha_g(\zeta_g-\zeta_t)
}.
\label{eq:cTwoRoots}
\end{equation}
Multiplication by the finite-range compact frame changes the
component-dependent residues but not the exponential roots. A selected
microscopic Nambu component therefore has the asymptotic form
\begin{equation}
\psi_m
\simeq
{\cal B}_t\zeta_t^m
+
{\cal B}_g\zeta_g^m.
\label{eq:PsiEdgeTwoRoots}
\end{equation}
The geometric contribution is present only when
$\ell_g^\dagger e_x\neq0$ and the selected microscopic component has
nonzero projection onto $W_d(\zeta_g)r_g$. When both amplitudes are
nonzero and $|\zeta_g|>|\zeta_t|$, the dimerization term controls the
shorter-range boundary structure while the geometric term controls the
asymptotic tail. Their crossover occurs approximately at
\[
m_\times
\simeq
\frac{\ln|{\cal B}_t/{\cal B}_g|}{\ln|\zeta_g/\zeta_t|}.
\]

\begin{figure}[t]
  \centering
  \includegraphics[width=0.98\linewidth]{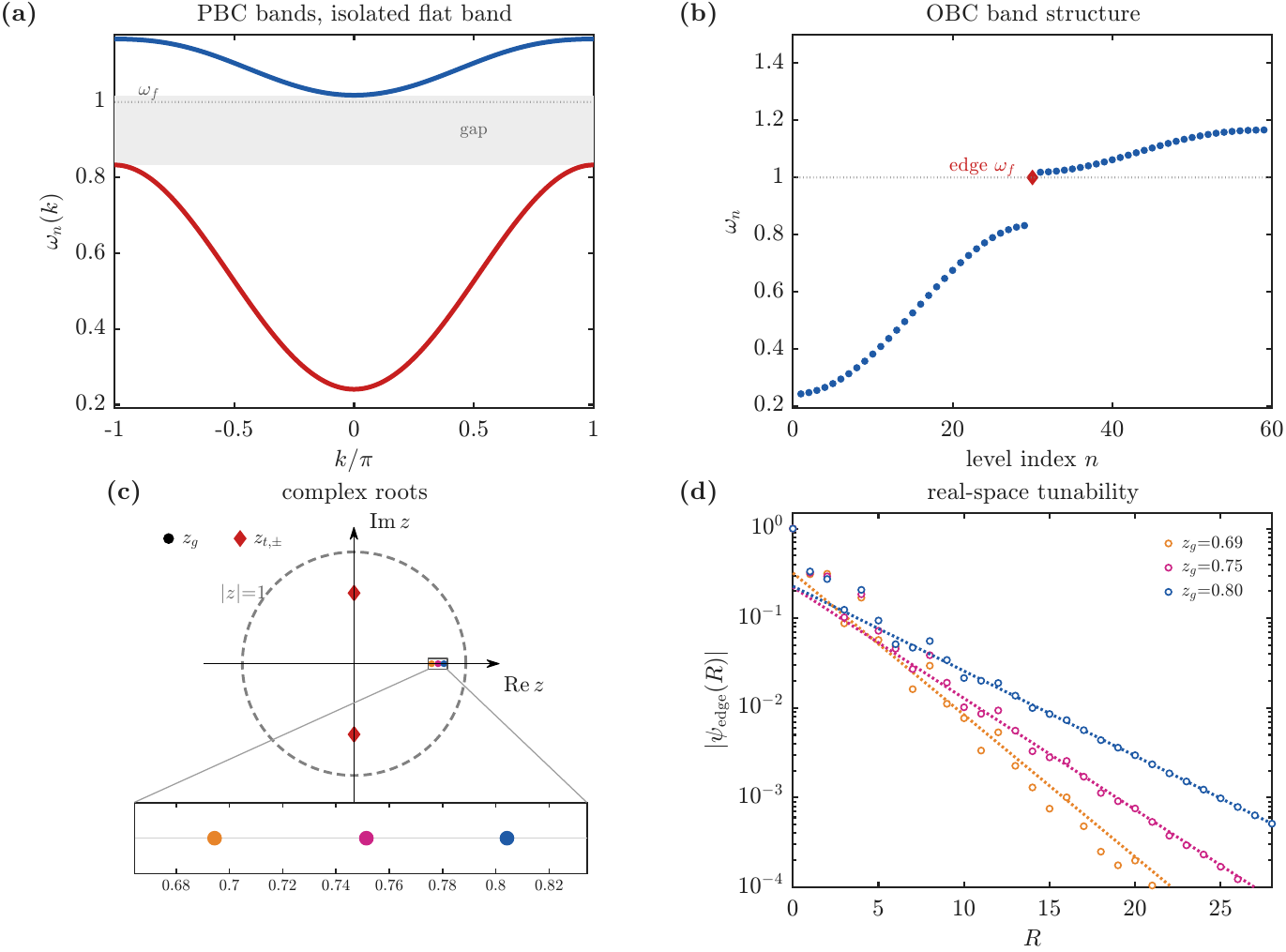}
  \caption{\textbf{Dimerized boundary state and decay lengths.}
A weak alternating (SSH-type) coupling of
strength $\epsilon=0.003$ between the compact flat-band generators keeps the
system stable and creates a boundary mode without any local defect.
(a) The bands in the PBC open a
gap around $\omega_f$.
(b) The OBC spectrum contains a boundary state at $\omega_f$ inside this
gap.
(c) In the complex-$z$ plane, the geometric roots
$z_g$ for the three parameter sets used in (d) are shown together
with the fixed dimerization pair
$z_{t,\pm}=\pm i\sqrt{t_1/t_2}$.
These correspond to the doubled-cell roots
$\zeta_g=z_g^2$ and $\zeta_t=-t_1/t_2=-0.4$ through
$\zeta=z^2$.
The dimerization length is
$\xi_t=-2a/\ln|\zeta_t|=2.18a$, and the three geometric roots
correspond to $\xi_P=2.74a,3.50a,4.59a$.
(d) Exact boundary profiles $|\psi_{\rm edge}(R)|$ and two-root
fits for $\xi_P=2.74a,3.50a,4.59a$.
The negative doubled-cell dimerization root $\zeta_t$ produces the
alternating near-boundary component, while the corresponding
geometric root controls the long-distance envelope.}
  \label{fig:Sdimerized}
\end{figure}

If $\zeta_t=\zeta_g\equiv\zeta_0$, the simple pole of $d(\zeta)$ and
the simple pole of $G_d^{-1}(\zeta)$ coincide. The physical coefficient function then has a second-order pole,
provided its numerator does not vanish, and
$\psi_m\simeq({\cal B}_0+m{\cal B}_1)\zeta_0^m$. The polynomial factor
changes the algebraic envelope but introduces no additional
exponential length. Figure~\ref{fig:Sdimerized} shows the boundary spectrum and the two
contributions to the spatial decay.

%---------------------------------------------------------------------

\section{Input-output formalism and filtered observables}
\label{sec:IOreadout}

The correlations derived above are equal-time properties of the
closed-system Bogoliubov vacuum. The experimentally measured output is
instead a steady-state open-system observable obtained after coupling
the lattice to external channels and collecting the outgoing fields in
a finite temporal mode. The same BdG modes and complex-momentum
structure enter both descriptions, but the output moments also depend
on damping, port geometry, and temporal filtering. We therefore keep
the closed-system covariance and the filtered output covariance
separate throughout.

A coherent probe and a noise measurement access different observables.
The coherent response is contained in first moments of the output
fields and resolves the driven signal together with its phase-conjugate
idler. Squeezing and Gaussian entanglement are properties of connected
second moments of the output fluctuations. Both are governed by the
same open-system resolvent, but only connected second moments can
certify squeezing or entanglement.

\subsection{Open bosonic BdG dynamics and port coupling}

Consider a finite device containing $D=NL$ physical annihilation
operators collected in the vector $\hat{\bm a}$, where $N$ is the
number of modes per unit cell and $L$ is the number of cells. We use
the Nambu field
$\hat{\bm\Psi}=(\hat{\bm a},\hat{\bm a}^{\dagger})^{\rm T}$ and the
metric $\Sigma_z=\operatorname{diag}(\mathbb I_D,-\mathbb I_D)$. The
quadratic Hamiltonian is
$\hat H=\tfrac12\hat{\bm\Psi}^{\dagger}H_{\rm BdG}\hat{\bm\Psi}+E_0$,
with dynamical matrix $M_{\rm dev}=\Sigma_zH_{\rm BdG}$. The label
``dev'' denotes either the translationally invariant parent or a finite
open device containing a defect.

Each physical mode is coupled locally to a monitored Markovian bath at
rate $\kappa_{\rm ext}$. An independent unmonitored vacuum bath with
rate $\kappa_{\rm int}$ accounts for internal loss, giving the total
linewidth $\kappa=\kappa_{\rm ext}+\kappa_{\rm int}$. We take these rates to be uniform over the lattice, while the measured
signal and idler are selected from the resulting output channels.

For each bath, define the Nambu input field
$\hat{\bm\Xi}_{\nu}^{\rm in}[\omega]
=(\hat{\bm b}_{\nu}^{\rm in}[\omega],
\hat{\bm b}_{\nu}^{{\rm in}\dagger}[-\omega])^{\rm T}$,
$\nu\in\{{\rm ext},{\rm int}\}$, using the Fourier convention
$\hat O(t)=\int(d\omega/2\pi)e^{-i\omega t}\hat O[\omega]$.
The upper Nambu block contains annihilation operators at detuning
$+\omega$, whereas the lower block contains creation operators at the
conjugate detuning $-\omega$. In a parametrically driven realization,
these are the signal and idler sidebands in the rotating frame in which
the BdG Hamiltonian is time independent.

The Heisenberg-Langevin equation is~\cite{Gardiner1985,Clerk2010}
\begin{equation}
\partial_t\hat{\bm\Psi}
=
\left(
-iM_{\rm dev}
-\frac{\kappa}{2}\mathbb I_{2D}
\right)
\hat{\bm\Psi}
+
\sqrt{\kappa_{\rm ext}}\,
\hat{\bm\Xi}_{\rm ext}^{\rm in}
+
\sqrt{\kappa_{\rm int}}\,
\hat{\bm\Xi}_{\rm int}^{\rm in}.
\label{eq:IOLangevinEquation}
\end{equation}
The monitored output satisfies the input-output boundary condition
\begin{equation}
\hat{\bm\Xi}_{\rm out}
=
\hat{\bm\Xi}_{\rm ext}^{\rm in}
-
\sqrt{\kappa_{\rm ext}}\,
\hat{\bm\Psi}.
\label{eq:IOBoundaryCondition}
\end{equation}
The drift matrix is
$\mathsf A=-iM_{\rm dev}-(\kappa/2)\mathbb I_{2D}$. Dynamical stability
requires every eigenvalue of $\mathsf A$ to have negative real part.
This open-system condition is distinct from energetic stability of the
closed quadratic Hamiltonian, $H_{\rm BdG}>0$.

Fourier transforming Eq.~\eqref{eq:IOLangevinEquation} gives
$\hat{\bm\Psi}[\omega]=\chi(\omega)
[\sqrt{\kappa_{\rm ext}}\hat{\bm\Xi}_{\rm ext}^{\rm in}[\omega]
+\sqrt{\kappa_{\rm int}}\hat{\bm\Xi}_{\rm int}^{\rm in}[\omega]]$, with
\begin{equation}
\chi(\omega)
=
\left[
i\left(M_{\rm dev}-\omega\mathbb I_{2D}\right)
+\frac{\kappa}{2}\mathbb I_{2D}
\right]^{-1}.
\label{eq:IOSusceptibility}
\end{equation}
This damped BdG resolvent contains both the normal response within the
annihilation sector and the anomalous response connecting annihilation
and creation sectors.

Substitution into Eq.~\eqref{eq:IOBoundaryCondition} yields
\begin{equation}
\hat{\bm\Xi}_{\rm out}
=
\mathcal S_{\rm ext}\hat{\bm\Xi}_{\rm ext}^{\rm in}
+
\mathcal S_{\rm int}\hat{\bm\Xi}_{\rm int}^{\rm in},
\label{eq:IOScatteringRelation}
\end{equation}
where
\begin{equation}
\mathcal S_{\rm ext}(\omega)
=
\mathbb I_{2D}-\kappa_{\rm ext}\chi(\omega),
\qquad
\mathcal S_{\rm int}(\omega)
=
-\sqrt{\kappa_{\rm ext}\kappa_{\rm int}}\,\chi(\omega).
\label{eq:IOScatteringMatrices}
\end{equation}
The first term contains direct reflection together with radiation from
the device. The second carries vacuum fluctuations entering through
unmonitored loss channels into the monitored output.

\subsection{Output covariance and filtered signal-idler modes}
\subsubsection{Vacuum output covariance}

For vacuum input, the only nonzero unsymmetrized bath correlation is
$\langle\hat{\bm\Xi}_{\nu}^{\rm in}[\omega]
\hat{\bm\Xi}_{\nu}^{\rm in\dagger}[\omega']\rangle
=2\pi\delta(\omega-\omega')J$, with
$J=\operatorname{diag}(\mathbb I_D,0)$. The upper block follows from
$\langle\hat b\hat b^\dagger\rangle=1$, while the lower block vanishes
because $\langle\hat b^\dagger\hat b\rangle=0$ in vacuum. External and
internal baths are independent.

Suppressing the overall frequency delta function, the monitored output
spectral covariance is
\begin{equation}
\mathcal G_{\rm out}(\omega)
=
\mathcal S_{\rm ext}(\omega)
J
\mathcal S_{\rm ext}^{\dagger}(\omega)
+
\mathcal S_{\rm int}(\omega)
J
\mathcal S_{\rm int}^{\dagger}(\omega).
\label{eq:IOOutputCovariance}
\end{equation}
The upper annihilation block contains the vacuum identity in addition
to the normally ordered occupation, so this identity must be subtracted
when extracting $N$. The off-diagonal Nambu block gives the anomalous
signal-idler correlations generated by the parametric couplings. In
this unsymmetrized convention the vacuum term is explicit here and
reappears as the $1/2$ reference level of the quadrature variance below.

\subsubsection{Conjugate filtered signal and idler modes}

A measurement selects a normalized temporal mode rather than an
infinitely sharp frequency. Let $\omega_0>0$ denote the center of a
bulk or defect resonance and $\Omega$ its detuning. The filtered signal
at the $A$ output in cell $0$ is
\begin{equation}
\bar b_s
=
\int\frac{d\Omega}{2\pi}\,
h(\Omega)
b_{A,0}^{\rm out}[\omega_0+\Omega],
\label{eq:IOFilteredSignal}
\end{equation}
and the conjugate idler at the $B$ output in cell $R$ is
\begin{equation}
\bar b_i
=
\int\frac{d\Omega}{2\pi}\,
h^{*}(\Omega)
b_{B,R}^{\rm out}[-\omega_0-\Omega].
\label{eq:IOFilteredIdler}
\end{equation}
The normalization
$\int(d\Omega/2\pi)|h(\Omega)|^2=1$ ensures canonical commutation
relations for the filtered modes. For the filtered results in Fig.~3, we use the normalized Gaussian
frequency weight
\[
w(\Omega)
=
\frac{1}{\sqrt{2\pi}\sigma_f}
\exp\!\left[-\frac{\Omega^2}{2\sigma_f^2}\right].
\]

The opposite sideband and complex-conjugate filter in
Eq.~\eqref{eq:IOFilteredIdler} follow directly from the Nambu
convention. At Nambu frequency $\omega_0+\Omega$, the selected lower
component is
$b_{B,R}^{{\rm out}\dagger}[-\omega_0-\Omega]$, which represents
$\bar b_i^\dagger$. The corresponding two-component filtered Nambu
field is therefore
$\hat{\bm B}_{si}=(\bar b_s,\bar b_i^\dagger)^{\rm T}$.

We define connected moments
$N_{\mu\nu}=\langle\delta\bar b_\mu^\dagger
\delta\bar b_\nu\rangle_{\rm ss}$ and
$M_{\mu\nu}^{\rm out}=\langle\delta\bar b_\mu
\delta\bar b_\nu\rangle_{\rm ss}$, where
$\delta\bar b_\mu=\bar b_\mu-\langle\bar b_\mu\rangle_{\rm ss}$ and
$\mu,\nu\in\{s,i\}$. In particular,
$M_{si}^{\rm out}(R)=\langle\delta\bar b_s
\delta\bar b_i\rangle_{\rm ss}$ is the anomalous output moment used in
the main text. The normalized frequency weight is
$w(\Omega)=|h(\Omega)|^2/(2\pi)$, with
$\int d\Omega\,w(\Omega)=1$.

Projecting Eq.~\eqref{eq:IOOutputCovariance} onto the selected signal
annihilation and idler creation components gives
$\mathcal C_{si}^{\rm out}(R)
=\langle\delta\hat{\bm B}_{si}
\delta\hat{\bm B}_{si}^{\dagger}\rangle
=\left(\begin{smallmatrix}1+N_{ss}&M_{si}^{\rm out}(R)\\
[M_{si}^{\rm out}(R)]^*&N_{ii}(R)\end{smallmatrix}\right)$, with
$N_{ss}=\int d\Omega\,w(\Omega)
[\mathcal G_{{\rm out},11}(\omega_0+\Omega)-1]$,
$N_{ii}(R)=\int d\Omega\,w(\Omega)
\mathcal G_{{\rm out},22}^{(R)}(\omega_0+\Omega)$, and
$M_{si}^{\rm out}(R)=\int d\Omega\,w(\Omega)
\mathcal G_{{\rm out},12}^{(R)}(\omega_0+\Omega)$. 

For the model considered here, the BdG dynamical matrix separates into
$(A\text{-annihilation},B\text{-creation})$ and
$(B\text{-annihilation},A\text{-creation})$ sectors, while the signal
and idler filters select nonoverlapping conjugate sidebands. These
properties give
\begin{equation}
N_{si}=N_{is}=0,
\qquad
M_{ss}^{\rm out}=M_{ii}^{\rm out}=0.
\label{eq:IOSelectionRules}
\end{equation}
Thus the selected pair is characterized by $N_{ss}$, $N_{ii}$, and
$M_{si}^{\rm out}$. These are selection rules of the present
block-decoupled model, not generic identities for Gaussian output
fields.

\subsection{Collective squeezing and two-mode entanglement}
\subsubsection{Collective quadrature squeezing}

To probe the nonlocal anomalous moment, combine the signal and idler
into the equal-weight mode
$\bar b_{\rm col}(\phi)=(\bar b_s+e^{i\phi}\bar b_i)/\sqrt2$.
Its normal and anomalous moments are
$N_{\rm col}(\phi)=\tfrac12[N_{ss}+N_{ii}+e^{i\phi}N_{si}
+e^{-i\phi}N_{is}]$ and
$M_{\rm col}^{\rm out}(\phi)=\tfrac12[M_{ss}^{\rm out}
+e^{2i\phi}M_{ii}^{\rm out}
+e^{i\phi}(M_{si}^{\rm out}+M_{is}^{\rm out})]$. For the quadrature
$X_{\vartheta,\phi}=[e^{-i\vartheta}\bar b_{\rm col}(\phi)
+e^{i\vartheta}\bar b_{\rm col}^{\dagger}(\phi)]/\sqrt2$, the connected
variance is
$\langle\delta X_{\vartheta,\phi}^{2}\rangle
=\tfrac12+N_{\rm col}(\phi)
+\operatorname{Re}[e^{-2i\vartheta}M_{\rm col}^{\rm out}(\phi)]$.

Using Eq.~\eqref{eq:IOSelectionRules} and optimizing the relative
collection phase and the homodyne phase gives
\begin{equation}
V_{\min}^{\rm out}(R)
=
\frac12
+
\frac{N_{ss}+N_{ii}(R)}{2}
-
|M_{si}^{\rm out}(R)|.
\label{eq:IOOptimizedSqueezing}
\end{equation}
The vacuum variance is $1/2$, so the filtered collective mode is
sub-vacuum squeezed when
\begin{equation}
|M_{si}^{\rm out}(R)|
>
\frac{N_{ss}+N_{ii}(R)}{2}.
\label{eq:IOSqueezingCondition}
\end{equation}
With $\bar N(R)=[N_{ss}+N_{ii}(R)]/2$, the below-vacuum depth is
$D_{\rm out}(R)=\tfrac12-V_{\min}^{\rm out}(R)
=|M_{si}^{\rm out}(R)|-\bar N(R)$. For the translationally invariant
parent, $N_{ss}=N_{ii}\equiv N_0$, so all spatial dependence of the
optimized variance is carried by the anomalous intercell moment. In
the defect geometry, the signal port is fixed at the defect cell while
$N_{ii}(R)$ can vary with the position of the idler port.

\subsubsection{Two-mode Gaussian entanglement}

Collective-mode squeezing and bipartite entanglement are distinct. For
the selected signal and idler modes, define
$X_\mu=(\bar b_\mu+\bar b_\mu^\dagger)/\sqrt2$ and
$P_\mu=(\bar b_\mu-\bar b_\mu^\dagger)/(i\sqrt2)$.
After local phase rotations that make $M_{si}^{\rm out}$ real and
positive, the covariance matrix in the ordering
$(X_s,P_s,X_i,P_i)$ is
\begin{equation}
V
=
\begin{pmatrix}
a&0&c&0\\
0&a&0&-c\\
c&0&b&0\\
0&-c&0&b
\end{pmatrix},
\label{eq:IOStandardCovariance}
\end{equation}
with $a=N_{ss}+\tfrac12$, $b=N_{ii}+\tfrac12$, and
$c=|M_{si}^{\rm out}|$. The opposite signs in the correlation block
follow from the anomalous correlation between the two annihilation
operators.

Partial transposition reverses the idler momentum quadrature. With
$\Delta^{\rm PT}=a^2+b^2+2c^2$ and
$\det V=(ab-c^2)^2$, the smallest symplectic eigenvalue of the partially
transposed covariance is
\begin{equation}
\nu_-^{\rm PT}
=
\sqrt{
\frac{
\Delta^{\rm PT}
-
\sqrt{(\Delta^{\rm PT})^2-4\det V}
}{2}
}.
\label{eq:IOPTSymplecticEigenvalue}
\end{equation}
The two-mode Gaussian PPT criterion~\cite{Weedbrook2012} is
$\nu_-^{\rm PT}<1/2$. For the standard form above, the threshold
factorizes as
$(N_{ss}N_{ii}-|M_{si}^{\rm out}|^2)
[N_{ss}N_{ii}+N_{ss}+N_{ii}+1-|M_{si}^{\rm out}|^2]=0$.
The second factor is nonnegative for a physical covariance matrix, so
the entanglement condition reduces to
\begin{equation}
|M_{si}^{\rm out}|>\sqrt{N_{ss}N_{ii}}.
\label{eq:IOSimplePPTCriterion}
\end{equation}

For $N_{ss}=N_{ii}$, as in the translationally invariant bulk, the PPT
and equal-weight collective-squeezing thresholds coincide. In the
defect geometry the occupations need not be equal, and
$[N_{ss}+N_{ii}]/2\geq\sqrt{N_{ss}N_{ii}}$. Two-mode entanglement can
therefore survive after equal-weight collective squeezing is lost. The
PPT criterion refers only to the selected filtered signal-idler pair
and does not witness multipartite entanglement of the full lattice.
\section{Bulk and defect output response}

We next apply the input-output formalism to the flat parent and defect
geometries. The intrinsic flat-band pole and the defect roots derived in
Secs.~\ref{sec:pairing_flat_band} and~\ref{sec:defect_state} enter
through the damped resolvent, while linewidth, port coupling, and
filtering determine how they appear in the measured output.

\subsection{Exactly flat bulk response}

For the translationally invariant flat-band model, the output moments
can be evaluated analytically. In the active Nambu sector
$\hat{\bm\Phi}_k=(\hat a_{A,k},\hat a_{B,-k}^{\dagger})^{\rm T}$, the
dynamical matrix is
\begin{equation}
M_{AB}(k)
=
\begin{pmatrix}
\omega_f+P_k&q_k\\
-q_k^{*}&\omega_f-T_k
\end{pmatrix},
\label{eq:IOActiveBulkMatrix}
\end{equation}
where $P_k=|p_k|^2$, $T_k=|s_k|^2$, and $q_k=p_ks_k$. For the general
four-coefficient compact frame,
$p_k=\sqrt g\,(\lambda_0+\lambda_1e^{ik})$ and
$s_k=\sqrt g\,(c_0+c_1e^{-ik})$. The commutator overlap is
\begin{equation}
S_k=T_k-P_k=S_0+2S_1\cos k,
\label{eq:IOOverlapRealMomentum}
\end{equation}
with
$S_0=g(c_0^2+c_1^2-\lambda_0^2-\lambda_1^2)$ and
$S_1=g(c_0c_1-\lambda_0\lambda_1)$. The two eigenvalues of this active
block are $\omega_f$ and $\omega_f-S_k$. The second is the
negative-frequency Nambu partner of the positive dispersive band
$S_k-\omega_f$ in the complementary block.

At the center of the flat-band resonance, $\omega=\omega_f$, set
$\gamma=\kappa/2$. The inverse susceptibility in the active sector is
$\chi_{AB}^{-1}(k,\omega_f)=\left(\begin{smallmatrix}
\gamma+iP_k&iq_k\\-iq_k^{*}&\gamma-iT_k
\end{smallmatrix}\right)$. Since $|q_k|^2=P_kT_k$,
$\det\chi_{AB}^{-1}=\gamma(\gamma-iS_k)$, and hence
$\chi_{AB}(k,\omega_f)=\frac{1}{\gamma(\gamma-iS_k)}
\left(\begin{smallmatrix}\gamma-iT_k&-iq_k\\
iq_k^{*}&\gamma+iP_k\end{smallmatrix}\right)$.

Substitution into Eq.~\eqref{eq:IOScatteringRelation} gives the exact
momentum-resolved output moments. With escape efficiency
$\eta_{\rm esc}=\kappa_{\rm ext}/\kappa$,
\begin{equation}
N_s^{\rm out}(k)
=
N_i^{\rm out}(k)
=
\eta_{\rm esc}\frac{4P_kT_k}{S_k^2+\gamma^2}.
\label{eq:IOExactBulkOccupation}
\end{equation}
The equality follows from the Bogoliubov scattering structure for
vacuum input. Internal loss leaves a single factor $\eta_{\rm esc}$, rather than
$\eta_{\rm esc}^2$, because vacuum entering through the unmonitored bath is also
converted by the device and contributes to the monitored output.

The anomalous output moment is
\begin{equation}
M_{si}^{\rm out}(k)
=
-\eta_{\rm esc}
\frac{[2H_k+i\kappa]q_k}{S_k^2+\gamma^2},
\qquad
H_k=T_k+P_k.
\label{eq:IOExactBulkAnomalousMoment}
\end{equation}
The two terms in the numerator arise from the resonant Bogoliubov
response together with the direct-reflection term in the input-output
boundary condition. The term proportional to $i\kappa$ vanishes as
$\kappa\to0$.

The filtered real-space moment follows from the full frequency-dependent
spectrum,
\[
M_{si}^{{\rm out},\,\rm bulk}(R)
=
\int_{-\pi}^{\pi}\frac{dk}{2\pi}e^{-ikR}
\int d\Omega\,w(\Omega)
M_{si}^{\rm out}(k,\omega_f+\Omega).
\]

\subsection{Complex-momentum poles of the resonant bulk output}

With the continuation $z=e^{-ik}$, the closed-system anomalous
covariance has a pole at the interior zero $z_S$ of
$S(z)=S_0+S_1(z+z^{-1})$ introduced in
Sec.~\ref{sec:pairing_flat_band}. Its exponential decay length is
$\xi_S=-a/\ln|z_S|$.

At finite linewidth, the resonant kernel contains
$[S^2(z)+\gamma^2]^{-1}
=\{[S(z)-i\gamma][S(z)+i\gamma]\}^{-1}$. The two nearby output poles
therefore satisfy
$S(z_\pm)=\pm i\gamma$. For $S'(z_S)\neq0$ and small linewidth,
$z_\pm=z_S\pm i\gamma/S'(z_S)+O(\gamma^2)$. The corresponding output
decay length thus approaches $\xi_S$ as $\kappa\rightarrow0$.

The measured finite-bandwidth covariance averages the
frequency-dependent poles near $\omega_f$ over $w(\Omega)$. Its decay
therefore approaches the intrinsic closed-system length only in the
joint weak-linewidth and narrow-filter limit. At finite $\kappa$ or
finite filter bandwidth, the output covariance is a distinct
open-system observable and need not have exactly the length $\xi_S$.

A pole contributes to a selected output correlation only when its
residue in that channel is nonzero. For the $A$-annihilation to
$B$-creation correlation of
Eq.~\eqref{eq:IOExactBulkAnomalousMoment}, the poles at $z_\pm$
contribute provided
\[
\left[2H(z_\pm)+i\kappa\right]q(z_\pm)\neq0 .
\]
When this condition holds, the long-distance response is determined by
the contributing poles nearest the unit circle. In the weak-linewidth
limit these poles approach $z_S$.

\subsection{Weak-pairing analytical estimates}

Analytical estimates for the output squeezing are particularly simple
in the reduced parametrization
$\lambda_0=\lambda_1\equiv\lambda$, $c_0\equiv c$, and $c_1=1$.
We take $c<-1$, $0<\lambda\ll1$, and
$\gamma\ll S_{\min}\equiv\min_kS_k$. Then $T_k=O(g)$ and
$P_k=O(g\lambda^2)$. These formulas isolate the competition between the
nonlocal anomalous moment and the local occupation. 

To leading nonvanishing order,
Eq.~\eqref{eq:IOExactBulkOccupation} gives
$N_s^{\rm out}(k)=N_i^{\rm out}(k)
=4\eta_{\rm esc} P_k/T_k
+O(\lambda^4,\lambda^2\gamma^2/S_{\min}^2)$, or
$4\eta_{\rm esc}\lambda^2(1+z)(1+z^{-1})/[(c+z)(c+z^{-1})]$ at leading order.
The local occupation is the zeroth Fourier coefficient. For $|c|>1$,
the enclosed poles are at $z=0$ and $z=-1/c$, which gives
\begin{equation}
N_0
=
\frac{8\eta_{\rm esc}\lambda^2}{c(c+1)}
+
O(\lambda^4,\lambda^2\gamma^2/S_{\min}^2).
\label{eq:IOWeakOccupation}
\end{equation}

At the same order,
Eq.~\eqref{eq:IOExactBulkAnomalousMoment} becomes
$M_{si}^{\rm out}(k)=-2\eta_{\rm esc} q_k/T_k
+O(\lambda^3,\lambda\gamma/S_{\min})$. With $z=e^{-ik}$,
$M_{si}^{\rm out}(z)=-2\eta_{\rm esc}\lambda(1+z)/(1+cz)$ at leading order.
For $c<-1$, the interior pole is $z_S=-1/c+O(\lambda^2)$, and contour
integration gives, for $R\geq1$,
\begin{equation}
M_{si}^{{\rm out},\,\rm bulk}(R)
=
2\eta_{\rm esc}\lambda(-1)^R\frac{c-1}{c^{R+1}}
=
2\eta_{\rm esc}\lambda\frac{|c|+1}{|c|^{R+1}}
+
O(\lambda^3,\lambda\gamma/S_{\min}).
\label{eq:IOWeakAnomalousRealSpace}
\end{equation}
The leading term is positive on this branch and decays with the
projector root $z_S$.

Combining Eqs.~\eqref{eq:IOWeakOccupation} and
\eqref{eq:IOWeakAnomalousRealSpace}, the below-vacuum depth at a fixed
separation is
\begin{align}
D_{\rm out}(R)
&\equiv
\frac12-V_{\min}^{\rm out}(R)
\nonumber\\
&\simeq
\eta_{\rm esc}\left[
\frac{2\lambda(|c|+1)}{|c|^{R+1}}
-
\frac{8\lambda^2}{|c|(|c|-1)}
\right],
\qquad c<-1.
\label{eq:IOWeakSqueezingDepth}
\end{align}
The anomalous contribution is linear in $\lambda$, whereas the local
occupation begins at order $\lambda^2$. Pairing therefore initially
increases the squeezing depth, but the accompanying occupation
ultimately reduces it.

At fixed $c$ and $R$, maximizing
Eq.~\eqref{eq:IOWeakSqueezingDepth} gives
\begin{equation}
\lambda_{\rm opt}
=
\frac{|c|^2-1}{8|c|^R},
\label{eq:IOLambdaOptimum}
\end{equation}
and
\begin{equation}
D_{\rm out}^{\max}
=
\eta_{\rm esc}
\frac{(|c|+1)^2(|c|-1)}{8|c|^{2R+1}}.
\label{eq:IOMaximumDepth}
\end{equation}
The parameter $c$ controls both the decay length and the residue.
Moving $|c|$ toward unity increases the geometric length but reduces
the separation from the remaining positive-frequency band. Before
imposing spectral isolation, the stationary condition for
Eq.~\eqref{eq:IOMaximumDepth} is
$2/(|c|+1)+1/(|c|-1)-(2R+1)/|c|=0$. For $R>1$ the physical solution is
\begin{equation}
|c_*|
=
\frac{-1+\sqrt{16R^2-8R-7}}{4(R-1)},
\qquad
c_*=-|c_*|.
\label{eq:IOUnconstrainedCOptimum}
\end{equation}

The positive frequencies are $\omega_f$ and $S_k-\omega_f$, so the
minimum spectral separation is
$\Delta_{\rm gap}=\min_k[S_k-2\omega_f]$. On the $c<-1$ branch the
minimum occurs at $k=0$, where
$\min_kS_k=g[(c+1)^2-4\lambda^2]$. Requiring
$\Delta_{\rm gap}\geq\Delta_{\rm res}$ gives the exact reduced-model
constraint
$|c|\geq1+\sqrt{4\lambda^2+(2\omega_f+\Delta_{\rm res})/g}$.
At leading order in the weak-pairing limit this becomes
\begin{equation}
|c|
\geq
|c_{\rm iso}|
=
1+
\sqrt{\frac{2\omega_f+\Delta_{\rm res}}{g}},
\qquad
c_{\rm iso}=-|c_{\rm iso}|.
\label{eq:IOIsolationBound}
\end{equation}
Here $\Delta_{\rm res}$ is the required spectral margin relative to the
linewidth and filter bandwidth. Combining this leading-order bound with
the unconstrained optimum gives
\begin{equation}
|c|
=
\max(|c_*|,|c_{\rm iso}|),
\qquad
c=-|c|,
\qquad
\lambda=\lambda_{\rm opt}(c).
\label{eq:IOConstrainedDesign}
\end{equation}

\subsection{Defect output response}

The defect frequency and its spatial roots were derived in
Sec.~\ref{sec:defect_state}. At $t_f=0$, a localized mode at frequency
$\omega_d$ has one interior bulk root $z_d$. Weak nonzero $t_f$ can
produce two contributing interior roots,
$z_{\rm short}$ and $z_{\rm long}$, obtained from
$D_2(\omega_d,z)=0$, with
$|z_{\rm short}|<|z_{\rm long}|<1$. Their lengths are
$\xi_j=-a/\ln|z_j|$. Which roots appear in a measured channel is fixed
by their residues rather than by the characteristic equation alone.

The defect susceptibility is
\[
\chi_{\rm def}(\omega)
=
\left[
i(M_{\rm def}-\omega\mathbb I)
+\frac{\kappa}{2}\mathbb I
\right]^{-1}.
\]
The signal is selected from the $A$-annihilation component at the
defect cell and the idler from the $B$-creation component at a cell
displaced by $R$.

Writing
$M_{\rm eff}=M_{\rm def}-i(\kappa/2)\mathbb I$ gives
$\chi_{\rm def}(\omega)=i[\omega\mathbb I-M_{\rm eff}]^{-1}$.
Because the damping is a scalar shift, $M_{\rm eff}$ has the same
biorthogonal eigenvectors as $M_{\rm def}$. For
$M_{\rm def}|R_n\rangle=\omega_n|R_n\rangle$,
$\langle L_n|M_{\rm def}=\omega_n\langle L_n|$, and
$\langle L_n|R_m\rangle=\delta_{nm}$, an isolated defect resonance has
\[
\chi_{\rm def}(\omega)
\simeq
\frac{i|R_d\rangle\langle L_d|}
{\omega-\omega_d+i\kappa/2}
+
\chi_{\rm reg}(\omega).
\]
When the signal port is fixed near the defect and the idler port is
moved to cell $R$, the spatial dependence follows the corresponding
component of the right defect eigenvector. In the unsymmetrized
covariance of Eq.~\eqref{eq:IOOutputCovariance}, the idler
creation-sector row enters complex conjugated. The anomalous output
moment therefore has the asymptotic form
\begin{equation}
M_{si}^{{\rm out},\,\rm def}(R)
\simeq
\sum_{j\in\mathcal Z_d}\mathcal A_j (z_j^*)^R
+
M_{\rm reg}(R),
\qquad
\mathcal Z_d=
\begin{cases}
\{z_d\},&t_f=0,\\
\{z_{\rm short},z_{\rm long}\},&t_f\neq0,
\end{cases}
\label{eq:IODefectAnomalousTail}
\end{equation}
for the cases considered here. The coefficients $\mathcal A_j$ contain the mode overlaps with the
input and output channels together with the input-output factors and
the temporal filter. A root contributes to the selected correlation
only when the corresponding coefficient $\mathcal A_j$ is nonzero.
Complex conjugation changes the spatial phase but not the exponential
decay length since $|z_j^*|=|z_j|$.

The isolated-pole form is accurate when the defect resonance is
separated from the other damped poles by more than the linewidth and
the filter bandwidth. 

\subsection{Coherent signal and phase-conjugate idler}

A coherent first-moment measurement probes the same susceptibility but
not the vacuum covariance. Drive the annihilation component of the
$A$ port in cell $0$ with a weak monochromatic amplitude $\alpha$ at
frequency $\omega_0$. After subtracting
the direct reflection at the driven port, the device-mediated output is
$-\kappa_{\rm ext}\chi(\omega_0)\alpha$.

We denote its $A$-annihilation component at cell $R$ by
$\beta_{\rm sig}(R)$ and its $B$-creation component by
$\beta_{\rm id}^{*}(R)$. The latter is the Nambu representation of the
phase-conjugate physical idler amplitude. For the translationally
invariant bulk,
\begin{align}
\beta_{\rm sig}(R)
={}&
-\kappa_{\rm ext}\alpha
\int_{-\pi}^{\pi}\frac{dk}{2\pi}e^{ikR}
[\chi(k,\omega_0)]_{A_{\rm ann},A_{\rm ann}},
\nonumber\\
\beta_{\rm id}^{*}(R)
={}&
-\kappa_{\rm ext}\alpha
\int_{-\pi}^{\pi}\frac{dk}{2\pi}e^{ikR}
[\chi(k,\omega_0)]_{B_{\rm cre},A_{\rm ann}}.
\label{eq:IOBulkFirstMoments}
\end{align}
For a finite defect system, the corresponding amplitudes are the
real-space matrix elements of $\chi_{\rm def}(\omega_d)$.

Near an isolated defect resonance, the same spatial roots appear in
both projections,
\begin{align}
\beta_{\rm sig}(R)
&\simeq
\sum_{j\in\mathcal Z_d}B_{{\rm sig},j}z_j^R,
\nonumber\\
\beta_{\rm id}^{*}(R)
&\simeq
\sum_{j\in\mathcal Z_d}B_{{\rm id},j}z_j^R.
\label{eq:IODefectFirstMomentTail}
\end{align}
The coefficients need not be equal, and a root can be absent from one
projection when its coefficient vanishes. For the exactly flat defect
used in the main text, $\mathcal Z_d=\{z_d\}$ and both projections
therefore share the same single exponential length.

\begin{figure}[t]
  \centering
  \includegraphics[width=0.98\linewidth]{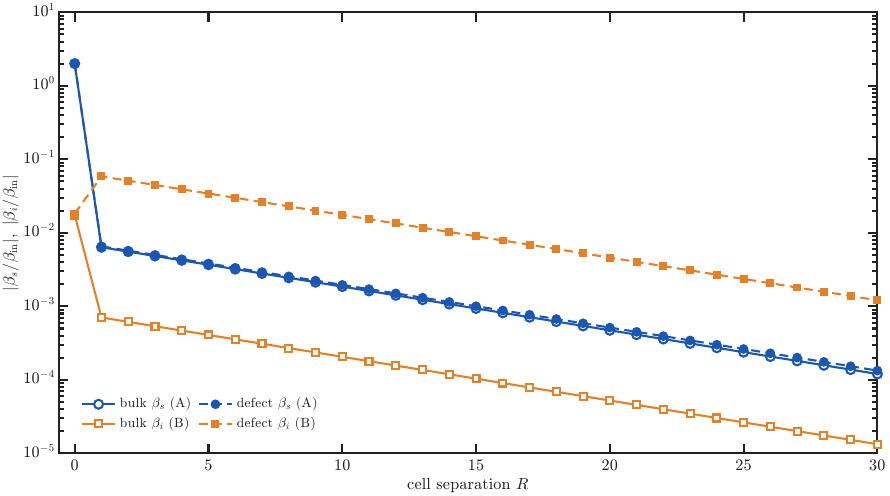}
 \caption{\textbf{Coherent signal and phase-conjugate idler response.}
Device-mediated first-moment amplitudes
$|\beta_{\rm sig}(R)/\alpha|$ in the $A$-annihilation sector and
$|\beta_{\rm id}(R)/\alpha|$ in the $B$-creation sector for a coherent
drive applied to the $A,0$ port. The direct contribution at the driven
port is subtracted. Bulk and defect responses are evaluated at
$\omega_f$ and $\omega_d$, respectively. At $t_f=0$, the bulk response
is governed by the finite-linewidth susceptibility poles continuously
connected to $z_S$, while the defect response follows the spatial root
$z_d$. The corresponding decay lengths are
$\xi_{\rm bulk}=7.30a$ and $\xi_d=7.48a$. The annihilation and creation
components in each geometry have the same exponential decay length.
The parameters are $g=32.5$, $\omega_f=1$, $\lambda_0=0.017$,
$\lambda_1=0.043$, $c_0=-2.037$, $c_1=1.77$,
$\kappa_{\rm ext}=0.02$, $\kappa_{\rm int}=0$, $t_f=0$,
$v_d=-0.10$, and $\omega_d\simeq0.900$. These first moments probe the
same spatial structure as the second-moment response in main-text
Fig.~3(b), but do not certify squeezing or entanglement.}
\label{fig:S4coherent}
\end{figure}

These coherent amplitudes probe the driven mode profile and its
phase-conjugate component, but do not by themselves certify squeezing
or entanglement. Figure~\ref{fig:S4coherent} shows the coherent signal and idler
responses for the bulk and defect geometries.

\end{document}